\documentclass[structabstract]{aa}  
\usepackage{graphicx}
\usepackage{aalongtable}
\usepackage{txfonts}
\usepackage{subfigure}
\usepackage{natbib}
\begin{document}
   \title{The nuclear star cluster of the Milky Way: Proper motions
     and mass}

   \subtitle{}

   \author{R. Sch\"odel
          \inst{1}
          \and
          D. Merritt\inst{2}
          \and
          A. Eckart\inst{3}
          }

   \institute{Instituto de Astrof\'isica de Andaluc\'ia (IAA) - CSIC,
              Camino Bajo de Hu\'etor 50, E-18008 Granada, Spain\\
              \email{rainer@iaa.es}
         \and
             Department of Physics and Center for Computational
             Relativity and Gravitation, Rochester Institute of Technology,
             Rochester, NY 14623, USA\\
             \email{merritt@astro.rit.edu}
         \and
             I.Physikalisches Institut, Universit\"at zu K\"oln, Z\"ulpicher Str.77, 50937 K\"oln, Germany\\
             \email{eckart@ph1.uni-koeln.de}
             }

   \date{}

% \abstract{}{}{}{}{} 
% 5 {} token are mandatory
   \abstract
{Nuclear star clusters (NSCs) are located at the photometric and
  dynamical centers of the majority of galaxies. They are among the densest
  star clusters in the Universe. The NSC in the Milky Way is the
  only object of this class that can be resolved into individual
  stars. The massive black hole Sagittarius\,A* is located at the
  dynamical center of the Milky Way NSC.}
{In this work we examine the proper motions of stars out to distances
  of 1.0\,pc from Sgr\,A*. The aim is to examine the velocity
  structure of the MW NSC and acquire a reliable estimate of the
  stellar mass in the central parsec of the MW NSC, in addition to the
  well-known black hole mass.}
{We use multi-epoch adaptive optics assisted near-infrared
  observations of the central parsec of the Galaxy obtained with
  NACO/CONICA at the ESO VLT. Stellar positions are measured via PSF
  fitting in the individual images and transformed into a common
  reference frame via suitable sets of reference stars.}
{We measured the proper motions of
more than 6000 stars within $\sim$1.0\,pc of Sagittarius\,A*. The full
data set is provided in this work.  We largely exclude the known
early-type stars with their peculiar dynamical properties from the
dynamical analysis.  The cluster is found to rotate parallel to
Galactic rotation, while the velocity dispersion appears isotropic
{(or anisotropy may be masked by the cluster rotation)}. The
Keplerian fall-off of the velocity dispersion due to the point mass of
Sgr\,A* is clearly detectable only at
$R\lesssim0.3$\,pc. Nonparametric isotropic and anisotropic Jeans
models are applied to the data.  They imply a best-fit black hole mass
of $3.6^{+0.2}_{-0.4}\times10^{6}$\,M$_{\odot}$. Although this value
is slightly lower than the current canonical value of
$4.0\times10^{6}$\,M$_{\odot}$, this is the first time that a proper
motion analysis provides a mass for Sagittarius\,A* that is consistent
with the mass inferred from orbits of individual stars. The point mass
of Sagittarius\,A* is not sufficient to explain the velocity data.  In
addition to the black hole, the models require the presence of an
extended mass of $0.5-1.5\times 10^{6}\,M_{\odot}$ in the central
parsec. This is the first time that the extended mass of the nuclear
star cluster is unambiguously detected. The influence of the extended
mass on the gravitational potential becomes notable at distances
$\gtrsim0.4$\,pc from Sgr\,A*. Constraints on the distribution of this
extended mass are weak. The extended mass can be explained well by the
mass of the stars that make up the cluster.}  {}


   \maketitle
%
%________________________________________________________________

\section{Introduction}

After a decade of sensitive high resolution imaging with the
\emph{Hubble Space Telescope} the presence of nuclear star clusters
(NSCs) at the centers of most galaxies has become a
well established observational fact
\citep{Phillips1996AJ,Carollo1998AJ,Matthews1999AJ,Cote2006ApJS}. NSCs
have typical effective radii of a few pc, luminosities of
$10^6-10^7\,L_{\odot}$, and masses of a few times $10^{5}$ to
$10^{7}\,M_{\odot}$ \citep
[e.g.~][]{Walcher2005ApJ,Ferrarese2006ApJ}. NSCs are the densest known
star clusters in the Universe. Most NSCs contain a mixed stellar
population with signs of repeated episodes of star formation
\citep{Walcher2006ApJ}. Recent research suggests that there exists a
fundamental relation between NSCs, supermassive black holes, and their
host galaxies
\citep{WehnerHarris2006ApJ,Ferrarese2006ApJ,Balcells2007ApJ,Seth2008ApJ},
similar to the relations between bulge luminosity, mass, or velocity
dispersion and supermassive black hole masses
\citep{Kormendy1995ARA&A,FerrareseMerritt2000ApJ,Gebhardt2000ApJ,Tremaine2002ApJ,HaeringRix2004ApJ}.
The causes for these correlations are not understood, which emphasizes
our need to obtain a better understanding of these objects,
Unfortunately, NSCs are compact sources and therefore barely resolved
in external galaxies at the diffraction limit of current 8-10\,m-class
and even future 30-50\,m-class telescopes. Any conclusions on the
structure and mass of extragalactic NSCs therefore have to be based on
the properties of the integrated light of millions of stars.

Located at a distance of only 8\,kpc
\citep{Reid1993ARA&A,Eisenhauer2005ApJ,Groenewegen2008A&A,Ghez2008ApJ,Trippe2008A&A,Gillessen2008arXiv},
the center of the Milky Way (Galactic Center, GC) offers the best
possibility to study an NSC in detail.  The GC is obscured by about
30\,magnitudes of visual extinction and can therefore only be studied
in infrared wavelengths \citep[first pioneering observations
  by][]{BecklinNeugebauer1968ApJ}. \citet{Launhardt2002A&A} studied
the nuclear bulge of the Milky Way using COBE DIRBE data. They
identified the NSC of the Milky Way (MW) and estimated its mass as
$3.5\pm1.5\times10^{7}\,M_{\odot}$. The MW NSC is close to isothermal,
with a power-law index around $1.8$
\citep{BecklinNeugebauer1968ApJ,Catchpole1990MNRAS,Haller1996ApJ,Eckart1993ApJ}.
\citet{Genzel2003ApJ} showed that the MW NSC contains a central
stellar \emph{cusp} and no flat core. \citet{Schoedel2007A&A} found
that the cusp is very small (with a projected cusp radius of
$0.22\pm0.04$\,pc) and rather flat, with a power-law index of just
$1.2\pm0.05$. It would be extremely difficult -- if not impossible -
to resolve this small cusp in any extragalactic system, even with
50m-class telescopes. The cusp region is observationally dominated by
the presence of a population of young, massive stars. Their surface
density follows a power-law with $\Sigma\propto R^{-2}$
\citep{Paumard2006ApJ,Lu2008arXiv}, while the surface density of the
late-type stellar population is almost constant in the cusp region
(Buchholz, Sch\"odel, \& Eckart, submitted to A\&A).

Like NSCs in external galaxies the Milky Way NSC consists of a mixed,
old and young stellar population. Several periods of star formation
have occurred in the MW NSC. The most recent star burst happened just a few
million years ago
\citep[e.g.,][]{Allen1990MNRAS,Krabbe1995ApJ,Paumard2006ApJ,Maness2007ApJ}.

Studies of stellar dynamics have provided striking evidence for the
existence of a supermassive black hole at the dynamical center of the
MW NSC
\citep[see][]{Eckart1996Natur,Ghez2000Natur,Genzel2000MNRAS}. The
measurements of stellar orbits have provided, so far, the best
evidence for its nature \citep[e.g.,][]{Schoedel2003ApJ,Ghez2003ApJ}.
Recent work on the orbit of the star S2/S02 gives, so far, the most
accurate measurement of the mass of the black hole
($\sim4\times10^{6}\,M_{\odot}$)
\citep{Eisenhauer2005ApJ,Ghez2005ApJ,Ghez2008ApJ,Gillessen2008arXiv}.  While some
extragalactic surveys can give the impression that NSCs and
supermassive black holes may be mutually exclusive
\citep{Ferrarese2006ApJ}, the case of the GC and of galaxies
  containing both AGN and NSCs \citep{Seth2008aApJ} demonstrates that
an NSC and a supermassive black hole can co-exist.

While the mass of the supermassive black hole, Sagittarius\,A*
(Sgr\,A*), at the GC has been determined with high accuracy,
this is not the case for the mass and mass density of the star cluster
around Sgr\,A*, for which there exists a large uncertainty. For
example, the data presented by \citet{Haller1996ApJ} are consistent
with a stellar mass between $0$ and a few $10^{6}\,M_{\odot}$ within
1\,pc of Sgr\,A*. The main problem here is the lack of sufficiently
large samples of stellar proper motion or line-of-sight (LOS) velocity
measurements at distances sufficiently far from Sgr\,A* so that the
velocity dispersion is not completely dominated by its mass
($r\gtrsim0.5$\,pc), but sufficiently close to Sgr\,A* in order to
measure stars well within the NSC ($r\lesssim2-3$\,pc) and thus to
avoid significant contamination by stars in the nuclear disk, bulge,
or foreground.

Due to the lack of data, estimates of the enclosed mass profile at the
GC were up to now heavily influenced by modeling assumptions, such as
adopting some \emph{ad hoc} value for the velocity dispersion at large
distances, or by estimates of the enclosed stellar and BH mass based
on measurements of gas velocities
\citep[e.g.,][]{Genzel1996ApJ,Schoedel2002Natur}. \citet{Schoedel2007A&A}
have re-analyzed this issue and concluded that the mass of the star
cluster in the central parsec is possibly significantly higher than
previously assumed. Their claim is based on observational data that
indicate that the measured line-of-sight velocity dispersion of late
type stars within $\sim0.8$\,pc of Sgr\,A* remains apparently
constant, with a value around 100\,km\,s$^{-1}$
\citep{Figer2003ApJ,Zhu2008ApJ}. This contrasts with the expectation
that the velocity dispersion would show a Keplerian decrease over the
entire central parsec if only the BH point mass were
important. Additionally, \citet{Reid2007ApJ} found that the mass of
Sgr\,A* is not sufficient to keep the maser star IRS\,9 on a bound
orbit and that this may imply the existence of several
$10^{5}\,M_{\odot}$ of extended mass within $r\approx0.3$\,pc of the
black hole.  This appears to contradict earlier mass estimates,
  like the ones mentioned above, that indicate a negligible amount of
extended mass within 1\,pc of Sgr\,A*.  Using the measured radial
velocity dispersions of late type stars in the central parsec in
combination with the density profile of the NSC
\citet{Schoedel2007A&A} provide a  simple model (using the
  Bahcall-Tremaine mass estimator and assuming no rotation, isotropy,
  and that the velocity dispersion stays constant beyond the central
  parsec, where it was measured,) of enclosed mass vs.\ distance from
Sgr\,* that is consistent with up to a few $10^{6}\,M_{\odot}$  of
  extended mass within 1\,pc of Sgr\,A*. The key difference to mass
profiles presented in earlier works is the realization that the
projected velocity dispersion follows a clear Kepler-law only out to projected
distances $R\lesssim0.3$\,pc from Sgr\,A*, but remains apparently
constant at $R \gtrsim0.5$\,pc.

\begin{figure}[!htb]
\includegraphics[width=\columnwidth]{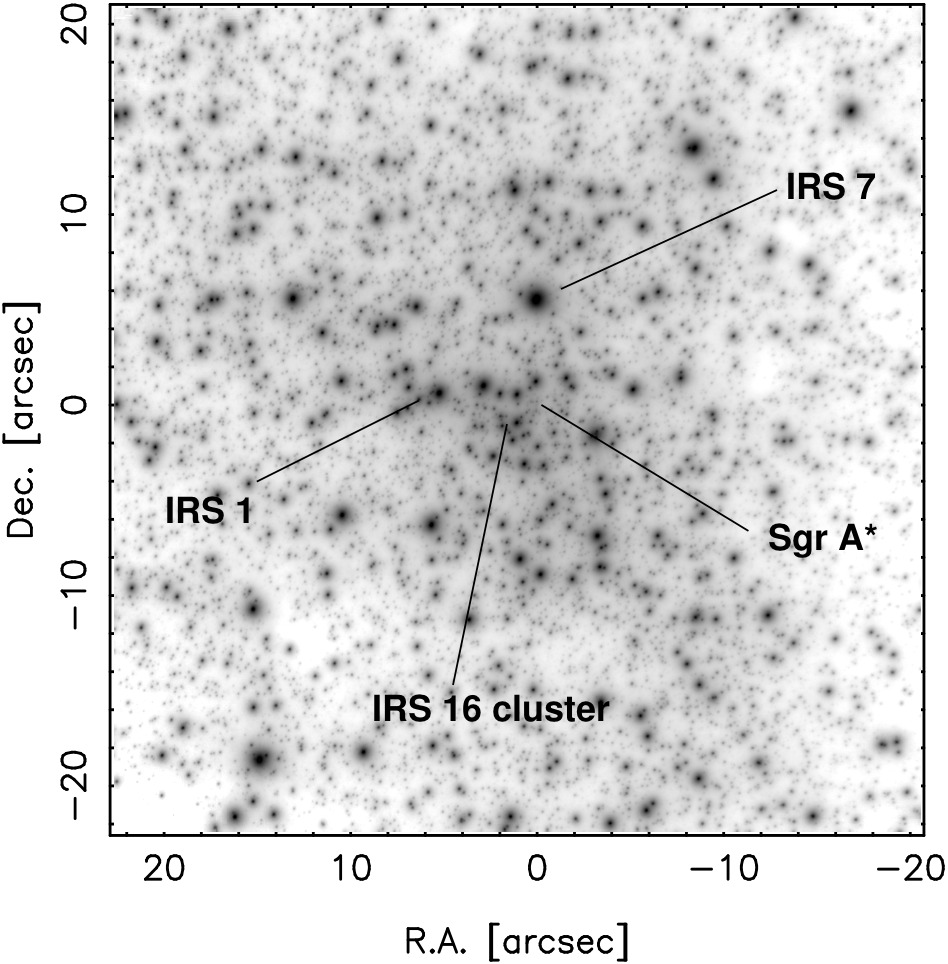}
\caption{\label{Fig:mosaic} Mosaic image of the observations from 1
  June 2006. A logarithmic gray scale has been adopted. The positions
  of IRS 7 and Sgr~A* are indicated. North is up and east is to the
  left. Offsets in arcseconds from Sgr\,A* are indicated. Note
  that this mosaic image is only roughly astrometric. A constant
    pixel scale of $0.027''$ per pixel has been assumed.  Any residual
    net rotation of the image derotator has not been determined, i.e.\ the
    image may have a non-zero rotation angle ($<1$\,deg). This may
    lead to offsets of up to a few tenths of arcseconds near the edge
    of the field.} 
\end{figure}

The mass and mass density of the nuclear star cluster around Sgr\,A*
is of great importance for understanding the dynamics of this complex
system. It can have strong implications for topics such as star
formation, the rate of stellar collisions, the relaxation time, cusp and
black hole growth, and the rate of gravitational wave emission events
\citep[for an overview of stellar processes near massive black holes,
  see, e.g.~][]{Alexander2007arxiv}.  In order to provide reliable
estimates of the mass of the MW NSC we therefore present in this work
a comprehensive sample of stellar proper motions within $\sim1$\,pc of
Sgr\,A*. The new data allow us to present accurate estimates of the
enclosed mass vs.~distance at the center of the Milky Way.

Sections 2, 3, 4, and 5 are largely technical and describe the data
processing and how the proper motions of stars in the GC were derived.
Readers who are primarily interested in the main results of our
analysis, can go directly to section\,6.

\section{Observations and data reduction}

\begin{figure}[!htb]
\includegraphics[width=\columnwidth]{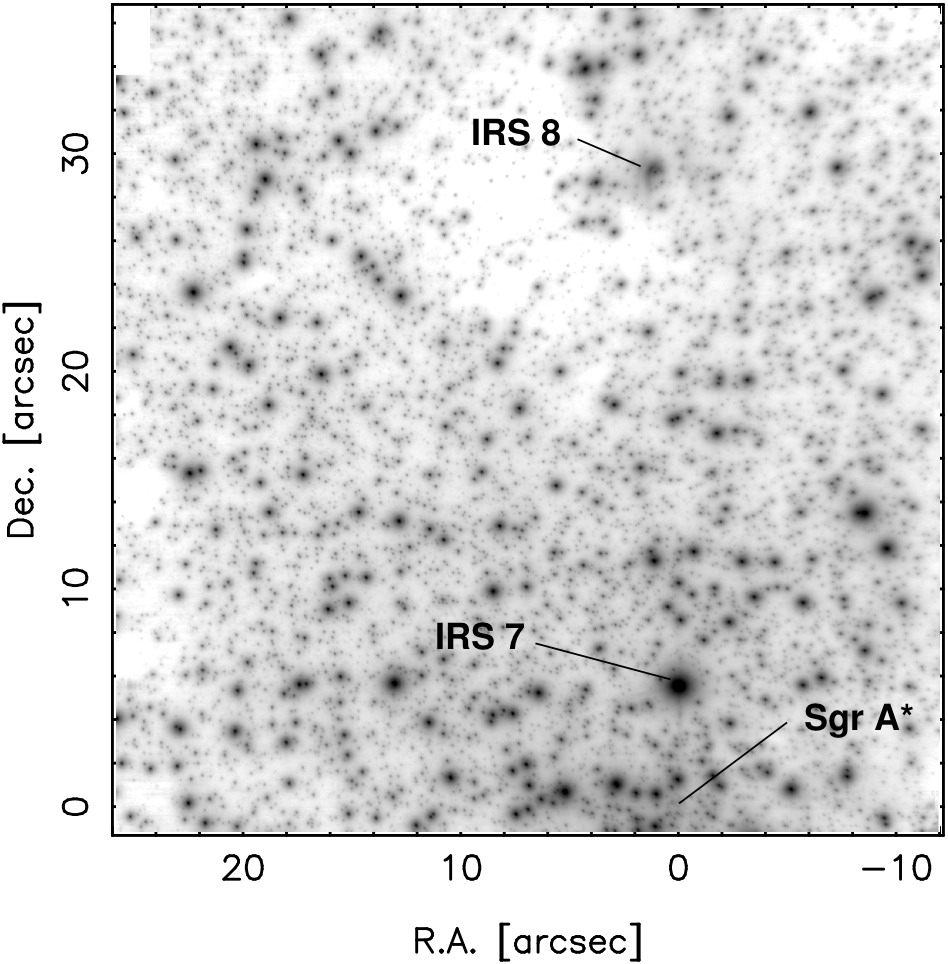}
\caption{\label{Fig:mosaicoff} Mosaic image of the observations from
  28 May 2008. The center of the image is offset about $20''$ NE from
  Sgr\,A*. A logarithmic gray scale has been adopted. The positions of
  IRS\,7, Sgr\,A*, and of the young and massive star IRS\,8, that is
  surrounded by a bow-shock \citep[][]{Geballe2006ApJ}, are
  indicated. North is up and east is to the left. Offsets in
  arcseconds from Sgr\,A* are indicated. Please note that this mosaic
  image is only roughly astrometric  (see comments in caption of
    Fig.\,\ref{Fig:mosaic})}.
\end{figure}

The imaging data used in this work were obtained with the
near-infrared (NIR) camera and adaptive optics (AO) system CONICA/NAOS
(short: NaCo) at the ESO VLT unit telescope~4 \footnote{Based on
  observations collected at the European Southern Observatory, Chile,
  programs 071.B-0077, 073.B-0085, 073.B-0745, 073.B-0775, 075.B-0093,
  075.C-0138, 077.B-0552, 081.B-0648}. For images centered on Sgr\,A*
the $\rm mag_{Ks}\approx6.5-7.0$ supergiant IRS~7 was used to close
the loop of the AO, using the unique NIR wavefront sensor NAOS is
equipped with. A guide star with $\rm mag_{V}\approx14.0$ located
$\sim$19$''$ NE of Sgr\,A* was used as reference for the AO for the
images offset from Sgr\,A*. The sky background was measured on a
largely empty patch of sky, a dark cloud about $400''$ north and
$713''$ east of the GC. Data reduction was standard, with sky
subtraction, bad pixel correction, and flat fielding. The
field-of-view (FOV) of a single exposure is $28''\times28''$. The
observations were dithered (either applying a fixed rectangular
pattern or a random pattern) in order to cover a FOV of
about $40''\times40''$.

The majority of the images were taken by dithering around a position
roughly centered on Sgr\,A*. These data will be referred to in this
article as the \emph{center data set}.  There are three observations
that were centered on a field roughly $20''$ NE of Sgr\,A*, the
\emph{offset data set}. Those images were used for determining the
velocity dispersion in the MW NSC at projected distances from Sgr\,A*
out to $1.15$\,pc. The pixel scale of all NaCo data used in this work
is $0.027''$ per pixel.  Details of the observations are listed in
Table\,\ref{Tab:Obs}. Mosaic images of the two observed fields are
shown in Figs.\,\ref{Fig:mosaic} and \ref{Fig:mosaicoff}.

\begin{table}
\caption{\label{Tab:Obs} Details of the imaging observations used in this
  work. DIT is the detector integration time, NDIT is the
  number of integrations that were averaged on-line by the read-out
  electronics, N is the number of (dithered) exposures (terminology of
  ESO observations). The total integration time of each observation
  amounts to N$\times$NDIT$\times$DIT. The pixel scale of all
  observations is $0.027''$ per pixel.}
\begin{tabular}{llllll}
\hline
\hline
Date & $\lambda_{\rm central}$ [$\mu$m] & $\Delta\lambda$  [$\mu$m] & N & NDIT & DIT [s]\\
\hline
03 May 2002 & 2.18 & 0.35 & 20 & 3 & 20 \\
10 May 2003 & 2.18 & 0.35 & 19 & 120 & 0.5\\
12 June 2004 & 2.06 & 0.06 & 96 & 1 & 30 \\
12 June 2004 & 2.24 & 0.06 & 99 & 1 & 30 \\
13 June 2004 & 2.33 & 0.06 & 119 & 1 & 30 \\
13 May 2005 & 2.18 & 0.35 & 103 & 60 & 0.5 \\
29 April 2006 & 1.66 & 0.33 & 32 & 28 & 2 \\
29 April 2006 & 2.18 & 0.35 & 32 & 28 & 2 \\
01 June 2006 & 2.18 & 0.35 & 80 & 3 & 10 \\
28 May 2008 & 2.18 & 0.35 & 35 & 4 & 10 \\
28 May 2008 & 2.18 & 0.35 & 20 & 2 & 20 \\
\hline
offset fields & & & & & \\
\hline
11 Aug 2004 & 2.18 & 0.35 & 16 & 3 & 40\\
27 July 2005 & 2.18 & 0.35 & 8 & 4 & 15\\
28 May 2008 & 2.18 & 0.35 & 20 & 20 & 2\\
\hline
\end{tabular}
\end{table}

\section{Photometry and astrometry \label{sec:astrometry}}

For accurate error assessment and in order to avoid any additional
errors introduced by the mosaicing process, we did not combine the
individual frames into mosaic images. Photometry and astrometry were
instead done on  individual exposures. This allowed us to compare
multiple independent measurements for each star at each epoch. The
number of individual frames per epoch is listed in column~4 of
Table\,\ref{Tab:Obs}. The PSF fitting program package
\emph{StarFinder} \citep{Diolaiti2000A&AS} was used. Since the
$28''\times28''$ ($1024\times1024$\,pixel) field-of-view (FOV) of the
NaCo S27 camera, that was used for all observations, is larger than
the isoplanatic angle of near-infrared adaptive optics observations
($\leq10-15''$), the images were divided into sub-frames of
$\sim7''\times7''$ size, i.e.\ with angular diameters smaller than the
isoplanatic angle.  PSF extraction, followed by astrometry and
photometry was done on each of the individual sub-frames. In order to
minimize any uncertainties related to PSF extraction, each image was
divided by a rectangular pattern in many overlapping sub-frames. The
step size between the mid-points of the sub-frames was chosen as half
the width of the sub-frames.

PSF extraction was done by identifying all suitable PSF reference
stars within each sub-frame (all potential PSF reference stars for the
entire FOV were marked previously by hand on a large mosaic
image). The noise for each sub-frame was determined from the read-out
and photon noise (algorithm provided by \emph{StarFinder}).  In order
to improve the PSF, the \emph{StarFinder} algorithm was run once on
the sub-frame with a detection threshold of $20\,\sigma$. PSF
extraction was then repeated. Since the quality of the PSF
deteriorates in the wings, the PSF had to be truncated. We used a
circular mask with radius 20\,pixel (about 6 times the FWHM of the
PSF).  The \emph{StarFinder} algorithm was then applied to the
sub-frame, using two iterations with a $3\,\sigma$ threshold and a
correlation threshold of $0.7$.

Measurements of stars in overlapping sub-frames were averaged. With the
exception of the stars near the edge of the field, there were 4
measurements of each source. The uncertainties derived from these
  measurements were smaller than the formal uncertainties of the PSF
  fitting routine.

 The cores of saturated sources were repaired during the PSF
 extraction process. The detected stars have magnitudes $ \rm mag_{Ks}
 \gtrsim 17.5$. Any spurious sources that may still be present in the
 data at this point were eliminated later when merging the source
 lists of the various exposures for each epoch after alignment with
 the reference frame. Each star had to be detected in multiple
 exposures (see appendix\,\ref{app:alignment}).

\section{Transformation into a common reference frame}

Since there is no absolute frame of reference available for
determining the proper motions of stars at the GC, one has to
transform the stellar positions into a common reference frame using a
large number of stars with either known proper motions or with the
assumption that their motions cancel on average. The problem has been
described previously in various publications, e.g.,
\citet{Eckart1997MNRAS} or \citet{Ghez1998ApJ}.

In this work, we present proper motions on a much larger FOV than what
has been published on the GC before  \citep[but see also][ which appeared
  shortly before this work]{Trippe2008A&A}. The large FOV means that
significant dither offsets from the initial pointing (of the order
$6''-8''$) had to be applied. Since this implies that stars come to
lie on different areas of the detector, camera distortions may become
of considerable importance in this case. Therefore, special care has
to be taken when aligning all stellar positions to a common reference
frame. This procedure is key to obtaining accurate proper motion
measurements. For the sake of repeatability of the experiment we
consider it important to describe our applied methodology in
detail. However, since this is a largely technical issue that may not
be of interest to many readers, we describe this procedure in
appendix\,\ref{app:alignment}.

\section{Proper motions}

\subsection{Center field}

\begin{figure}[!tbh]
\includegraphics[width=\columnwidth]{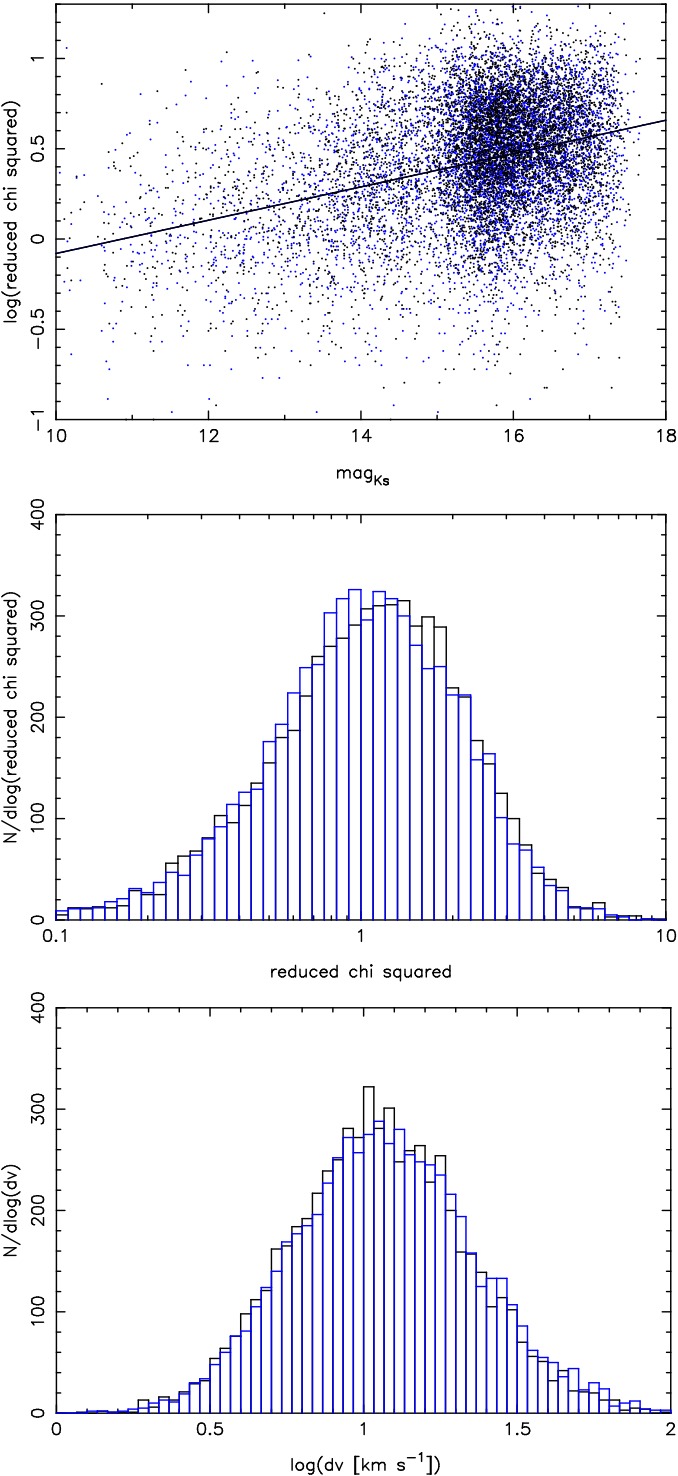}
\caption{\label{Fig:chi2dv} Error analysis for center field data. Top:
  Plot of log($\chi^{2}_{red}$) vs.\ Ks-band magnitude. The straight line is
  a least squares linear fit. It is used to re-calibrate the
  $\chi^{2}_{red}$-values so that the stars of different magnitude can be
  compared.  Middle: Distribution of the re-calibrated, reduced
  $\chi^{2}_{red}$ values for the linear fits of the data of position
  vs.\ time. The black histogram is for the fits in right ascension,
  the blue histogram for the fits in declination. Bottom:
  Distribution of the velocity uncertainties, after scaling
  with the reduced $\chi^{2}$ values. Black for right ascension and
  blue for declination.}
\end{figure}

\begin{figure*}[!htb]
\includegraphics[width=\textwidth]{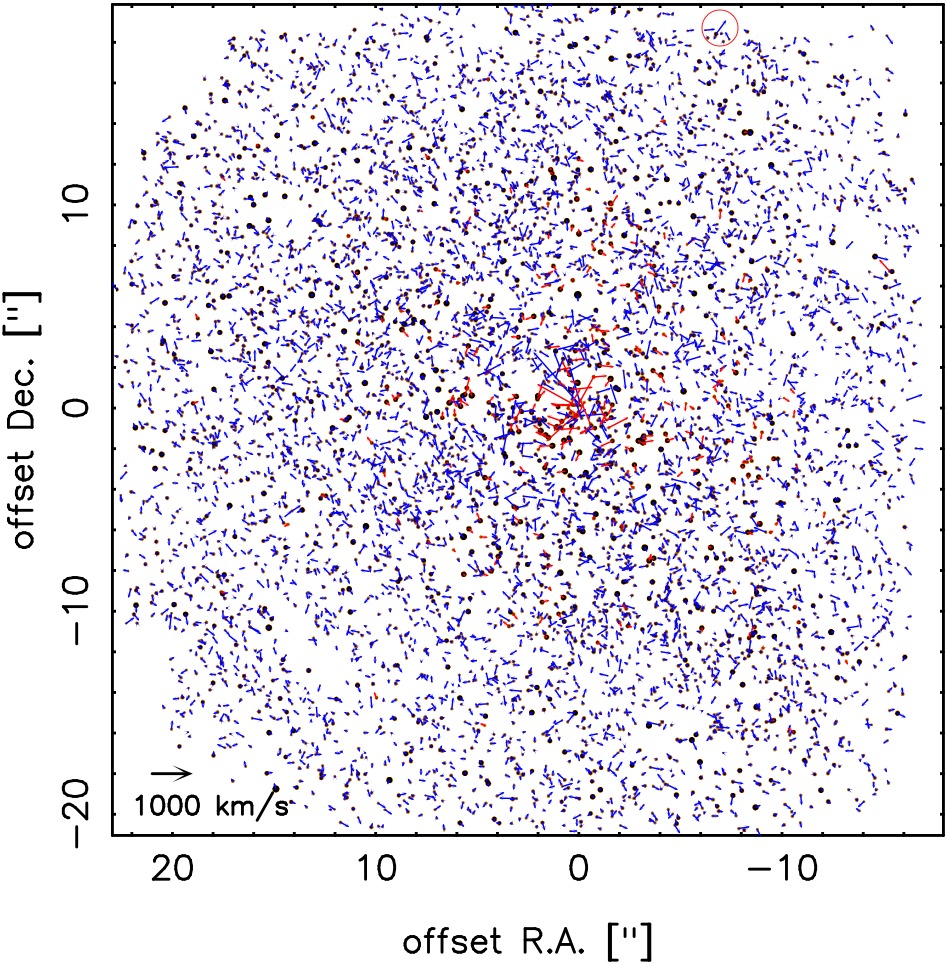}
\caption{\label{Fig:velmap} Map of stars with measured proper motions
  in the GC central field. North is up and east is to the left. Arrows
  indicate magnitude and direction of the proper motion
  velocities. The black arrow in the lower left corner indicates the
  length of a 1000\,km\,s$^{-1}$ arrow. Positions and velocities of
  the stars are listed in in Tab.\,\ref{Tab:list}. Early-type stars,
  identified by \citet{Paumard2006ApJ} and Buchholz, Sch\"odel \&
  Eckart (2009, submitted to A\&A) are indicated by red arrows.}
  The small red circle marks the best candidate 
for a star escaping the cluster.
\end{figure*}

After alignment of the stars into a common reference frame, stars were
matched by searching within a circle of 2 pixels radius around the
position in the reference frame. This radius is smaller than the FWHM
of the AO data, which avoids mismatching. It is large enough in order
to detect stars with proper motion velocities up to
$500$\,km\,s$^{-1}$ in each coordinate. Sources with proper
  motions exceeding this value would be missed by this analysis, but
  should be extremely rare (see histogram in
  Fig.\,\ref{Fig:siglb}). Sources with such high velocities can easily
  be detected by visual inspection of images taken about 2 years
  apart. We did not find such fast moving sources in a visual
  inspection (limited to $\rm mag_{K}\approx16$) of the images.
Matching of the stars within $1''$ of Sgr\,A*, where source density is
highest and proper motion velocities exceed several hundred
km\,s$^{-1}$ was done manually.  Proper motions were subsequently
determined by linear fits to the measured positions vs.\ time.  In
order to be included in our proper motion sample, a star had to be
detected in at least 4~different years. This assures an adequate
sampling of its proper motion. The reasons why certain stars are not
present in all the epochs are, among others, variable quality of the
data (Strehl ratio, exposure time, noise introduced by read-out
  electronics, etc.), a variable FOV of the observations, or, in some
exceptional cases, strong orbital acceleration of stars close to
Sgr~A* ($R<0.5''$). More than 80\% of all stars in the common
  overlap area that are not present at all epochs (corresponding to
  $\approx$10\% of all stars) are faint stars ($\rm mag_{K} < 15.5$),
  whose detection and measurement is particularly affected by the data
  quality.

%\clearpage

Outliers in the data were removed by first applying an unweighted fit
(in order to avoid to be biased by spurious erroneous measurements)
and rejecting any data point with a deviation $>5\,\sigma$ from the
fitted position. Then the linear fit was repeated by weighting the
positions according to their $1\,\sigma$ uncertainties. Although
  we have not investigated the exact causes of these outliers -- there
  may be various causes -- we believe that the main reason for the
  spurious data points is confusion with unresolved (maybe in a few
  cases also resolved) sources in the dense NSC. This source of
  systematic uncertainty has been investigated in detail by
  \citet{Ghez2008ApJ}. This hypothesis is supported by the fact that
  outliers are largely associated with faint stars. More than 75\%
  (97\%) of the outliers are associated with stars fainter than $\rm
  mag_{K}=15.5~(14.0)$. As concerns the numbers of removed data
  points, 22\% of the stars had one data point removed, 10\% two, and
  8\% three or more. The above mentioned criterion that position
  measurements had to be available for at least 4~different years was
  applied only {\it after} removing the outliers.

A distance of 8.0\,kpc to the GC was assumed. A pixel scale of
$0.027''/$pixel was adopted for the camera detector (see ESO manual con
NAOS/CONICA, available at the ESO web site). The adopted pixel scale
is somewhat smaller than the value given in the manual
($0.02715''/$pixel). However, this difference will lead to a systematic
error of less than $0.6\%$ on the measured proper motions .

In the top panel of Fig.\,\ref{Fig:chi2dv} we show a plot of reduced
$\chi^{2}$ vs.\ Ks-band magnitude for the proper motion fits. As can
be seen, $\chi^{2}_{red}$ is close to $1.0$ for stars brighter than
mag$_{Ks}\approx14.0$, but increases toward fainter magnitudes. This
indicates that there are measurement uncertainties that we have not
taken into account and that become increasingly important for faint
stars.  Since the positional uncertainty for each star is derived from
multiple images at each epoch, the positional uncertainty for a given
epoch can be expected to be correctly determined. This implies that
the missing source of uncertainty must be due to systematic deviations
between the epochs.  This means that while a stellar position may be
measured with high \emph{precision} in one epoch, its \emph{accuracy}
may in fact be subject to additional errors. We believe that the most
probable source of error that has not been taken into account in our
analysis consists of systematic deviations of the positions of faint
stars due to their motion through an extremely dense stellar field and
bright background due to unresolved stars. This source of error is
described and analyzed in detail for the star S2/S0-2 in a recent paper
by \citet{Ghez2008ApJ}.

The systematic increase of $\chi^{2}_{red}$ vs.\ brightness can be
fitted well with a simple line in log-log space. This fit is used to
re-normalize the $\chi^{2}_{red}$-values of the stars before determining the
overall distribution of $\chi^{2}_{red}$-values. The latter is shown in the
middle panel of Fig.\,\ref{Fig:chi2dv} and can be seen to peak
close to $1.0$ as expected. Please note that we do not deal with an
\emph{ideal} $\chi^{2}_{red}$-distribution because the number of degrees of
freedom varies, depending on the number of measurements available per
star.

In order to avoid under- (in the majority of cases, see top panel
  of Fig.\,\ref{Fig:chi2dv}) or over-estimating the uncertainties of
the inferred velocities, the uncertainties were re-scaled with the
corresponding $\chi^{2}_{red}$ of the fit. This procedure is viable
because we can assume that a linear fit is in fact a reasonable model
for our data. The distribution of the velocity uncertainties after
re-scaling is shown in the bottom panel of Fig.\,\ref{Fig:chi2dv}. The
percentage of sources with $dv_{x,y}>50$\,km\,s$^{-1}$ is
$\leq2\%$. More than $80\%$ of the stars have velocity uncertainties
$dv_{x,y}<25$\,km\,s$^{-1}$.

After application of the methodology described above, we obtained a
list of 6124 stars with measured proper motions for the central
field. The positions and velocities of the measured stars are
illustrated in Fig.\,\ref{Fig:velmap}. Some important features can be
seen at first glance: the velocities are highest near Sgr\,A* (at the
origin of the coordinate system) and decrease with distance from the
black hole; some apparently coherently moving groups of stars can be
seen at a few arcseconds distance from Sgr\,A*, related to known
groups (IRS\,13, IRS\,16) and/or the disk(s) of young stars in the
central half parsec
\citep[see][]{Levin2003ApJ,Genzel2003ApJ,Lu2005ApJ,Schoedel2005ApJ,Paumard2006ApJ};
at larger distances the directions of the proper motions appear to be
random.

The influence of the proper motions of the reference stars on the
accuracy of the coordinate transformation was checked by iterating the
described procedure to obtain proper motions. The initially measured
proper motions of the reference stars were used to calculate their
correct positions for each epoch. The resulting change in the measured
stellar velocities for all sources is insignificant
($\ll1\,\sigma$). Hence, the approach of using a dense grid of evenly
sampled reference stars delivers very stable solutions for the
coordinate transformation.  As a further test, we repeated the above
procedure by applying just a second order transformation
($i,j_{max}=2$ in eqs.\,\ref{eq:polyX} and \ref{eq:polyY}) of the
stellar positions into the reference frame. Again, the deviations were
insignificant.

\begin{figure}[!htb]
\includegraphics[width=\columnwidth]{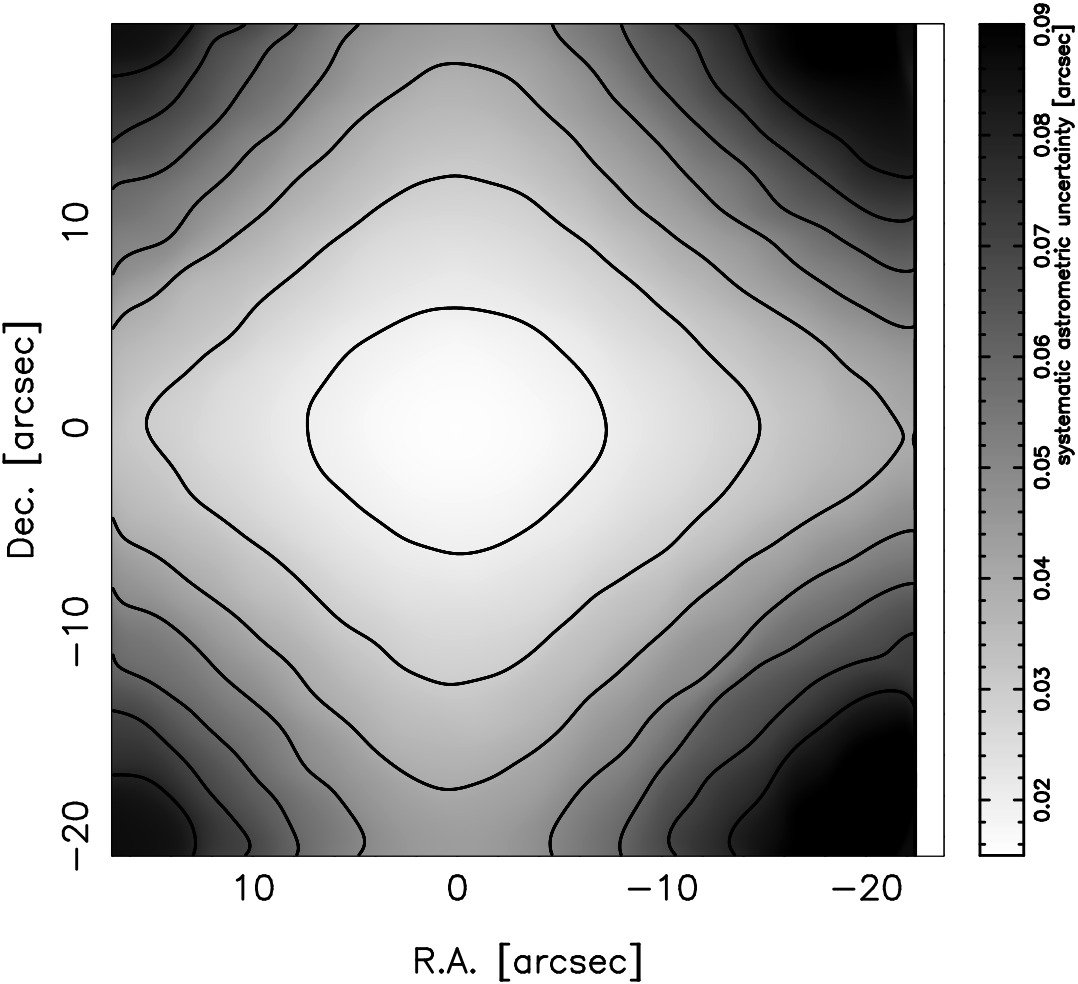}
\caption{\label{Fig:radiosigma} Map of the systematic positional
  uncertainty of the astrometric positions of the stars listed in
  Tab.\,\ref{Tab:list}.  Contours are plotted in steps of 10\,mas from 20
  to 80\,mas.}
\end{figure}

In a final step, the pixel positions of the stars were transformed
into the radio reference frame as established by maser stars. The
positions and proper motions of the SiO masers IRS\,15NE, IRS\,7,
IRS\,17, IRS\,10EE, IRS\,28, IRS\,9, and IRS\,12N are taken from the
values measured by \citet{Reid2007ApJ}, while their IR positions and
proper motions are taken from the linear fits derived from our data
set.  A first order transformation was applied ($i,j_{max}=1$ in
Eqs.\,\ref{eq:polyX} and \ref{eq:polyY}). In order to estimate the
uncertainty of the transformation into the radio frames, the
transformation parameters were estimated repeatedly with subsets of 6
out of the 7 maser stars.  A smoothed (by applying a Gaussian filter
of $2.0''$ FWHM) map of the systematic uncertainty of the astrometric
position in the radio frame was created and is shown in
Fig.\,\ref{Fig:radiosigma}. Near Sgr\,A* an absolute positional rms
accuracy of $\sim$15\,milli-arcseconds is reached. Probably the most
important effects that limit the accuracy of the alignment with the
astrometric reference frame are the uncertainty of the proper motions
of the maser stars in the IR frame and  residual distortions in
the IR reference frame  because we did not determine a distortion
  solution for our 2006 reference frame (see
  section\,\ref{app:align}). The maser stars are saturated in the
data of most epochs, with the positional information therefore being
based mainly on the PSF wings of these stars. The proper motions of
the maser stars in the infrared and radio frames agree very well (see
section\,\ref{sec:IRradio} below).

The intrinsic $H-K$ colour of the stars in the GC field is almost (to
within $\sim$0.1\,mag) independent of their stellar type \citep[see
  discussion in][]{Schoedel2007A&A}. Therefore $H-K$-colours can be
used to get a fairly accurate value of the extinction toward
individual stars. We identified about $30$~foreground stars in the
proper motion sample by their low extinction ($A_{K}<2.0$) and removed
them from the sample. Here, we used the $H-K$ values and extinction
measurements of Sch\"odel et al.\ (in preparation).

The final list of stars with their fitted positions in $2004.44$,
their measured proper motions, and magnitudes is presented in
Table\,\ref{Tab:list}. Since the few million year-old
  population of early-type stars in the central parsec has particular
  kinematic properties\citep[see][]{Paumard2006ApJ,Lu2008arXiv}, we
  identified early-type stars in the sample. We find 79 early-type
  stars from the spectroscopic analysis by \citet{Paumard2006ApJ}
  (their table~2) and 202 additional early-type candidates from the
  photometric analysis of Buchholz, Sch\"odel \& Eckart (2009, submitted to
  A\&A). Spectroscopically identified early-type stars are marked with
  a ``1'' and photometrically identified ones with a ``2'' in the last
  column of Table\,\ref{Tab:list}.  We recommend future users of this
  list to cross-check the identifications with the latest available
  publications. Note also that identification of spectral type
  was only available for a fraction of the stars, leaving a large
  number of stars unidentified (near the edge of the FOV and all stars
  fainter than $mag_{Ks}\approx15.5$).  Note also that the
  number of early-type stars is much smaller than the overall number
  of stars. Also, their surface density decreases rapidly beyond a few
  arcseconds distance from Sgr\,A* \citep{Paumard2006ApJ,
    Lu2008arXiv}. Therefore, their weight on the measured statistical
  properties of the entire cluster is almost negligible, except in the
  innermost arcseconds. Nevertheless, we will largely exclude the identified
  early-type stars from our analysis due to
  their special dynamical properties.

The reader should keep in mind that orbital  accelerations
have not been taken into account in the present analysis. This is the
reason why the position of the star S2 in Table\,\ref{Tab:list}
(second line) has an offset from its actual position in 2004.44 that
is $\sim20$\,mas larger than the astrometric uncertainty at its
position. This kind of additional uncertainty only affects very few
($<5$) stars. We did not exclude them from our sample because they
provide important information on the velocity dispersion near Sgr\,A*.

\subsection{Offset field}

The results of the proper motion analysis for the offset field serve
as cross-validation of our results, but were not used for the
subsequent modeling because the quality of the proper motions is
  lower (only three epochs compared to 11 for the center field) amd
  the radial range of the proper motions is not extended
  significantly. The data and results are described in
appendix\,\ref{app:offset}.

\subsection{Relative motion between radio and infrared frame of
  reference \label{sec:IRradio}}

A question of great interest is whether there exists any relative
motion between the radio and near-infrared reference frames. The
  black hole, Sgr\,A* is at rest in the so-called {\it radio reference
    frame}. The positions and velocities of the maser stars have been
  measured in the radio frame via VLBI observations \citep[see,
    e.g.,][]{Reid2007ApJ}. When deriving the positions and proper motions
  of stars in the {\it infrared frame}, we assume that the cluster of
  stars has an average velocity of zero  (see also
  appendix\,\ref{app:align}). Hence, comparing the velocities of the
  maser stars measured in the radio and in the infrared frames is an
  important cross-check. Their velocities are expected to be identical
  in the two reference frames. Any significant non-zero average value
  of the difference between the masers' radio and infrared proper
  motions would imply a relative movement between Sgr \,A* and the
  star cluster. 

We list the measured proper motions in the infrared frame of the
7 maser stars used for the astrometric alignment in
Table\,\ref{Tab:masers}.  All infrared proper motions of the maser
stars agree within $<3\,\sigma$ with their proper motions as measured
by VLBI \citep[][]{Reid2007ApJ}.

\begin{table*}
\caption{\label{Tab:masers} Measured proper motions, in the infrared
  frame, of the maser stars used for astrometric alignment. Positions
  are given for 2004.44 as offsets from Sgr\,A*, positive toward the
  north and east. Proper motions are given in milli-arcseconds per
  year, positive toward north and east. The last two columns list the
  differences between the radio \citep{Reid2007ApJ} and IR proper
  motions.}  \centering
\begin{tabular}{lrrrrrr}
\hline
\hline
Name & R.A. [arcsec] & Dec. [arcsec] & v$_{R.A:}$ [mas\,yr$^{-1}$] & v$_{Dec}$ [mas\,yr$^{-1}$] & $\Delta_{\rm radio,IR}$v$_{R.A:}$ [mas\,yr$^{-1}$] & $\Delta_{\rm radio,IR}$v$_{Dec}$ [mas\,yr$^{-1}$]\\
 \hline
IRS~15NE &  $1.226\pm0.015$ &  $11.327\pm0.024$ & $-1.51\pm0.15$  &
$-5.81\pm0.15$ & $-0.46\pm0.17$ &  $0.13\pm0.19$\\
IRS~7    &  $0.046\pm0.010$ &  $5.569\pm0.016$ &  $-0.05\pm0.23$  &
$-4.59\pm0.23$ & $-0.53\pm0.55$ & $1.07\pm0.59$\\
IRS~17   &  $13.143\pm0.019$ & $5.566\pm0.030$ &  $-1.70\pm0.12$ &
$-1.04\pm0.12$ & $0.09\pm1.09$ & $0.29\pm1.23$\\
IRS~10EE &  $7.688\pm0.012$ &  $4.220\pm0.020$ & $-0.40\pm0.18$ &
$-1.55\pm0.18$& $0.44\pm0.19$ & $-0.55\pm0.19$\\
IRS~28   &  $10.477\pm0.016$ &  $-5.825\pm0.026$ &  $1.62\pm0.23$ &
$-4.71\pm0.23$ & $0.38\pm0.44$ & $-0.57\pm0.48$\\
IRS~9    &  $5.672\pm0.013$  & $-6.335\pm0.021$  & $3.30\pm0.25$  &
$3.03\pm0.25$ & $-0.24\pm0.26$ & $-0.92\pm0.31$\\
IRS~12N  &  $-3.250\pm0.012$ & $-6.876\pm0.019$ &  $-1.62\pm0.22$ &
$-2.78\pm0.22$ & $0.56\pm0.24$ & $0.07\pm0.28$\\
\hline
\end{tabular}
\end{table*}

The weighted mean difference between the velocities of the seven maser
stars in the radio and the IR frames is in right ascension
$0.01\pm0.10$\,mas\,yr$^{-1}$ (unweighted:
$0.04\pm0.45$\,mas\,yr$^{-1}$) toward the east and in declination
$0.22\pm0.11$\,mas\,yr$^{-1}$ (unweighted:
$0.07\pm0.67$\,mas\,yr$^{-1}$) toward the south. At a distance of
8.0\,kpc a proper motion of 1\,mas\,yr$^{-1}$ corresponds to
$\sim$38\,km\,s$^{-1}$.  Converting the weighted mean motions to
km\,s$^{-1}$ this means that the radio reference frame moves relative
to the IR frame with $0.4\pm3.8$\,km\,s$^{-1}$ toward east and
$8.4\pm3.8$\,km\,s$^{-1}$ toward south.  Following \citet{Reid2007ApJ}
an additional systematic error of 5\,km\,s$^{-1}$ should be added to
these values in quadrature in order to take into account the removal
of the average motion of the IR frame (IR-motions are derived by
assuming that the net motion of the reference stars is 0). This
results in a $1\,\sigma$ uncertainty of $6.4$\,km\,s$^{-1}$ on the
relative motion between the radio and IR reference frames. This
reduces the significance of the southward motion to less than
$2\,\sigma$. Considering additionally that the weighted relative mean
velocity between the radio and IR frames in \citet{Reid2007ApJ} is a
net \emph{northward} motion, while we measure a \emph{southward}
motion here, we can safely conclude that the result is consistent with
no detectable motion.  We conclude that within the accuracy of the
presented measurements there is no detectable relative motion between
the infrared and the radio reference frames and consequently between
the stellar cluster and the central supermassive black hole Sgr\,A*.
This is consistent with our expectations. The expected rms
  velocity of the BH due to gravitational perturbations from stars is
  $\sim$0.2\,km\,s$^{-1}$ \citep{Merritt2007AJ}.

\section{Velocity structure of the NSC}

%\begin{figure}[!htb]
%\includegraphics[width=\columnwidth]{v2D.jpg}
%\caption{\label{Fig:vm2D} Map of the projected mean velocities
%  computed on grid points for the center field. A Gaussian smoothing
%  kernel has been applied (see text for details). The arrows are
%  therefore not statistically independent.  The clockwise
%  rotating disc of early-type, i.e., young stars can be seen clearly
%  as a vortex-like feature around Sgr\,A*.}
%\end{figure}

\begin{figure*}[!htb]
\centering
\includegraphics[width=.9\textwidth]{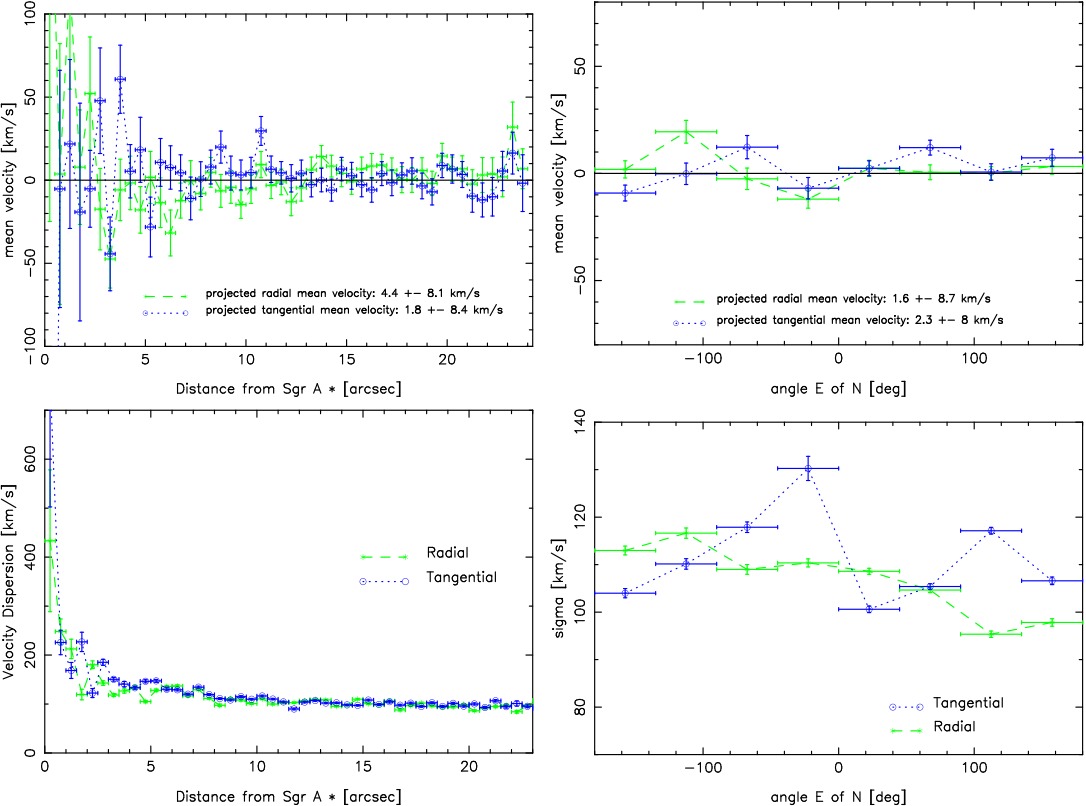}
\caption{\label{Fig:sigma} Top: Mean projected radial and tangential
  velocities vs.\ projected distance from Sgr\,A* (left) and vs. angle
  east of north (right). Mean values and the standard deviations
  annotated in the plots were calculated from the unweighted data
  points. Bottom: Projected radial (green) and tangential (blue)
  velocity dispersions in the GC nuclear star cluster vs.\ projected
  distance from Sgr\,A* (left) and vs. angle east of north
  (right). These plots are based on the center data set (see
  Fig.\,\ref{Fig:mosaic}) under the exclusion of identified early-type stars.}
\end{figure*}

%\clearpage

\subsection{First look}

%%Fig.\,\ref{Fig:vm2D}. The mean velocities have been computed at grid
%points and a weighting scheme using a Gaussian kernel has been
%applied. The FWHM of this kernel was chosen as $1"$ at $R=1"$. It
%varies as $R^{0.5}$. Some coherent motion appears to be present in
%some regions of the field: The vortex-like feature near the center is
%related to the well-known clockwise rotating disc (seen almost
%face-on) of young-stars in this region
%\citep[see][]{Genzel2003ApJ,Paumard2006ApJ,Lu2008arXiv}. Areas of
%apparently coherent motion are also visible at projected distances of
%$R\approx8''$ in the NW and SE quadrants. We tentatively interpret
%this as signs of the rotation of the cluster in the Galactic plane
%(see \citet{Trippe2008A&A} and subsection\,\ref{sec:rotation}
%below). Due to extinction the probability of detecting stars at the
%front-side of the cluster can be expected to be higher than for stars
%at the back side. Taking Galactic rotation into account, this would
%lead to the observed streaming motions in the NW and SE quadrants.

The velocity data were converted from directions along right ascension
and declination into physically more meaningful projected radial and
tangential velocities with respect to Sgr\,A*. The mean projected
radial and tangential velocities and velocity dispersions for the
center data set, excluding identified early-type stars, are shown in
Fig.\,\ref{Fig:sigma}. The top left panel shows the mean velocities
vs.\ distance from Sgr\,A* in radial bins. The mean velocities are
generally close to zero. There are some deviations from zero, but also
larger error bars, at $R\lesssim6''$. In these innermost bins stellar
numbers are lower because of the correspondingly smaller surface
areas. Also, the relative number of early-type stars increases toward
Sgr\,*. This tends to decrease additionally the number of stars in the
bins (early-type stars were excluded from this analysis). There are no
significant indications of net expansion, contraction, or rotation of
parts of the cluster in the plane of the sky. An overall net rotation
of the entire cluster {\it in the plane of the sky with a constant
  angular velocity} cannot be detected by our method. The reason is
that the coordinate transformation between the epochs is based on the
assumption that the overall motion of the stars is zero. Rotation of
{\it parts} of the cluster, such as the disk of early-type stars, can
be detected, however (see Fig.\,\ref{Fig:velmap}).

A comparison between proper motions of maser stars in the radio and
infrared reference frames allows some check on the rotation of the
cluster in the plane of the sky. It does not provide any evidence for
rotation in the plane of the sky (see
subsection\,\ref{subsec:skyrot}). The top right panel of
Fig.\,\ref{Fig:sigma} shows the mean velocities of the stars
vs.\ angle on the sky, measured east of north. They are close to zero,
but a possibly significant sinusoidal pattern can be discerned,
especially in the tangential mean velocities. This is probably the
imprint of an overall rotation of the cluster (see below) {\it in the
  Galactic plane}.

The bottom left panel of Fig.\,\ref{Fig:sigma} shows a plot of the
projected radial and tangential velocity dispersion vs.\ distance from
Sgr\,A*. With the exception of the region at $R<6"$ the data suggest
isotropy with considerable accuracy. Any deviations from isotropy at
$R<6"$ may be due to either worse statistics (smaller surface area and
exclusion of early-type stars) and / or the presence of not identified
early-type stars in the sample, which may follow a coherent rotation
pattern \citep[see][ and
  Fig.\,\ref{Fig:velmap}]{Genzel2003ApJ,Paumard2006ApJ,Lu2006JPhCS}. The
plot of the velocity dispersion vs.\ angle on the sky (bottom right)
shows a sinusoidal pattern for the projected tangential velocity
dispersion, which is probably due to rotation of the cluster in the
Galactic plane (see sub-section\,\ref{rotation} below).

\subsection{Rotation in the plane of the sky \label{subsec:skyrot}}

The proper motions were derived by a third order polynomial alignment
of the stellar positions with the reference epoch. This procedure
excludes detecting rotation of the {\it entire} cluster with a {\it
  constant} angular velocity in the plane of the sky. It does,
however, not exclude detecting rotation of sub-groups of star, such as
the early type stars within a few arcseconds from Sgr\,A* (see
Fig.\,\ref{Fig:velmap}) that is described in detail, e.g., in
\citet{Paumard2006ApJ,Lu2008arXiv}.  

\begin{figure}[!htb]
\includegraphics[width=\columnwidth]{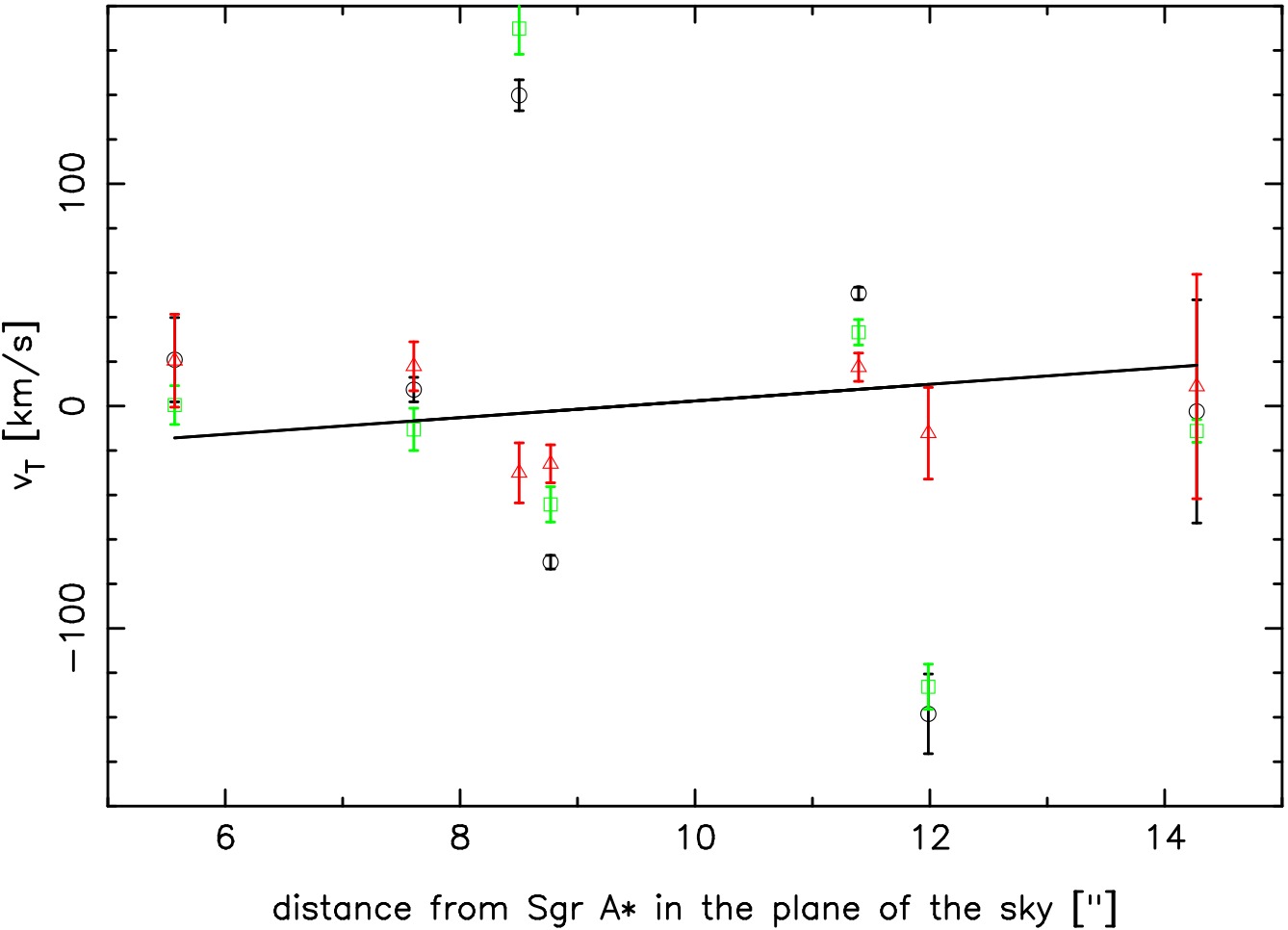}
\caption{\label{Fig:maserrot}  Projected tangential proper motion
  of the maser stars vs.\ projected distance from Sgr\,A*. Black
  circles: radio proper motions; green boxes: infrared proper motions;
  red triangles: difference between radio and infrared proper
  motions. The straight line has been fitted to the difference
  data. No relative rotation between radio and infrared frames can be
  detected within the uncertainties of this analysis.}
\end{figure}

The maser stars offer, in principle, the possibility of a cross check
because their velocity have been measured independently in the radio
reference frame. In Fig.\,\ref{Fig:maserrot} we show a plot of the
projected tangential proper motions of the maser stars as measured in
the radio and infrared frames, as well as the respective
differences. While 7 stars are far too less for any statistically
meaningful test of the absolute rotation of the cluster in the plane
of the sky, it is at least possible to check for any {\it relative}
rotation between the radio and infrared frames.

The mean difference (radio minus infrared) tangential proper motion is
$-0.5$\,km\,s$^{-1}$ with a standard deviation of
$21.8$\,km\,s$^{-1}$. A linear fit to the difference values results in
a slope of $3.8\pm11.1$\,km\,s$^{-1}$\,arcsec$^{-1}$ (uncertainty
re-scaled to a reduced $\chi2$ of 1). We conclude that there is no
detectable relative rotation between the radio and infrared reference
frames. Future measurements with longer time baselines and more maser
stars can help to reduce the still large uncertainty of this
comparison.

\subsection{Rotation in the Galactic plane \label{rotation}}

The analyses of the radial and tangential projected mean velocities
and velocity dispersions show that there may be an overall rotation in
the cluster. Overall rotation of the Galactic center star cluster has
been reported recently by \citet{Trippe2008A&A}, based on
similar proper motion measurements as in this work and as well on
spectroscopic measurements of the line-of-sight velocity of late
type-stars.

\begin{figure}[!htb]
\includegraphics[width=\columnwidth]{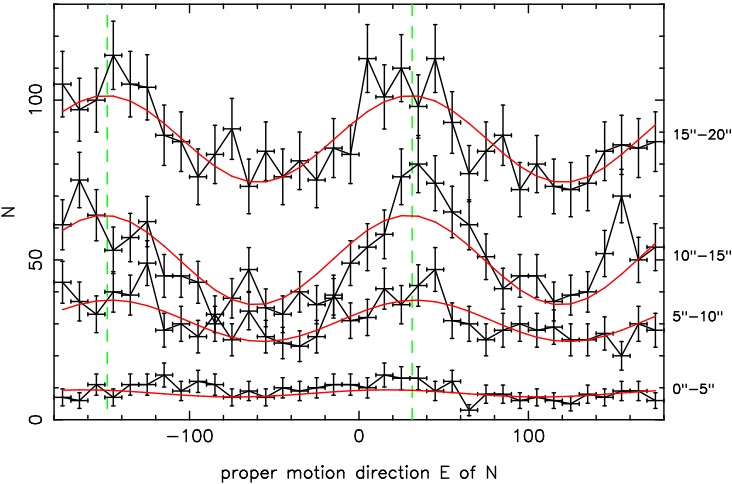}
\caption{\label{Fig:dirs} Histograms of the direction angle,
  measured east of north, of measured stellar proper motions, for the
  distance bins $0''-5''$, $''5''-10''$, $10''-15''$, and
  $15''-20''$.}
\end{figure}

The top panel of Figure\,\ref{Fig:dirs} shows histograms of the
directions of the proper motions of stars \citep[methodology adapted
  from][]{Trippe2008A&A}, measured east of north, in four different
projected radial distance bins. At $R\gtrsim5''$ a sinusoidal pattern
emerges. Cosine functions were fitted to the data. The angles and
their formal fit uncertainties of the corresponding rotation axes are
$34\pm11$\,deg ($5''\leq R\leq10''$), $29\pm7$\,deg ($10''\leq
R\leq15''$), and $30\pm6$\,deg ($15''\leq R\leq20''$).

\begin{figure*}[!htb]
\centering
\includegraphics[width=.9\textwidth]{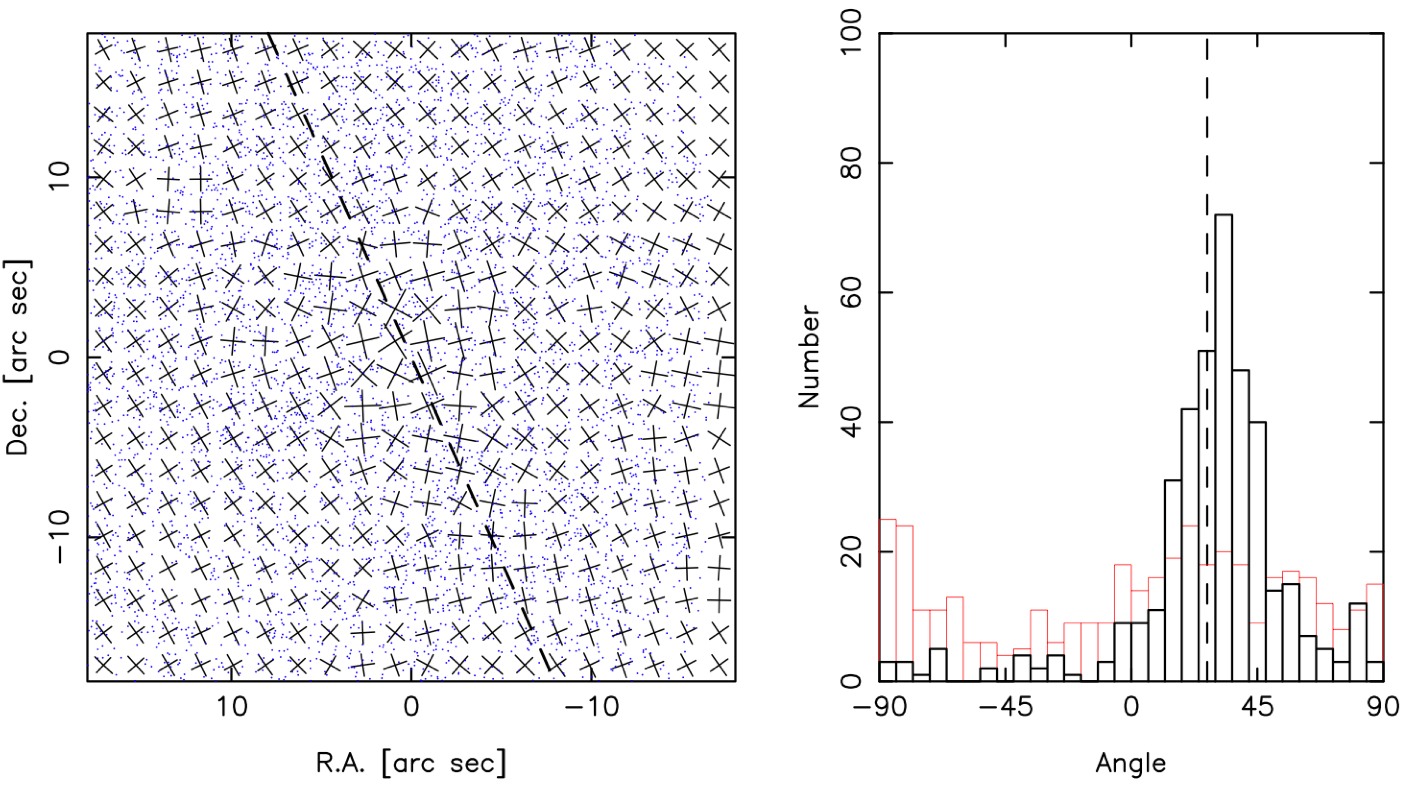}
\caption{ \label{Fig:axes} Left: Map of the principal axes of the 2D
  velocity ellipses of the late-type stars on the plane of the
  sky. Right: Histograms of the distribution of angles (measured north
  of east) defined by the longest axis of the velocity ellipses. The
  red histogram was calculated using only stars within $\pm6''$ from
  Sgr\,A*, the black one includes all stars. The dashed line indicates
  the angle of the Galactic plane.}
\end{figure*}

Insight into preferred directions of motion can also be obtained from
the velocity dispersion. In the general, three-dimensional case, the
local velocity dispersion is described by an ellipsoid with three
principal axes. Here we have evaluated the projected velocity
dispersion on grid points over the FOV. Locally, the projected
velocity dispersion is then described by ellipses. A map of the
principal axes of the 2D velocity ellipses of the late-type stars on
the plane of the sky is shown in the left panel of
Fig.\,\ref{Fig:axes}. The histograms in the right panel of
Fig.\,\ref{Fig:axes} shows the distribution of angles (measured north
of east) defined by the longest axis of the velocity ellipses. The red
histogram was calculated using only stars within $\pm6''$ from
Sgr\,A*, the black one includes all stars. The histogram for all stars
shows a clear peak at $34\pm5$\,deg.  In the central arcseconds (red
histogram) there is no well-defined preferred direction.

The finding of a preferred axis in the plane of the sky can be
explained by assuming that the NSC shows a general rotation
pattern parallel to Galactic rotation.  The angle on the sky of the
Galactic plane is $31.4$\,deg east of north in J2000 coordinates
\citep[see][]{Reid2004ApJ}. This agrees very well with the preferred
axis directions found in our analysis. \citet{Trippe2008A&A}, who
include additionally spectroscopic data for their analysis, show that
the sense of the rotation of the cluster agrees with overall Galactic
rotation.

\subsection{Velocity dispersion}

\begin{figure*}[!htb]
\includegraphics[width=\textwidth]{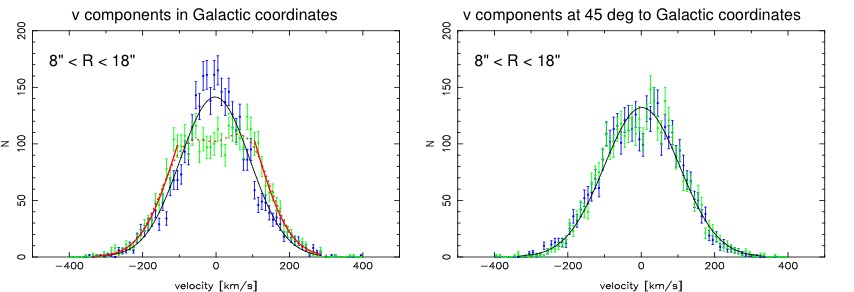}
\caption{\label{Fig:siglb} Left: Histograms of the velocity parallel
  (green), $v_{l}$, and perpendicular (blue), $v_{b}$, to the angle of
  the Galactic plane on the sky. The black solid lines is a fit to the
  histogram of $v_{b}$ with a Gaussian function. The red dotted line is
  a fit to the histogram of $v_{l}$ with the sum of two Gaussian
  function. The straight red lines are Gaussian fits to the data with
  $v_{l}> 100$\,km\,s$^{-1}$ and $v_{l}< 100$\,km\,s$^{-1}$,
  respectively, using the same velocity dispersion as for the Gaussian
  that fits the histogram of $v_{b}$. Right: Histograms of the velocity
  dispersion parallel (green) and perpendicular (blue) to a plane at
  45\,deg to the angle of the Galactic plane on the sky. The Gaussian
  functions fit to the two histograms are indistinguishable within the
  fit uncertainties. The plots include only data within $8''>R>18''$
  projected distance from Sgr\,A*.}
\end{figure*}

The measured proper motion velocities were transformed to velocities
parallel and perpendicular to the Galactic plane in order to examine
the influence of cluster rotation on the two-dimensional velocity
dispersion. Plots of the corresponding velocity ($v_{l}$ along
Galactic longitude and $v_{b}$ along Galactic latitude) histograms are
shown in the left panel of Fig.\,\ref{Fig:siglb}.  The plots include
only data within $8''>R>18''$ projected distance from Sgr\,A* because
inside of $R=8''$ the projected velocity dispersion shows a clear
Keplerian increase and at $R>18''$ the FOV becomes increasingly
asymmetric (see Fig.\,\ref{Fig:velmap}).  The histogram of the
velocity along Galactic latitude can be fit well with a Gaussian
function with a mean of $-2.6\pm1.7$\,km\,s$^{-1}$ and a standard
deviation of $95.2\pm1.3$\,km\,s$^{-1}$ ($1\,\sigma$
uncertainties). The histogram of the velocity along Galactic longitude
appears broadened, with an indication of two symmetric peaks, due to
the overall rotation of the cluster. The peak at negative $v_{l}$
appears somewhat smaller. This is to be expected since stars with
$v_{l} <0$ will have a larger probability to be located near the
backside of the cluster and thus a somewhat smaller chance to be
detected (e.g., because of extinction). The
histogram of $v_{l}$ can be formally fit with two Gaussians (dotted
red line in left panel of Fig.\,\ref{Fig:siglb}), having means of
$-77.2\pm9.9$\,km\,s$^{-1}$ and $82.1\pm9.0$\,km\,s$^{-1}$ and
standard deviations of $73.7\pm4.8$\,km\,s$^{-1}$ and
$69.3.7\pm4.3$\,km\,s$^{-1}$, respectively. However, while formally
correct, this fit does not reflect well the physical reality because
it assumes a ring of stars rotating with a constant velocity. The
stars are distributed over a range of distances in an approximately
spherical cluster and the rotation velocity can be expected to be a
function of distance from the center \citep[see
  also][]{Trippe2008A&A}.

The right panel of Fig.\,\ref{Fig:siglb} shows velocity histograms
after projecting the velocities parallel and perpendicular to an axis
that runs at 45\,deg to the angle of the Galactic plane. The effect of
Galactic rotation is in this case distributed evenly among the two
histograms. The Gaussian functions fitted to the two histograms are
indistinguishable within their uncertainties. They have have peak
values of $131.0$ and $132.1$, mean values of
$-1.8\pm1.8$\,km\,s$^{-1}$ and $2.1\pm1.8$\,km\,s$^{-1}$, and standard
deviations of $103.6\pm1.3$\,km\,s$^{-1}$ and
$102.9\pm1.4$\,km\,s$^{-1}$. We interpret this as evidence that no
anisotropy in the kinematics of the late-type stellar population is
detected that is larger or comparable to the rotation signature.

Assuming isotropy, we can obtain a zeroth order estimate of the maximum
rotation velocity of the NSC at the edge of our FOV from the histogram
of $v_{l}$. The histogram of $v_{l}$ can be assumed to result from the
convolution of a Gaussian with some function that describes the
rotation velocity.  The Gaussian is hereby assumed to have the same
standard deviation as the one that fits the histogram of
$v_{b}$. Fitting the flanks of the histogram of $v_{l}$ with such a
Gaussian will hence give us an estimate of the maximum rotation
velocity within our FOV via the shift of the mean. Such fits (straight
red lines in left panel of Fig.\,\ref{Fig:siglb}) result in velocities
of $20.3\pm2.8$\,km\,s$^{-1}$ and $-19.0\pm3.0$\,km\,s$^{-1}$. This is
in good agreement with the model of \citep{Trippe2008A&A}, which
would result in a rotation velocity of $25.6\pm6.5$\,km\,s$^{-1}$ at a
distance of $18''$. 

It is important to consider the systematic uncertainty in the proper
motion velocity dispersion. The uncertainties of the individual
stellar velocities will cause the velocity dispersion to be biased
toward larger values \citep[see also][]{Genzel2000MNRAS}. We tested
this effect by a MC simulation in which 5000 stars were drawn from a
distribution with an intrinsic $\sigma_{i}=100$\,km\,s$^{-1}$. If the
uncertainty of all individual velocity measurements is $\delta
v=$12\,km\,s$^{-1}$, the average resulting $\sigma$, $\sigma_{res}$, from the randomly
selected velocities in 100 runs is
$\sigma_{res}=$100.7$\pm1.0$\,km\,s$^{-1}$. For $\delta
v=$25\,km\,s$^{-1}$ we obtain
$\sigma_{res}=$103.1$\pm1.1$\,km\,s$^{-1}$, and for $\delta
v=$50\,km\,s$^{-1}$ the measured velocity dispersion is
$\sigma_{res}=$111.8$\pm1.1$\,km\,s$^{-1}$. As mentioned above, more
than $80\%$ of the stars in the center field have velocity
uncertainties $dv_{x,y}<25$\,km\,s$^{-1}$. Therefore we estimate that
the bias on the velocity dispersion due to the uncertainties of the
individual measurements in our data is $<$5\%. Nevertheless, the
uncertainties of the individual stellar velocities will be taken into
account in the modelling outlined in section\,\ref{sec:model}.

\subsection{Runaway stars}

There are a few stars with proper motion velocities in either axis
exceeding 400\,km\,s$^{-1}$. We have checked the corresponding data
and proper motion fits individually. Almost all of these stars are
among the faintest in the sample and show large $1\sigma$
uncertainties of their velocities.  They can therefore be only
regarded as candidates for extremely fast stars.  There is one notable
exception, however, a star at $(-6.75'',18.41'')$ offset from Sgr\,A*
(marked by a red circle in Fig.\,\ref{Fig:velmap}). Its projected
radial and tangential proper motion velocities are $407.4\pm16.5$ and
$119.0\pm22.3$\,km\,s$^{-1}$, respectively. Assuming that the star is
located on the plane of the sky and has zero velocity along the
line-of-sight, we compute a minimum mass of
$1.6\times10^{7}\pm1.3\times10^{6}$\,M$_{\odot}$ required to bind this
star to the cluster. Since this mass is unrealistically high, we
conclude that the star is a solid candidate for an object that escapes
the Milky Way nuclear star cluster.

\subsection{Offset data set}

The mean velocities and velocity dispersions for the offset data are
described in appendix\,\ref{app:offset}. The main result is that the
offset data show that the projected velocity dispersion stays
approximately constant
and isotropic within the measurement uncertainties  out to a
projected radius of $R=30''$.

%\clearpage

\section{Dynamics and masses}\label{sec:model}

\subsection{Assumptions}

An advantage of proper motions over spectroscopically determined, 
line-of-sight velocities is that one obtains a more complete picture 
of the internal kinematics \citep{Leonard1989ApJ}.
In a spherical nonrotating system, knowledge of the proper motion
velocities at all projected radii is equivalent to knowledge of the 
shape of the velocity ellipsoid at all internal radii.
The enclosed mass then follows uniquely from the Jeans equation.
If only line-of-sight velocities are available,
inferences about the mass will suffer from a (potentially extreme)
degeneracy due to the unknown shape of the velocity ellipsoid
and its variation with radius \citep[e.g.][]{Dejonghe1992ApJ}.

Based on the results in the previous section, we here model
the late-type stars in the nuclear cluster as a spherical,
nonrotating population.
The observed stars are assumed to move in the combined gravitational
field of the black hole, plus an additional distributed mass component,
also assumed to be spherically symmetric but otherwise undetermined.
We denote the observed (1d) velocity
dispersions parallel and tangential to the radius vector $\mathbf{R}$ 
in the plane of the sky as $\sigma_R(R), \sigma_T(R)$.
The velocity dispersions parallel and tangential to the spatial
(not projected) radius vector $\mathbf{r}$ are 
$\sigma_r(r), \sigma_t(r)$.

As discussed in \cite{Leonard1989ApJ}, there is a formally
unique relation between the (observed) functions ($\sigma_R, \sigma_T$)
and the (intrinsic) functions ($\sigma_r, \sigma_t$) if
the number density profile $n(r)$ is also known.
Given the intrinsic velocity dispersions, the enclosed mass is
\begin{equation}
GM(r) = -{r^2\over n} {d(n\sigma_r^2)\over dr} - 
2r\left(\sigma_r^2-\sigma_t^2\right).
\label{eq:Jeans}
\end{equation}

The uniqueness of the kinematical deprojection is a strong motivation
for modelling the nuclear cluster in this way.  One cost is that we
are not able to reproduce the gradual alignment of the velocity
vectors parallel to the Galactic plane that is observed outside of
$\sim 6''\approx 0.3$ pc (Fig.~\ref{Fig:axes}).  Reproducing this
feature would require adding more complexity to our model, which in
turn would require more kinematical data in order to constrain the
extra degrees of freedom.  For instance, the cluster could be
represented as an oblate spheroid in which the orbits respect three
independent integrals of the motion; the additional information needed
to constrain the model could come from line-of-sight velocities
measured over the 2d field.  We believe that our approach is the most
appropriate given the information currently available.  

Neglecting the rotation of the cluster parallel to Galactic rotation
in our analysis appears justified for two reasons. First, the actual
velocity of rotation and its radial dependence is not well constrained
(see discussion in sub-section\,\ref{sec:rotation}). Second, the
influence of rotation on the mass estimates for the central parsec
will be small. Even if the rotation velocity were as high as
30\,km\,s$^{-1}$ at 1\,pc, this would still be only about 30\% of the
velocity dispersion. Since both quantities enter the Jeans equation
quadratically, the error on the enclosed mass would only be of order
10\%.

\subsection{Moment estimators}

As a first step, we examine the moment mass estimator
defined by \cite{Leonard1989ApJ}: 
\begin{equation}
\langle M(r)\rangle = {16\over 3\pi G} 
\langle R\left(2V_R^2 + V_T^2\right)\rangle
\label{eq:LM}
\end{equation}
where angle brackets denote number-weighted averages over
the entire system.
In a cluster with constant mass-to-light ratio (i.e.,
$\rho(r)\propto n(r)$),
$\langle M(r)\rangle = M_T/2$ with $M_T$ the total mass;
while if all the mass is located in a central point of mass 
$M_{BH}$, $\langle M(r)\rangle=M_{BH}$.
Unlike the projected mass estimators of 
\cite{BahcallTremaine1981ApJ} and \cite{Heisler1985ApJ},
the LM estimator contains no undetermined parameter to compensate
for the unknown anisotropy.
However, like all moment estimators, equation~(\ref{eq:LM}) contains
only limited information about  the {\it distribution} of the mass.
In addition, as emphasized by 
\cite{Genzel2000MNRAS}, \cite{Figer2003ApJ} and \cite{Zhu2008ApJ},
moment estimators must be applied cautiously in cases where 
data do not extend over the entire system, or where the observed
sample is dominated by stars that are intrinsically far from
(but projected near to) the center.

\begin{figure}[!htb]
\includegraphics[width=0.9\columnwidth,angle=-90.]{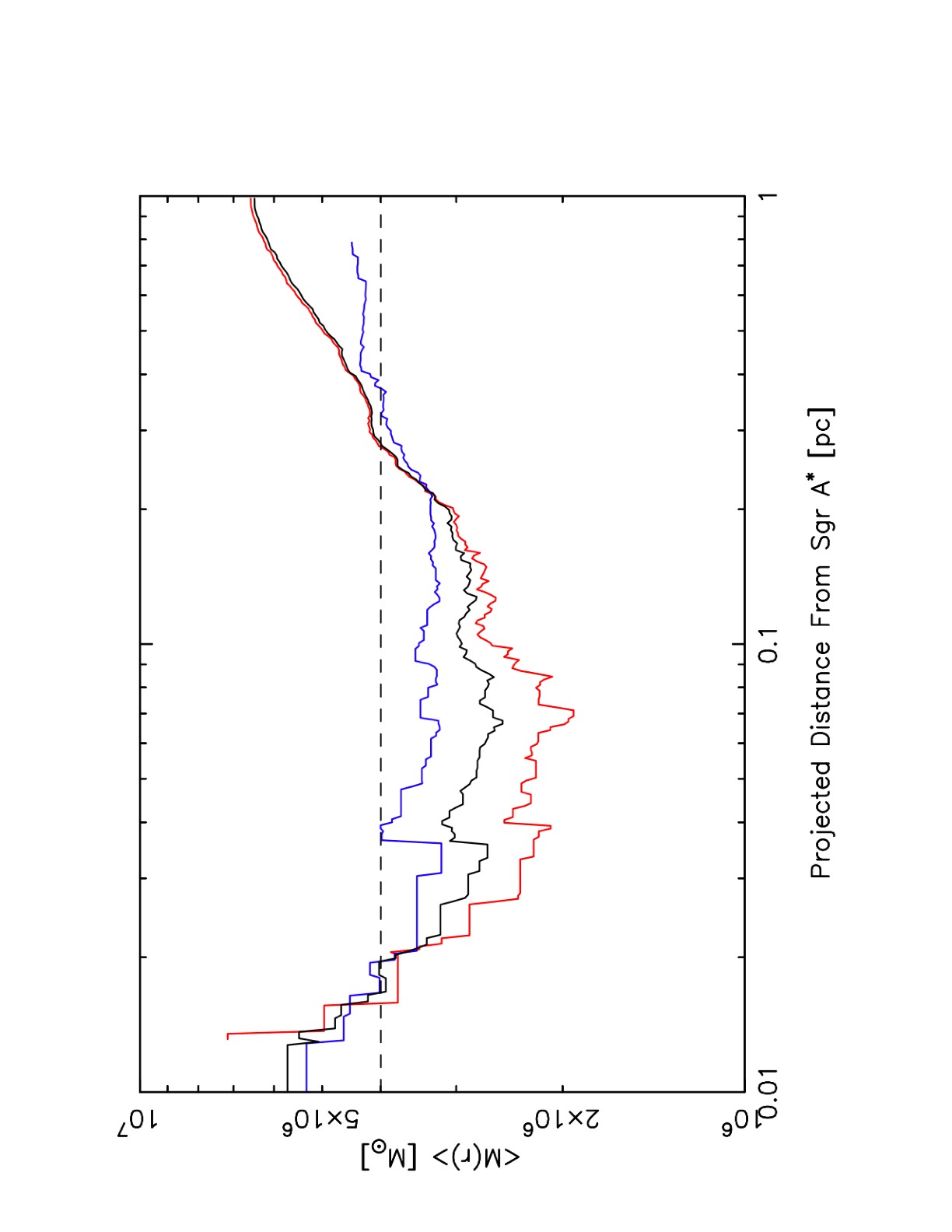}
\caption{\label{Fig:LM} LM mass estimator as a function of maximum
projected distance from Sgr\,A*, based on the proper motions in the 
central field. 
{\it Red:} late-type stars; {\it blue:} early-type stars:
{\it black:} combined sample.}
\end{figure}

Figure\,\ref{Fig:LM} shows the result of computing (\ref{eq:LM}) as a
function of the outer radius of the sample.  Results from both
late-type and early-type stars, considered separately, are shown; also
shown is the result using the combined sample.  The late-type stars
alone imply a mass of $\sim 2\times 10^6 M_\odot$ at small radii,
$R\lesssim 0.1$ pc, increasing gradually toward larger radii.  The
smaller sample of early-type stars yields a mass estimate that is more
nearly constant with radius, $M_{\rm BH}\approx 3-4\times
10^6M_\odot$.  As noted by \citet{Paumard2006ApJ} and
\citet{Lu2008arXiv}, the surface number density of early-type stars
decreases steeply with distance from Sgr\,A*, like a power-law with
index $\sim-2$. Therefore, more than $90\%$ of the early type stars
are contained within a projected radius $R<0.5$\,pc from Sgr\,A*. The
sample of the early-type stars can therefore be regarded as
complete. The LM mass estimator for these stars at the largest
projected radii can for this reason be regarded as an accurate
estimate of the BH mass.

Estimates of $M_{\rm BH}$ based on the assumption of Keplerian motion
for the closest stars to the central dark mass are generally
considered to be the most reliable.  For an assumed GC distance of
8.0\,kpc, the most recently published values are $M_{\rm
  Sgr\,A*}=4.1\pm0.6\times10^{6}\,M_{\odot}$ \citep{Ghez2003ApJ},
$4.1\pm0.4\times10^{6}\,M_{\odot}$ \citep{Eisenhauer2005ApJ},
$3.7\pm0.2\times10^{6}\,M_{\odot}$ \citep{Ghez2005ApJ},
$4.1\pm0.1\times10^{6}\,M_{\odot}$ \citep{Ghez2008ApJ}, and
$4.0\pm0.1\times10^{6}\,M_{\odot}$ \citep{Gillessen2008arXiv}.  The
value $4.0\times 10^6 M_\odot$ is shown as the dashed line on
Figure~\ref{Fig:LM}; it is consistent with the LM mass estimate
derived from the young stars, but lies above the estimate derived from
the old stars for $R\lesssim 0.3$ pc. 

The comparison is complicated by the fact that the NSC  extends
  probably far beyond the central parsec.  Samples that extend to
projected distances of $\sim $ 1 pc from Sgr A$^*$ contain many stars
that are moving far from the BH and that feel the gravitational force
from the distributed mass.  The gradual rise in $\langle M(r)\rangle$
at $R\gtrsim 0.3$ pc  -- in contrast to the near constant LM mass
  estimator for the early-type stars --  may reflect this.  On the
other hand, restricting the sample to small projected radii is
inconsistent with the assumptions made in deriving the LM estimator
and will also bias the estimate.

We note that many previous studies based on radial velocities or
proper motions of stars in the inner parsec have found values of
$M_{\rm BH}$ that were $\sim30-50\%$ lower than the currently accepted
value
\citep[e.g.][]{Eckart1997MNRAS,Ghez1998ApJ,Genzel2000MNRAS,Chakrabarty2001AJ}.
\cite{Figer2003ApJ} and \cite{Zhu2008ApJ} have argued that this bias
can be attributed to the low space density of late-type stars near Sgr
A$^*$.  Indeed there appears to be a ``hole'' in the distribution of
old stars: their number counts are flat inside of $\sim 0.25$ pc,
implying a space density that may even decline toward Sgr A$^*$
\cite[][and discussion therein]{Zhu2008ApJ}.  In the extreme case of
     {\it no} stars intrinsically close to the BH, mass estimates
     based on the moment equations would clearly be biased toward low
     values. The drop of the LM mass estimate for the late-type stars
     down to only about 50\% of the actual BH mass is most probably
     related to this deficit which leads to an underestimation of the
     true 3D distances of the stars from Sgr\,A*. The fact that the
     proper motion velocity dispersions of the dominant (late-type)
     population are observed to rise toward the projected center
     (e.g. Fig.~\ref{Fig:sigma}) suggests that at least some of the
     old stars are physically close to the BH.  Nevertheless,
     Figure~\ref{Fig:LM} reinforces the idea that estimates of the BH
     mass from proper motions might be significantly influenced by the spatial
     distribution of the kinematical sample.

\subsection{Isotropic modelling}

Proper motion velocities of the late-type stars
appear to be nearly isotropic at most radii (Fig.~\ref{Fig:sigma}).
As our next step, we therefore model the NSC assuming 
$\sigma_r(r)=\sigma_t(r)\equiv\sigma(r)$ at all radii.
Writing $\sigma_P^2(R)\equiv (1/2)\left[\sigma_R^2(R)+\sigma_T^2(R)\right]$,
it is easy to show \citep[e.g.][]{Genzel1996ApJ} that
\begin{equation}
\Sigma(R)\sigma_P^2(R) = 2\int_R^{\infty} {dr\ r\over\sqrt{r^2-R^2}} n(r)\sigma^2(r).
\label{eq:sproject}
\end{equation}
Here $\Sigma(R)$ is the surface number density of stars at a given
projected distance.  We can express $n\sigma^2$ in terms of the
enclosed mass (BH plus stars) via the Jeans equation (\ref{eq:Jeans})
with $\sigma_r=\sigma_t$:
\begin{equation}
n(r)\sigma^2(r) = \int_r^{\infty} dr' {G M(r') n(r')\over r'^2}
\label{eq:Jeansiso}
\end{equation}
and combining (\ref{eq:sproject}) and (\ref{eq:Jeansiso}),
\begin{eqnarray}
\Sigma(R)\sigma_P^2(R) &=& 2G\int_R^\infty 
{dr\ r\over\sqrt{r^2-R^2}}\int_r^\infty {dr'n(r')M(r')\over r'^2} \\
&=& 2G\int_R^\infty dr\ {\sqrt{r^2-R^2}n(r)M(r)\over r^2} .
\end{eqnarray}
Finally, writing $\Sigma(R)$ as the projection of $n(r)$,
\begin{equation}
\Sigma(R) = 2\int_R^\infty {dr\ r n(r)\over\sqrt{r^2-R^2}}
\end{equation}
yields
\begin{equation}
G^{-1}\sigma_P^2(R) = {\int_R^\infty dr\ r^{-2}\left(r^2-R^2\right)^{1/2} n(r) M(r) \over
\int_R^\infty dr\ r \left(r^2-R^2\right)^{-1/2}n(r)}.
\label{eq:sigmap}
\end{equation}

\begin{figure}
\includegraphics[width=0.85\columnwidth,angle=-90.]{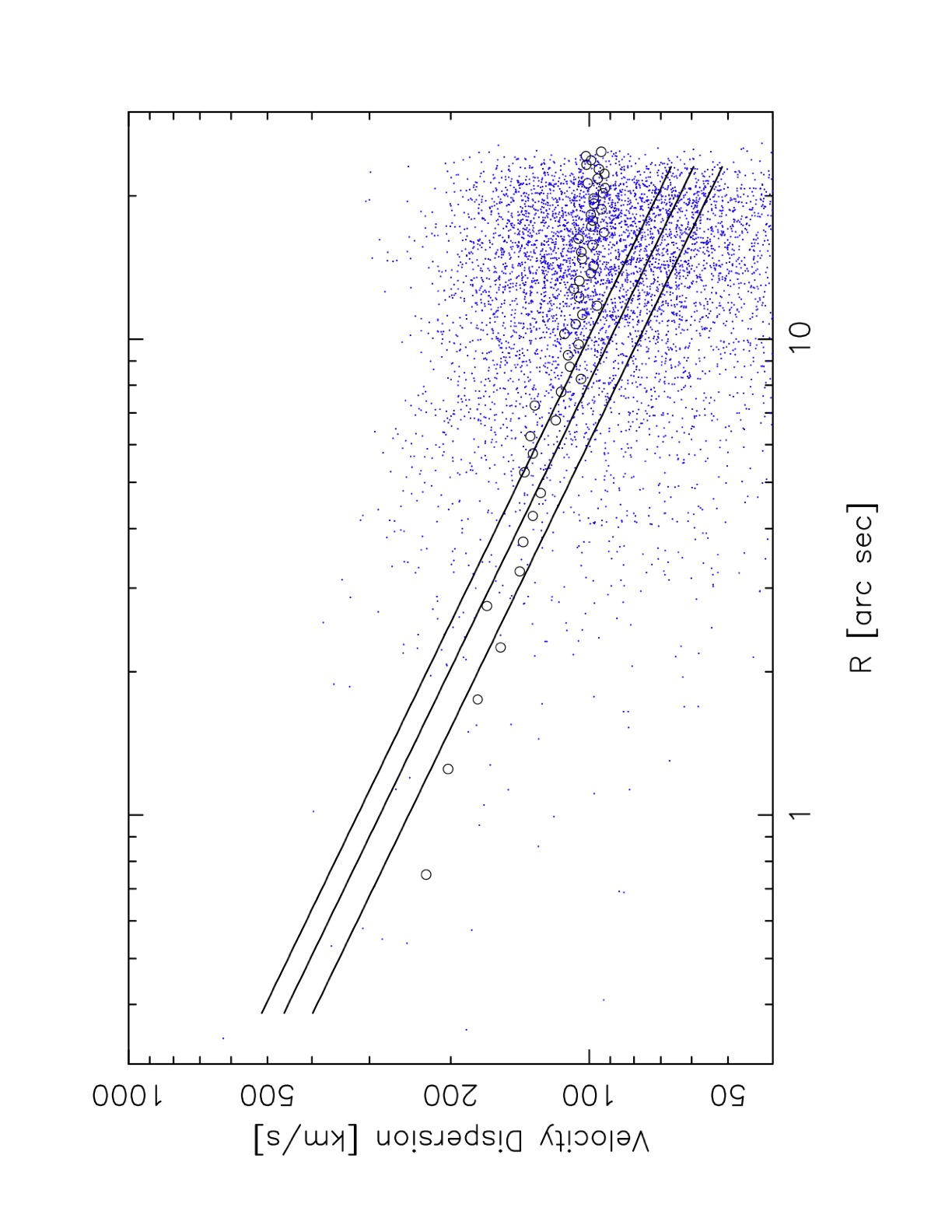}
\caption{\label{Fig:iso} Modelling the nuclear cluster assuming
that all the mass is contained in the black hole and that the velocities
are isotropic.
The number density profile of the kinematical sample was assumed
to be a single power law of the radius.
Open circles are velocity dispersions computed using fixed
radial bins; blue points are absolute values of the measured velocities.
Straight lines show model predictions 
for $M_{\rm BH}=(3,4,5)\times 10^6 M_\odot$.}
\end{figure}

The simplest case to consider is a model in which all of the mass is in 
the central black hole, $M(r)=M_{\rm BH}$.
If we also assume for simplicity that $n(r)=n_0(r/r_0)^{-\gamma}$,
then equation~(\ref{eq:sigmap}) predicts
\begin{equation}
\sigma_P^2(R) = F(\gamma) {GM_{\rm BH}\over R}, \ \ \ \ 
F(\gamma) = {1\over 2} {\left[\Gamma(\gamma/2)\right]^2\over
\Gamma({\gamma+3\over 2}) \Gamma({\gamma-1\over 2})}.
\end{equation}
The function $F$ is weakly dependent on $\gamma$; for
$1.5\le\gamma\le 3.5$, $0.18\le F(\gamma)\le 0.21$.
Figure~\ref{Fig:iso} shows the velocity dispersion profile predicted
by this model, assuming $F=0.18$ and 
$M_{\rm BH}=(3,4,5)\times 10^6M_\odot$.
It is clear that no value for $M_{\rm BH}$ can fit the data 
both at large and small radii.
For a BH mass in this range,
the observed velocities
begin to rise appreciably above the model
predictions at $R\gtrsim 5''\approx 0.25$ pc, suggesting that an additional
component of the mass becomes important at this radius.
This is consistent with the behavior noted above for the
moment mass estimator.
In addition, the observed velocities fall below the predicted
values inside of $\sim 4''\approx 0.2$ pc, again consistent
with the LM estimator, which implied a lower value for $M_{\rm BH}$
when only the inner data were used.

Continuing under the assumption of isotropy, we can
constrain a model of the stellar mass density by comparing
the predictions of equation~(\ref{eq:sigmap}) with the observed
velocities via
\begin{equation}
\chi^2 = 
\sum_{i=1}^N {\left[V_i^2-\sigma_P^2(R_i)\right]^2\over
\Delta^2(R_i)}
\label{eq:chisqiso}
\end{equation}
under various assumptions about $M(r)$.  Here, $N$ is the number of
measured velocities, $V_i^2=v_i^2 - {\rm error}^2(v_i)$ is the square
of the $i$th measured velocity corrected for measurement error,
$\sigma_P(R)$ is the model prediction, and $\Delta$ is an estimate of
the dispersion of $V^2$ about its mean value at radius $R$; the latter
was computed using an adaptive kernel estimate of the velocity
dispersion profile. The minimum reduced $\chi^2$ of our isotropic
models was 0.96.

We minimized $\chi^2$ over a set of parameters defining
the mass distribution:
\begin{equation}
M(r) = M_{\rm BH} + 4\pi \int_0^r dr\ r^2\rho(r).
\end{equation}
If we assume simple power laws for both $n(r)$ and $\rho(r)$,
\begin{equation}
n(r) \sim r^{-\gamma},\ \ \ \ \rho(r)\sim r^{-\Gamma},
\end{equation}
the integral in the numerator of 
equation~(\ref{eq:sigmap}) is divergent unless
$\gamma+\Gamma > 3$.
This creates difficulties since, in the region
of interest, $\gamma$ is small requiring large $\Gamma$.
To avoid these unphysical divergences, we represented both 
$n(r)$ and $\rho(r)$ as broken power laws:
\begin{equation}
n(r) = n_0 \left({r\over r_0}\right)^{-\gamma} 
\left(1+{r\over r_0}\right)^{\gamma-A}, 
\label{eq:nmodel}
\end{equation}
\begin{equation}
\rho(r) = \rho_0 \left({r\over r_M}\right)^{-\Gamma} 
\left(1+{r\over r_M}\right)^{\Gamma-B}.
\label{eq:rhomodel}
\end{equation}
A number of studies have found a large-radius ($r\gtrsim 1$pc)
dependence $n(r)\sim r^{-1.8}$ for the old stellar population
\citep[][and references therein]{Schoedel2007A&A}.  We accordingly set
$A=1.8$ in eq.~(\ref{eq:nmodel}).  The form of $n(r)$ at smaller radii
is less well determined.  As noted above, the number counts of
late-type stars appear to flatten or even decline inside $\sim 0.5$
pc, and this result was recently strengthened via  a new analysis
  by Buchholz, Sch\"odel, \& Eckart (2009, submitted to A\&A).  Based on
the latter paper, we set $\gamma=0.5$ and $r_0=20''$ in
eq.~(\ref{eq:nmodel}) ($r_0$ is larger than the break radius
  given in \citet{Schoedel2007A&A} because the latter must be
  de-projected.).  We note that $\gamma=0.5$ is the flattest slope
that is consistent with an isotropic phase-space density in a
point-mass potential.

For the mass density $\rho(r)$ we set $r_M=5$ pc $\approx 100''$ and
$B=4$, yielding essentially a single power-law dependence,
$\rho\sim r^{-\Gamma}$, over the 
radial range of our data.
For a given $n(r)$, the three remaining parameters are
the BH mass, the mass density slope $\Gamma$, and the mass density 
normalization $\rho_0$.
These can be written
\begin{equation}
M_{\rm BH}, \ \ M_\star(<1 {\rm pc}),  \ \ \Gamma
\end{equation}
where $M_\star (<1 {\rm pc})$ is the stellar mass within
one parsec.

\begin{figure}
\includegraphics[width=0.85\columnwidth,angle=-90.]{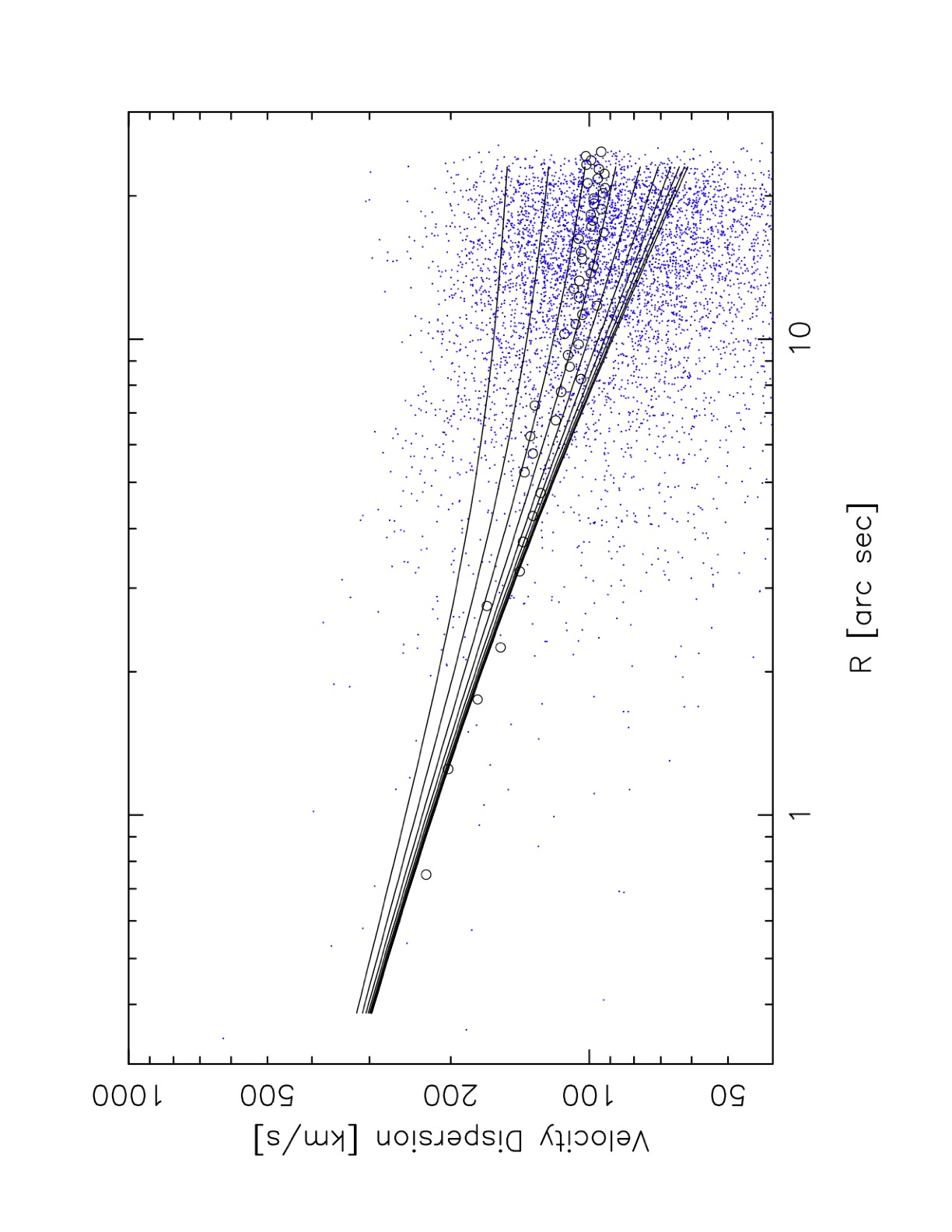}
\caption{\label{Fig:iso2} Isotropic modelling of the nuclear cluster 
including a contribution to the gravitational potential from
the stars, assumed to have a mass density that falls off
as $\rho\propto r^{-\Gamma}, \Gamma=1$
over the radial range covered by the data.
$M_{\rm BH}=3.6\times 10^6M_\odot$, and the various curves show
the predicted velocity dispersion profiles for a range of
normalizations of the stellar density, from
$5\times 10^4 M_\odot \le M_\star(<1 {\rm pc}) \le 
5\times 10^6 M_\odot$ with equal logarithmic steps.
Other symbols are as in Fig.~\ref{Fig:iso}.}
\end{figure}

Figure \ref{Fig:iso2} illustrates the effects of including
a non-zero stellar mass.
Each model shown there has $M_{\rm BH}=3.6\times 10^6M_\odot$
and $\Gamma=1$, i.e. the stellar mass density falls as $\sim r^{-1}$
in the region where there are velocity data.
The best fit from this series is obtained for 
$M_\star(<1 {\rm pc})\approx 1.5\times 10^6M_\odot$.
Because the central slope of $n(r)$ ($\gamma=0.5$) is flatter 
than was assumed for Fig.~\ref{Fig:iso} ($\gamma=1.5$),
the observable effect of the BH on the stellar motions near the projected
center is smaller and a more massive BH is required to reproduce
the inner velocity dispersions. 

\begin{figure}
\includegraphics[width=0.9\columnwidth,angle=-90.]{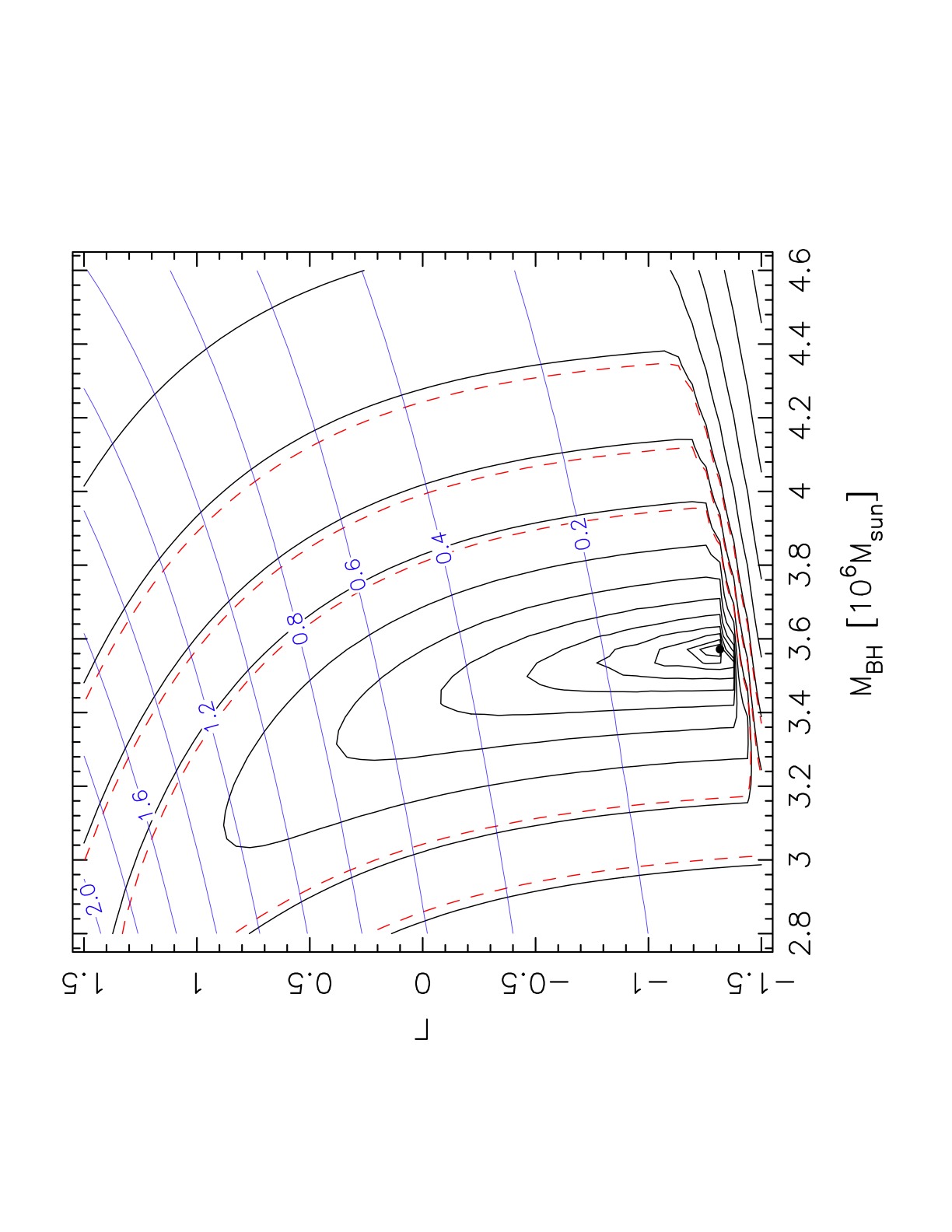}
\caption{\label{Fig:Isocont} Results of isotropic modelling of the 
NSC assuming the mass model of eq.~(\ref{eq:rhomodel}).
The three parameters ($M_{\rm BH}, M_\star, \Gamma$)
were varied in comparing the fit of the model to the velocity dispersion data,
eq.(\ref{eq:chisqiso}).
Black (thick) curves are contours of constant $\chi^2$, separated by a
constant factor of $10^{0.3}$;
dashed (red) curves indicate (68\%, 90\% and 99\%) confidence intervals.
Blue (thin) curves are contours of the best-fit value of
$M_\star(r<1 {\rm pc})$ at each value of ($M_{\rm BH}, \Gamma$);
these curves are labelled by $M_\star/10^6 M_\odot$.
The overall best-fit model is indicated by the filled circle.}
\end{figure}

\begin{figure}
\includegraphics[width=0.9\columnwidth,angle=-90.]{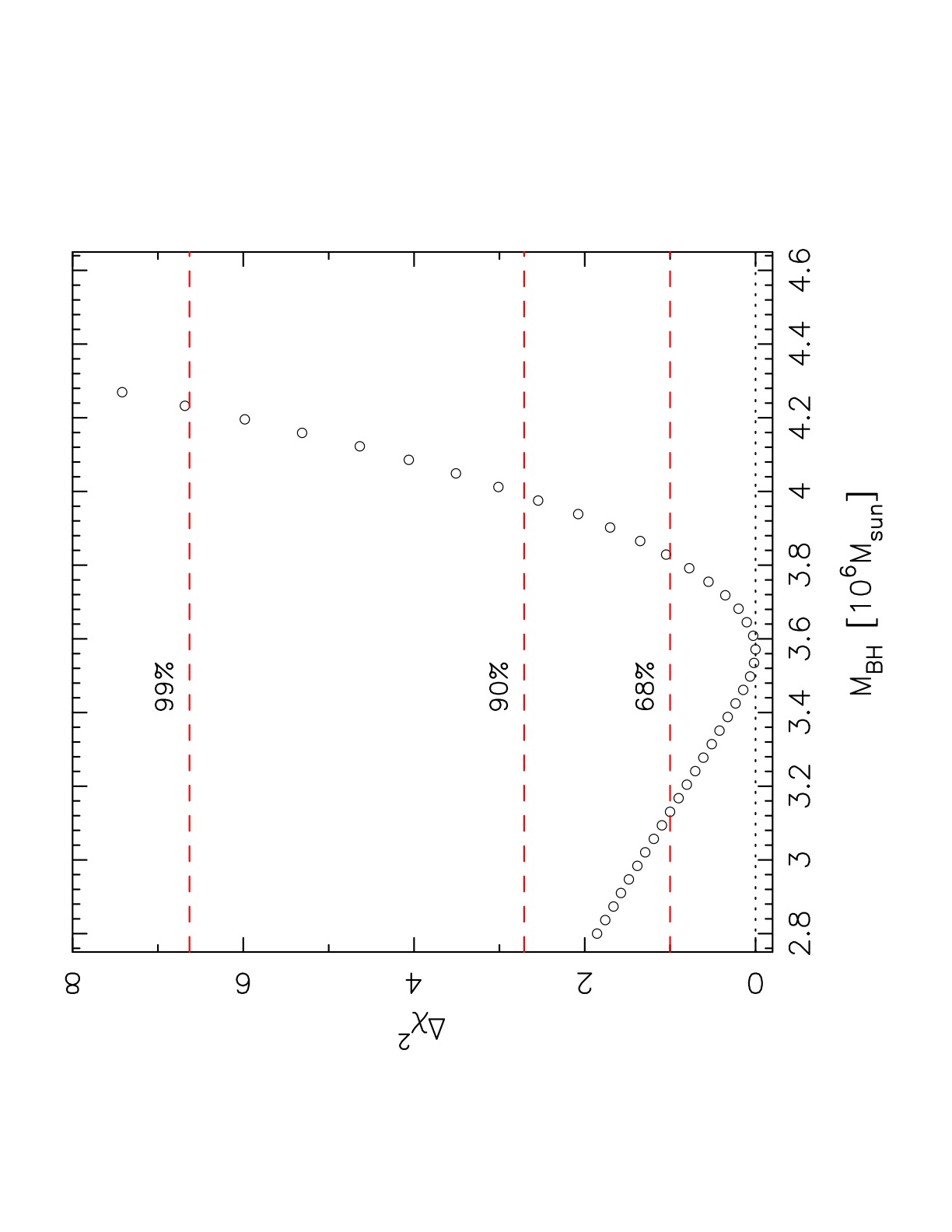}
\caption{\label{Fig:Isochisq} $\Delta\chi^2$ vs. BH mass for the
  isotropic models.
Plotted is the minimum $\chi^2$ value at each $M_{\rm BH}$ 
among the set of ($M_\star,\Gamma$) values considered in
Fig.~\ref{Fig:Isocont}.
Dashed (red) lines show values of $\Delta\chi^2$ corresponding
to (68\%,90\% and 99\%) confidence, as indicated.
}
\end{figure}

Figure \ref{Fig:Isocont} summarizes the fits of a set of mass models
computed on a 3d grid in parameter space.
For each ($M_{\rm BH},\Gamma$), the plot shows contours of two 
quantities associated with the model that best fits the kinematical
data: $M_\star(<1 {\rm pc})$ and $\chi^2$.
The red (dashed) contours show 
confidence intervals of 68\%,90\%, and 99\% \citep{Lampton1976ApJ}.
The overall best-fit model from this set has a negative $\Gamma$,
i.e., the mass density {\it increases} with radius.
However the value of $\Gamma$ is very weakly constrained, 
especially if $M_{\rm BH}$ is considered to be a free parameter.
Moving up along the near-plateau in $\chi^2$, a decreasing
BH mass can be compensated for by increasing the stellar mass
and by making the mass density profile more steep, mimicking
a central point mass.
Figure~\ref{Fig:Isochisq} plots the minimum $\chi^2$ value at each
$M_{\rm BH}$.  
The ``correct'' BH mass, $4.0\times 10^6M_\odot$, is consistent
at the 90\% level with the isotropic modelling.

The best-fitting isotropic models all require a non-zero
distributed mass.
If $\rho(r)$ is assumed to  decrease with radius,
i.e. $\Gamma>0$,
then the implied mass within $1$ pc is always greater than
$\sim 0.4\times 10^6 M_\odot$ for 
$3.5\times 10^6 \lesssim M_{\rm BH}/M_\odot
\lesssim 4.5\times 10^6$.
If $M_{\rm BH}=4.0\times 10^6M_\odot$ and $\Gamma>0$,
the distributed mass within 1 pc
must be greater than $\sim 0.5\times 10^6M_\odot$.

%\clearpage

\subsection{Anisotropic modelling}

We next consider models that allow the two independent
components ($\sigma_r,\sigma_t$) of the stellar velocity dispersion
to be different at each intrinsic radius $r$.
In principle, complete knowledge of the two proper-motion
velocity dispersion profiles $\sigma_R(R),\sigma_T(R)$,
together with the deprojected number-density profile $n(r)$,
is equivalent to complete knowledge of $\sigma_r(r)$ and $\sigma_t(r)$
\citep{Leonard1989ApJ}.
The enclosed mass would then follow uniquely from the Jeans equation.
In practice, we only measure the proper motions over a limited
range of radii, and inferences about the mass density will
in general depend on the degree of anisotropy beyond the last
measured point \citep[e.g.][]{Merritt1988AJ}.
In addition, as discussed above, the assumption of spherical
symmetry in the modelling is not completely consistent with the
observed behavior of the proper motions at large radii.
For these reasons, a direct deprojection of the proper-motion
velocity dispersions was deemed undesirable.
Instead we constructed anisotropic models and compared their
projected properties with the data.
We stress that our algorithm is completely nonparametric in terms
of its treatment of $\sigma_r(r)$ and $\sigma_t(r)$.

The relation between the intrinsic and projected velocity dispersions
is 
\begin{eqnarray}
&&\Sigma(R)\sigma_R^2(R) \nonumber \\
&& = 2\int_R^{R_{max}} {r dr\over\sqrt{r^2-R^2}} 
\bigg[{R^2\over r^2}n(r)\sigma_r^2(r) 
+ \left(1-{R^2\over r^2}\right) n(r)\sigma_t^2(r)\bigg], \nonumber \\
&&\Sigma(R)\sigma_T^2(R) = 2\int_R^{R_{max}} {rn(r)\sigma_t^2(r)\over 
\sqrt{r^2-R^2}}dr.
\label{Eq:Leonard}
\end{eqnarray}

We chose as our undetermined function $f(r)\equiv n(r)\sigma_r(r)^{2}$ 
and specified this function on a grid in radius.
Iterations consisted in varying the values $f_i$ on the grid.
At each iteration, $n(r)\sigma_t(r)^{2}$ was computed from the $f_i$
and from the assumed $M(r)$ and $n(r)$ using the Jeans equation
in the form
\begin{equation}
n(r)\sigma_t^2(r) = f(r) + {r\over 2} {df\over dr} 
+ {GM(r)n(r)\over 2r}.
\label{eq:anisoJeans}
\end{equation}
Because $f(r)$ appears as a derivative in this expression,
we needed to impose a constraint to ensure that $f$ remains 
a smooth differentiable function during the optimization.
We did this in the standard way by adding a penalty function to
$\chi^2$, of the form
\begin{equation}
P_\lambda(n\sigma_r^2) = 
\lambda \int_0^{\infty} \left[{d^2\log f\over d\log r^2}\right]^2 dr
\label{eq:pf}
\end{equation}
with $\lambda$ a parameter that controls the degree of smoothness.
The expression (\ref{eq:pf}) ``penalizes'' functions $n\sigma_r^2$
that fluctuate too rapidly in their dependence on $r$.  In addition,
this penalty function has the desirable property that any power-law
dependence of $n\sigma_r^2$ on $r$ is defined to be ``perfectly
smooth,'' i.e. $P=0$.  Since the form of $n\sigma_r^2(r)$ in the
vicinity of a SMBH is likely to be close to a power law, smoothing via
the penalty function (\ref{eq:pf}) is not likely to bias the results
substantially even if $\lambda$ is large.  In practice, $\lambda$ was
chosen to be as small as possible consistent with a reasonably smooth
result for $\sigma_r(r)$ (we took
$0.0001\leq\lambda\leq0.003$).

The goodness of fit of the model to the data was defined 
in a manner analogous to the isotropic case:
\begin{equation}
\chi^2 = 
\sum_{i=1}^N {\left[V_{R,i}^2-\sigma_R^2(R_i)\right]^2\over
\Delta_R^2(R_i^2)} + 
\sum_{i=1}^N {\left[V_{T,i}^2-\sigma_T^2(R_i)\right]^2\over
\Delta_T^2(R_i^2)}.
\label{eq:chisq}
\end{equation}
In this expression, the functions $\sigma_R(R)$ and $\sigma_T(R)$ are
understood to be related to $\sigma_r(r)$ and $\sigma_t(r)$ via
equations~(\ref{Eq:Leonard}).  Optimization (i.e. minimization of
$\chi^2+P_\lambda$) was carried out using the NAG routine E04FYF.
  The minimum reduced $\chi^2$ of our anisotropic models was 1.06.

\begin{figure}
\includegraphics[width=0.9\columnwidth,angle=-90.]{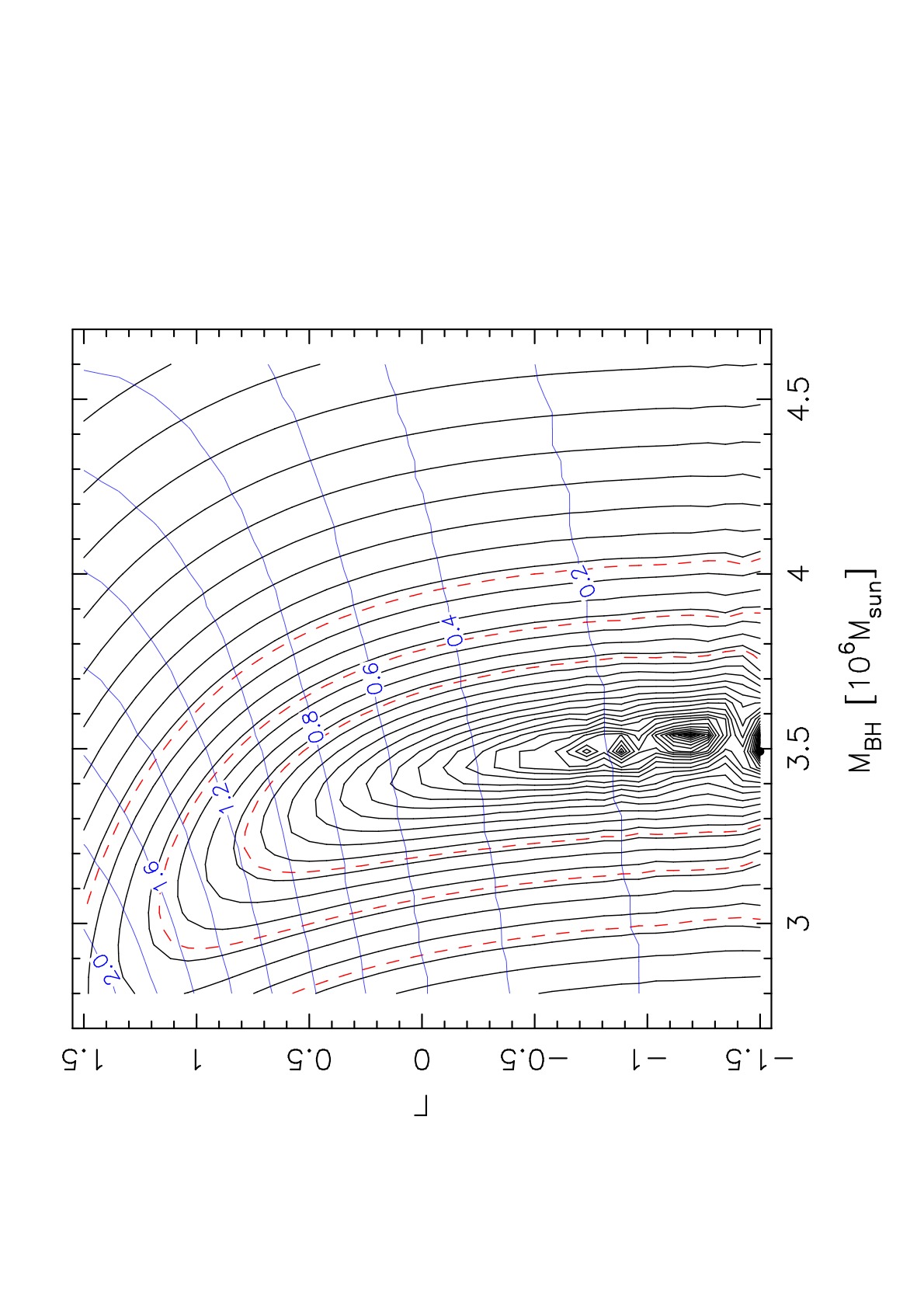}
\caption{\label{Fig:Anisocont} Like Fig.~\ref{Fig:Isocont}, but
for the anisotropic modelling.}
\end{figure}

\begin{figure}
\includegraphics[width=0.9\columnwidth,angle=-90.]{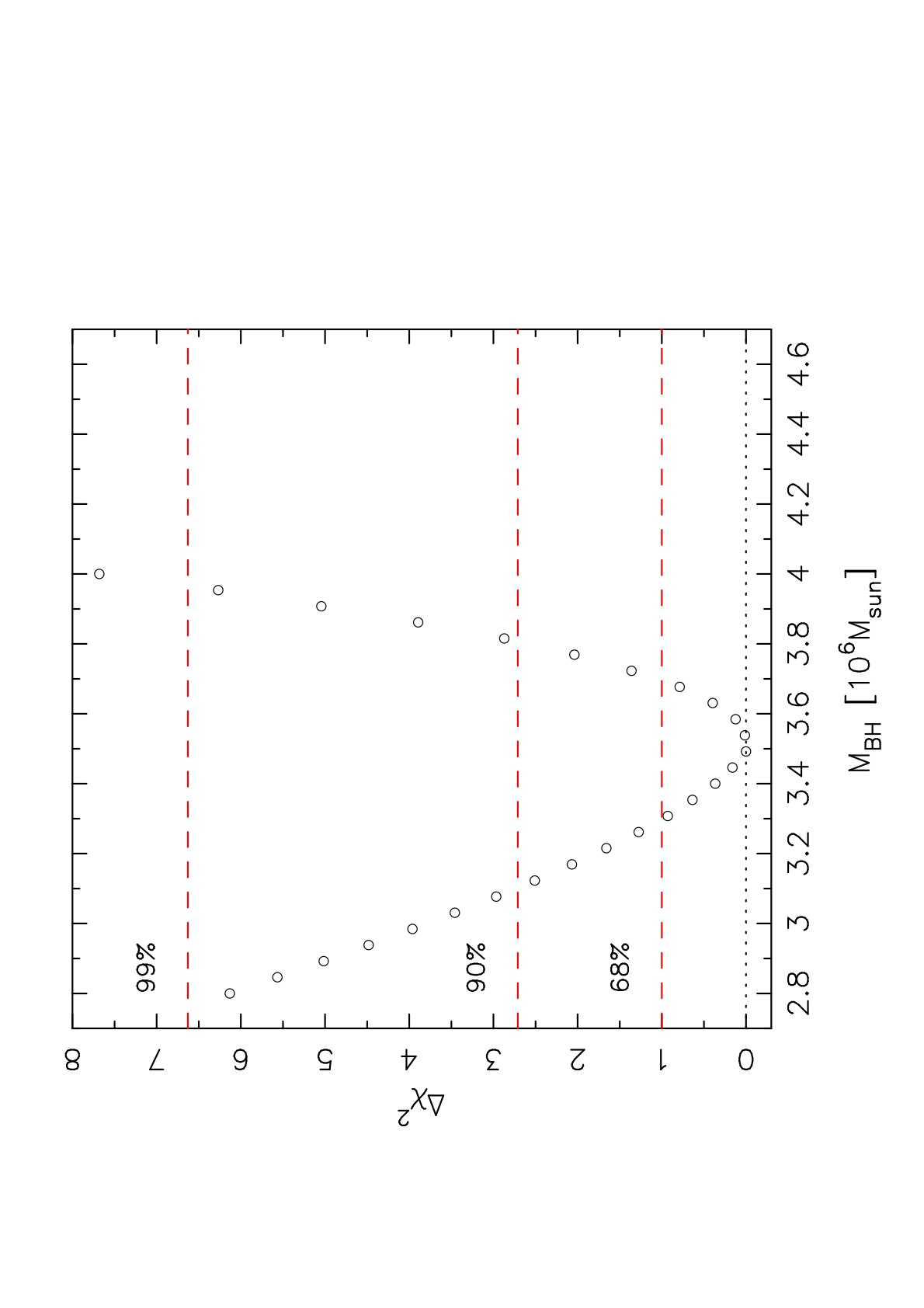}
\caption{\label{Fig:Anisochisq} Like Fig.~\ref{Fig:Isochisq}
but for the anisotropic modelling.
}
\end{figure}

\begin{figure}
\includegraphics[width=1.1\columnwidth]{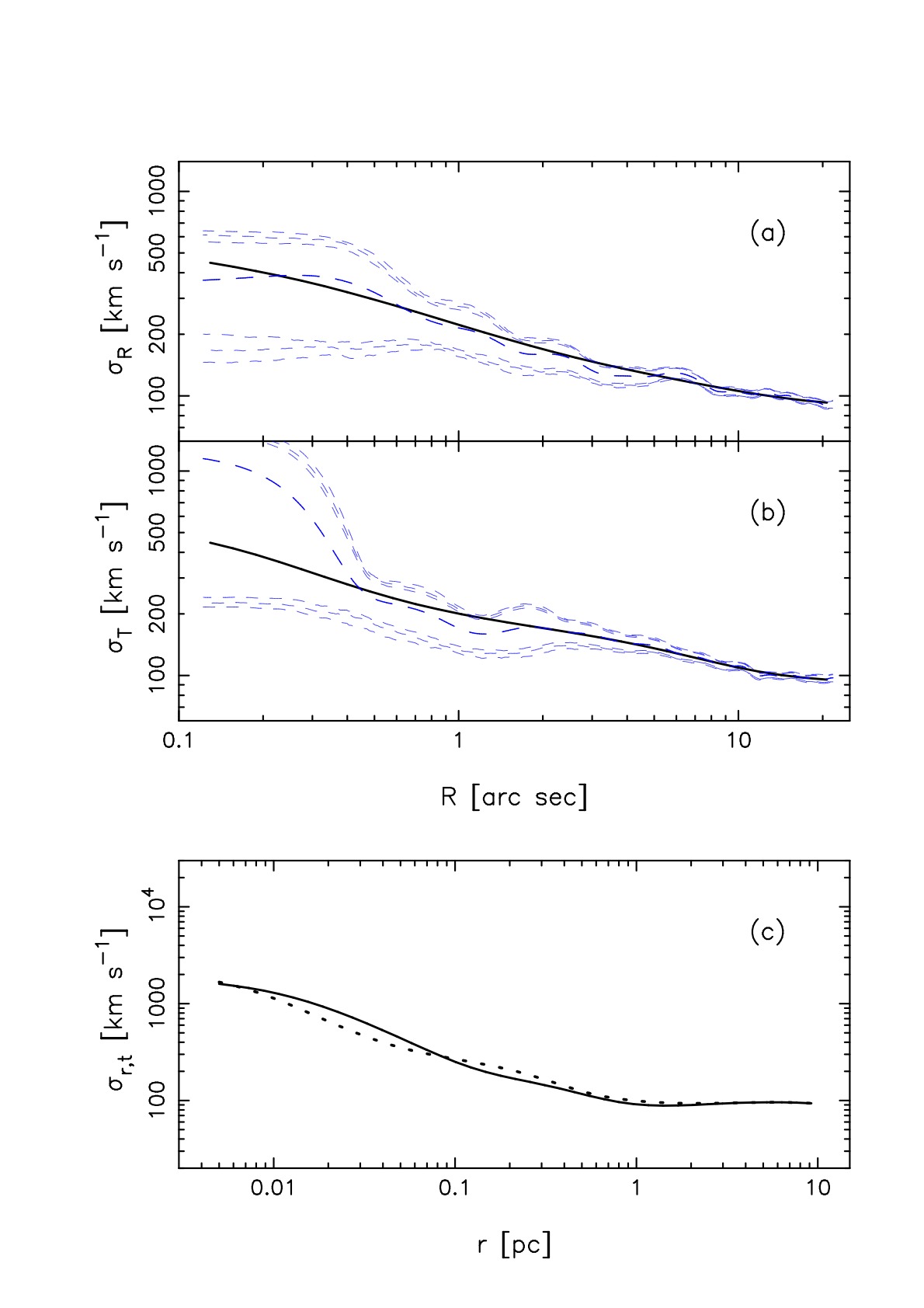}
\caption{\label{Fig:Best} Best-fit anisotropic model.
(a) and (b) are the projected, radial and tangential velocity
dispersions (black/solid lines), compared with a kernel-based velocity 
dispersion profile and associated (90\%,95\%,98\%) 
confidence intervals (blue/dashed lines).
(c) Intrinsic velocity dispersions: radial (solid) and tangential
(dashed) lines.  }
\end{figure}

\begin{figure}
\includegraphics[width=1.1\columnwidth]{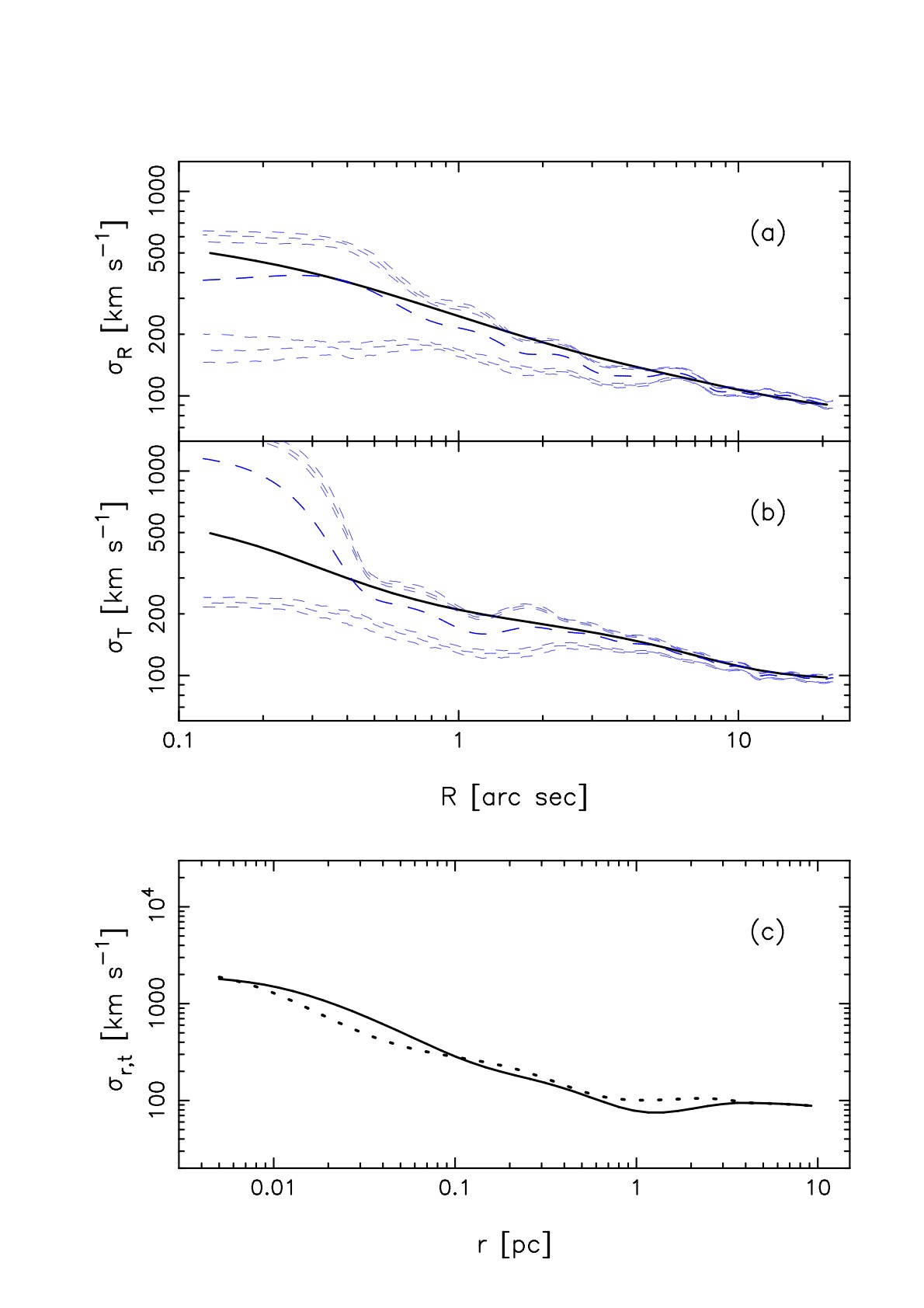}
\caption{\label{Fig:Best4} 
Anisotropic model with $M_{\rm BH}=4.0\times 10^6M_\odot$,
$\Gamma=0$, and $M_\star(r<1{\rm pc}) = 0.5\times 10^6M_\odot$.
Curves and symbols are defined as in Fig.~\ref{Fig:Best}.
}
\end{figure}

The radial grid on which $n\sigma_r^{2}$ was specified extended to
50 pc, well beyond the outermost measured velocity at $\sim 1$ pc.
It was discovered that allowing complete freedom in 
$n\sigma_r^2$ in the 
region $1 {\rm pc}\lesssim r
\lesssim 50 {\rm pc}$ led sometimes to solutions
in which the anisotropy was very large and/or increasing
at large radii.
Such models reproduce the observed 
departures from a Keplerian velocity falloff at $\sim 1$ pc
by placing stars on very eccentric orbits beyond the region
where the solution is strongly constrained by the data.
While such models are physically permissible, they seem rather
unlikely.
We focus here on models that were constrained at large radii
to be isotropic.
Specifically, we forced $\sigma_r(r)$ to be equal to the isotropic
$\sigma(r)$, equation~(\ref{eq:Jeansiso}), at all $r\ge r_{\rm iso}=
3 {\rm pc} \approx 60 ''$.
Imposing this constraint (which results in a different $\sigma_r(r)$
profile at $r\ge r_{\rm iso}$ pc for each assumed $M(r)$) guarantees
that $\sigma_t(r)=\sigma_r(r)=\sigma(r)$ at $r\ge r_{\rm iso}$.

The results are summarized in
Figures~\ref{Fig:Anisocont}-\ref{Fig:Best4}.  Overall the results are
similar to those obtained under the assumption of isotropy, except
that the confidence intervals are somewhat narrower, due to the
additional information contained within the two velocity dispersion
components.  The 90\% confidence bounds on $M_{\rm BH}$
(Figure~\ref{Fig:Anisochisq}) are $3.1\times 10^6
M_\odot\lesssim M_{\rm BH}/ \lesssim 3.8\times 10^6
M_\odot$.

Figure~\ref{Fig:Best} shows the kinematics of the best-fit solution,
with $M_{\rm BH}=3.6\times 10^6M_\odot$.
The stellar velocities are mildly radially anisotropic at $r\lesssim 0.1$ pc
and mildly tangentially anisotropic for $0.1 {\rm pc} \lesssim r \lesssim 1$ pc;
by construction, they are isotropic beyond $\sim 3$ pc.

Figure~\ref{Fig:Best4} shows the kinematics of a second solution with
$M_{\rm BH}=4.0\times 10^6M_\odot$ and $\Gamma=0$; the
stellar mass within $1$ pc is $0.5\times 10^6M_\odot$.
The differences with the global best-fit model are slight, consisting
mostly of a larger degree of anisotropy.

We note that both the isotropic and anisotropic modelling seem to rule
out securely a mass density profile as steep as $\rho\sim r^{-2}$.

\section{Discussion}

\subsection{Rotation of the NSC \label{sec:rotation}}

\begin{figure}
\includegraphics[width=\columnwidth]{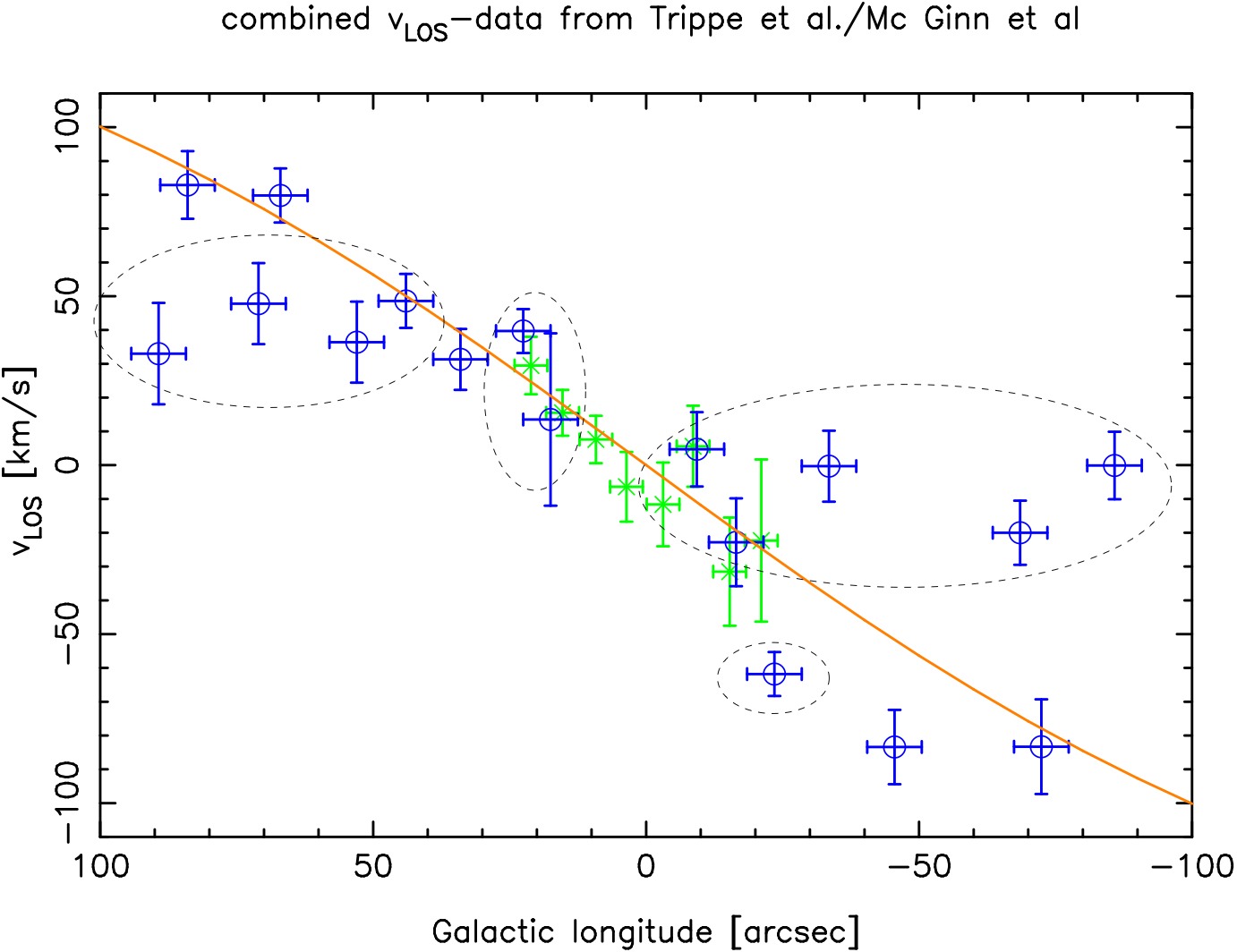}
\caption{\label{Fig:vlos} Line-of-sight velocity vs. Galactic
  longitude. Blue circles: measurements from \citet{McGinn1989ApJ};
  green stars: measurements from \citet{Trippe2008A&A}. The straight
  orange line indicates the model of \citet{Trippe2008A&A}. The
  dashed black ellipses show the data that are not shown in Fig.\,12 of
  \citet{Trippe2008A&A}. }
\end{figure}

An early spectroscopic study by \citet{McGinn1989ApJ} found evidence
for rotation of the GC NSC, at least beyond 1\,pc distance from
Sgr\,A*. Other earlier studies based on spectroscopic observations
have found no signs of net rotation in the late-type stars in the
nuclear star cluster \citep[][the latter present a detailed
  analysis]{Sellgren1990ApJ,Genzel2000MNRAS,Figer2003ApJ}. However,
these studies were based on small numbers of late-type stars in the
central parsec of the Milky Way and did therefore probably not reach
the necessary accuracy of a few km/s needed to detect rotation in the
innermost parsec.

\citet{Trippe2008A&A} found clear signs for an overall rotation of
the NSC parallel to galactic rotation through the analysis of proper
motions and new adaptive-optics assisted integral field spectroscopic
observations. The results of our work confirm their finding of an
overall rotation of the NSC in the Galactic. \citet{Trippe2008A&A}
provide a model for the rotation velocity that increases linearly with
Galactic longitude with a value of $1.42\pm0.36$\,km\,s$^{-1}$ per
arcsecond projected distance from Sgr\,A* along Galactic longitude.

We show the data of \citet{McGinn1989ApJ} and \citet{Trippe2008A&A}
in Fig.\,\ref{Fig:vlos}. Figure\,\ref{Fig:vlos} includes {\it all}
data from \citet{McGinn1989ApJ}.  For unknown reasons,
\citet{Trippe2008A&A} only show 5 of 17 data points of
\citet{McGinn1989ApJ}. The plot shows that while the model of
\citet{Trippe2008A&A} fits well all the data in the central few tens
of arcseconds, the rotation curve may be considerably flatter at
Galactic longitudes $|l|\gtrsim 40''$. Due to the strong and highly
variable extinction toward the GC, the data of \citet{McGinn1989ApJ}
may sample quite different depths and therefore stellar populations.
We may therefore speculate on the possibility that we actually see
{\it two} distinct rotating systems in the GC, one with a steeper and
one with a flatter rotation curve.  In this context it appears
worthwhile to point out the recent result of \citet{Seth2008ApJ} from
integral-field spectroscopy of the nuclear star cluster of NGC\,4244:
They identify a younger disk-like stellar population superposed on an
older spheroidal component. It may be that a similar situation
presents itself in the center of the Milky Way. A nuclear disk exists
the GC in addition to the spherical NSC by
\citet{Launhardt2002A&A}. However, an important caveat is the fact
that the size scale of the nuclear disk is orders of magnitudes larger
then the one of the NSC, unlike the situation in NGC\,4244.

As pointed out by \citet{Seth2008ApJ}, rotation of the NSC parallel to
Galactic rotation implies that it may at least in part have formed by
accretion of gas or star clusters from the galactic disk. This is of
great importance for understanding the origin of nuclear star clusters
in galaxies \citep[for a brief overview of some formation scenarios
  see, e.g.,][]{Boeker2008JPhCS}.

\subsection{Isotropy/anisotropy}

 Consistent with earlier work \citep[e.g.~][]{Genzel2000MNRAS} we find
 evidence for tangential mean motion in the central arcseconds (see
 Fig.\,\ref{Fig:velmap}). This anisotropy is related to the presence of young,
 massive stars, a significant fraction of which shows coherent motion
 within one (or possibly two) stellar disks
 \citep{Levin2003ApJ,Genzel2003ApJ,Lu2006JPhCS,Paumard2006ApJ}. \citet{Lu2008arXiv}
 discard the existence of the counter-clockwise disk and show that
 about 50\% of the young stars belong to the clockwise rotating disk,
 while the other 50\% appear to have more randomized motions.  Our
 analysis here shows that, after excluding the early-type stars, the
 cluster appears close to isotropic (see lower left panel in
 Fig.\,\ref{Fig:sigma}, right panel in Fig.\,\ref{Fig:siglb}, and left
 panel in Fig.\,\ref{Fig:sigmaoff}).  The situation may be different
 in the innermost $6''$, where the data are somewhat more ambiguous
 (see lower left panel of Fig.\,\ref{Fig:sigma}). Our anisotropic
 Jeans models also result in solutions are are close to isotropy (see
 Figs.\,\ref{Fig:Best} and \ref{Fig:Best4}). A caveat is,
   however, the rotation of the cluster, which could mask anisotropy
   with a signature smaller than or comparable to the rotation
   signature.

\subsection{Mass modeling: black hole mass}
Our modeling of the proper motion data yields a best-fit black hole
mass of $3.6^{+0.2}_{-0.4}\times10^{6}$\,M$_{\odot}$ (68\%) under the
isotropic assumption, and
$3.5^{+0.15}_{-0.35}\times10^{6}$\,M$_{\odot}$ if anisotropy is
allowed (for an assumed distance of 8\,kpc to the GC).  The
  smaller uncertainties of the anisotropic model may appear
  counter-intuitive. However, when we go from isotropic to anisotropic
  models, we also go from one observed function (the isotropized
  velocity dispersion profile) to two observed functions (the two,
  radial and tangential dispersion profiles).  So from a mathematical
  point of view, the ratio between the number of "model functions" and
  the number of "data functions" remains the same. There is no obvious
  reason why the error bars on BH mass obtained from the isotropic
  modelling should be tighter, or looser, than those obtained from the
  anisotropic modelling.

A BH mass of $4.0\times 10^6M_\odot$ is consistent with the modelling
at the 90\% (isotropic) and 99\% (anisotropic) levels.  Thus, while a
value for $M_{\rm BH}$ slightly lower than the currently canonical
value \citep{Ghez2003ApJ,Eisenhauer2005ApJ,Ghez2005ApJ,Ghez2008ApJ} is
preferred by our modelling, our results are still consistent with that
value.

Mass estimates of Sgr\,A* based on proper motions
\citep[e.g.][]{Ghez1998ApJ,Chakrabarty2001AJ,Eckart2002MNRAS,Genzel2000MNRAS}
have routinely provided lower values than what has been found by the
analysis of stellar orbits. The orbit of the star S2/S0-2 has
consistently provided higher black hole masses with high precision
(see above). Determining the mass of Sgr\,A* from a stellar orbit is
straightforward and relies solely on the assumption that the star
moves on a Keplerian orbit and that higher order effects can be
neglected \citep[see, e.g.,][and
  others]{Rubilar2001A&A,Weinberg2005ApJ,Zucker2006ApJ,Ghez2008ApJ}.
Mass estimates based on proper motions, on the other hand, have to
make a number of assumptions on the spatial and velocity structure of
the cluster that can bias the results. 

Source confusion and the fast proper motion of stars within
$R\approx0.5$\,pc of Sgr\,A* may have biased the proper motions of
stars near Sgr\,A* in early work
\citep[e.g.,][]{Genzel1997MNRAS,Ghez1998ApJ,Genzel2000MNRAS} toward
lower values of the velocity dispersion (identification of slow stars
is easier between epochs; faint and fast stars become easily confused
with brighter ones). In this work, we only use high-resolution AO
observations at an 8\,m-class telescope. The bias toward low velocity
dispersion therefore should be minimal. Also, the modeling uses proper
motions in the entire central parsec. This decreases considerably the
weight of the possibly biased proper motions within $R\approx0.5''$.

The black hole masses from both our isotropic and anisotropic models
are in reasonable agreement with the measurements from the orbit of
S2/S0-2. We believe that the most important source of bias for the
determination of the BH mass from proper motions is the assumed law
for the radial dependence of the tracer population from distance to
Sgr\,A*. Earlier work has usually assumed steep power-laws
\citep[e.g., $n(r)\propto r^{-2.5}$ in][]{Genzel2000MNRAS} for the
tracer population. This leads inevitably to an under-estimation of the
3D distances of the tracer stars from the BH because the number
density behaves rather like $n(r)\propto r^{-1.2}$ in the central
parsec \citep{Schoedel2007A&A}. The issue becomes even more important
when the analysis is limited to the late-type stars near Sgr\,A*,
whose density may be even decreasing toward the black hole (see, e.g.,
Figer et al.º 2003; Genzel et al.\ 2003; Buchholz, Sch\"odel,\&
Eckart, 2009, submitted to A\&A). In our analysis we have assumed
$n(r)\propto r^{0.5}$ for the late-type stars near Sgr\,A*. This is
the flattest power-law still consistent with isotropy. The issue is
worth further investigation.

\subsection{Mass modeling: cluster mass \label{sec:clustermass}}

Our basic modeling assumptions are stationarity and spherical symmetry
of the cluster.  We show that the proper motion data \emph{cannot be
  explained by just a point mass at the position of Sgr\,A*}. Both
isotropic and anisotropic Jeans models require a mass within
$r\leq1$\,pc of $>0.5\times10^{6}\,M_{\odot}$, in addition to the
point mass of the black hole, Sgr\,A* (under the assumption that
$\Gamma>0$, i.e. the mass density increases toward Sgr\,A*). Also, the
Leonard Merritt moment mass estimator shows
clear evidence for an increasing contribution of extended/stellar mass
to the gravitational potential at $R\gtrsim0.4$\,pc. The result for
the amount of extended mass within $r=1$\,pc is insensitive to the
exact assumed density-law for the tracer population in the immediate
vicinity of Sgr\,A* (see above).

\begin{figure}
\includegraphics[width=\columnwidth]{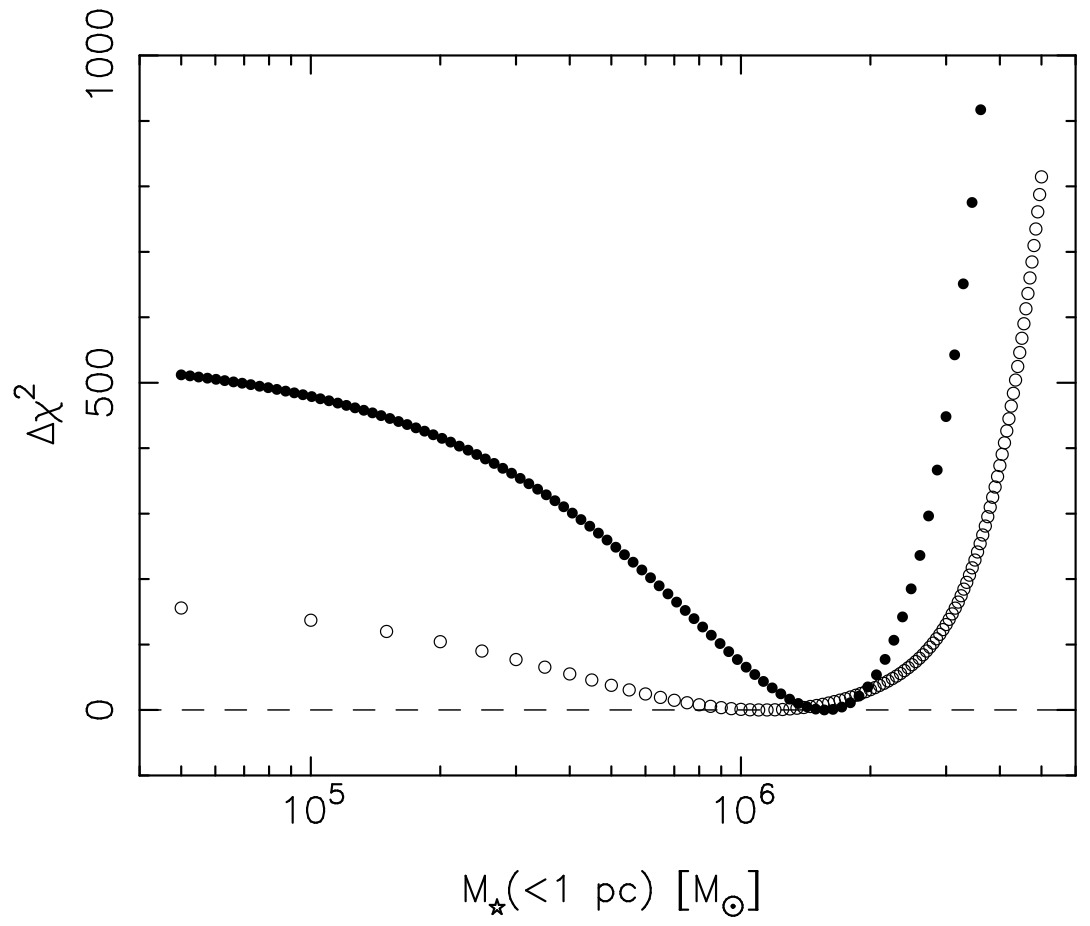}
\caption{\label{Fig:1d} 
Fits to the proper motion data when the BH mass is fixed to
$4.0\times 10^6M_\odot$ and the stellar mass density is
assumed proportional to the stellar number density.
{\it Filled circles:} isotropic modelling; 
{\it open circles:} anisotropic modelling.}
\end{figure}

As an alternative to the assumptions made above, we can model the
proper motions assuming that the stellar mass density is proportional
to the stellar number density, eq.~(\ref{eq:nmodel}), and that the
mass of the black hole is the value given by modelling of the S-star
orbits, $4.0\times 10^6M_\odot$.  The single remaining free parameter
is then the normalization of the stellar mass density, i.e, the
mass-to-light ratio of the stars.  Figure~\ref{Fig:1d} shows how the
fit to the proper motion data varies as a function of the stellar mass
under these assumptions, for both isotropic and anisotropic cases.
The preferred value for the stellar mass within 1 pc is seen to be
$\sim 1.5\times 10^6 M_\odot$ (isotropic) and $\sim 1.1\times 10^6
M_\odot$ (anisotropic).

While early work \citep[e.g.,][]{Sellgren1990ApJ,Eckart1993ApJ} found
enclosed masses at $R\approx0.5$\,pc that are comparable to what we
have derived in this work, almost negligible amounts of extended,
i.e.\ stellar, mass within $R<1$\,pc were reported in later work
\citep[see,e.g.,][]{Haller1996ApJ,Genzel2000MNRAS,Schoedel2003ApJ}. However,
with these low values it would be difficult to reconcile the high mass
of the NSC of $3.5\pm1.5\times10^{7}\,M_{\odot}$
\citep{Launhardt2002A&A} with the almost zero mass in the central
parsec: the old mass models would  not reach the integrated mass
of the NSC within its radius of $\lesssim10$\,pc. High precision
measurements of the velocity of the maser star IRS~9, located at
$R=0.33$\,pc may also indicate an enclosed mass of a few times
$10^{5}\,M_{\odot}$ in addition to the mass of Sgr~A*
\citep{Reid2007ApJ}.  \citet{Schoedel2007A&A} have roughly estimated
the enclosed extended mass in the central parsec by combining the
measured line-of-sight velocity dispersion of late-type stars in the
central pc, assuming isotropy, with the derived density structure of
the stellar cluster. They found that the enclosed mass in the GC may
start to rise significantly already at projected distances from Sgr~A*
as low as $R\approx0.3$\,pc.

What can be the reason why the enclosed stellar mass at the GC has
been underestimated in the past decade? We believe that there are
primarily two factors responsible. In the absence of adequately sampled
proper motion or line-of-sight velocity measurements at distances
$R\gtrsim0.5$\,pc, it was necessary to include measurements of gas
velocities in the circum nuclear disk
\citep[e.g.,][]{Guesten1987ApJ,Christopher2005ApJ} into the mass
estimates. However, gas can be subject to winds, magnetic fields, or
cloud collisions, contrary to stars, which are ideal test particles of
the gravitational potential. A second factor that influenced previous
mass estimates of the stellar cluster around Sgr\,A* were model
assumptions and the related combination data sets within a too simple
model (just a bulge, no NSC, no nuclear disk). In principle, earlier
assumptions ignored the existence of the MW NSC. The effect was to
include bulge velocity dispersions into the mass estimates
(\citet{Haller1996ApJ}, e.g., include measurements at distances of
$\sim$100\,pc from Sgr\,A* into their analysis, while \citet{Genzel1996ApJ}
assume $\sigma_{\infty}=55$\,km\,s$^{-1}$ in their Jeans
modelling. However, the velocity dispersion of the Milky Way bulge is
fairly uncertain \citep[see][]{Merritt2001ApJ,Tremaine2002ApJ}. Also,
the MW NSC may well be a system that is dynamically decoupled from the
bulge.

As described in the introduction, observations in the 1990s (mainly by
HST) revealed the existence of nuclear star clusters as entities that
are morphologically and dynamically separate from the bulges. This
cannot be neglected in the determination of the MW NSC mass. If the
velocity dispersion in the bulge is lower than in the NSC, then
spectroscopically measured velocity dispersions outside of the central
parsec may already be biased toward low values (due to superposition
of bulge stars along the line of sight toward the NSC). This may be
the reason why \citet{RiekeRieke1988ApJ} obtain a line-of-sight
velocity dispersion from late type stars in the region
$0.36<R<6.5$\,pc ($\sigma=75$\,km\,s$^{-1}$) that is lower than the
$\sigma\approx100$\,km\,s$^{-1}$ for late-type stars within
$R<0.8$\,pc measured by \citet{Genzel2000MNRAS}, \citet{Figer2003ApJ},
or \citet{Zhu2008ApJ}.

%Also, the nature of the NSC and the bulge \citep[or probably rather
%  nuclear stellar disk][]{Launhardt2002A&A} as separate entities
%should be taken into account in Jeans models that extend to projected
%radii of several 10s to 100s of parsecs.  To conclude, we think that
%the bias towards low MW NSC masses was mainly due to the (at that
%point unavoidable) inclusion of gas velocity measurements into the
%mass estimates and, above all, lack of knowledge of the existence of
%NSCs as separate entities.

What are the consequences of the -- compared to previously published
values -- significantly increased stellar mass in the cluster around
Sgr~A*? As concerns theoretical models of the dynamics of gas and
stars at the GC, future modelling must take into account that the
gravitational potential at the GC starts deviating from that of a
point mass significantly already at a distance of roughly 0.5\,pc from
Sgr~A*. As pointed out by \citet{Schoedel2007A&A} an important
implication of the non-negligible mass of the NSC implies that the CND
cannot be modeled as a simple disk in Keplerian rotation. This is also
supported by the mass models of \citet{Figer2003ApJ}. Also, the
dynamics of the mini-spiral
\citep[see][]{Lacy1991ApJ,VollmerDuschl2000NewA,Paumard2004A&A} will
probably have to be revised (also to take into account the
$\sim30-50\%$ increased black hole mass, compared to the values used
in some of these models).

Our models do not allow us to set strong constraints on the power-law
index of the assumed mass distribution. Assuming increasing mass
density toward the black hole, values of $0\leq\Gamma\leq1.5$
can be reconciled with the models.  The power-law index of the stellar
density profile reported by \citet{Schoedel2007A&A} is $1.19\pm0.05$ i
in the central parsec. This value lies within the range of the
$\Gamma$-values from our models. A caveat at this point is, however,
that \citet{Schoedel2007A&A} did not explicitly discard early-type
stars from their number counts and therefore report some sort of
average value for early and late-type populations. When only the
late-type stars are considered, the density profile may be
considerably flatter or even slightly inverted in the immediate
environment of Sgr\,A* \citep[see,e.g.,][, or Buchholz, Sch\"odel, \&
  Eckart, 2009, submitted to A\&A]{Figer2003ApJ,Genzel2003ApJ}. The
exact density profile of the late-type stars is an important issue
that needs to be addressed by future research.

A value as high as $\Gamma=2.0$ in the central parsec appears to be
safely ruled out (Figs.\ref{Fig:Isocont} and \ref{Fig:Anisocont}).
The parameter $\Gamma$ is of great importance because it determines
the radial profile of the mass density.  The latter has important
consequences for stellar dynamics because the high densities shorten
the relaxation time and increase the probability for close stellar
encounters or collisions. There may be a non-negligible probability of
stellar collisions at radii $r<0.2$\,pc during the lifetimes of stars,
causing -- among other effects -- the destruction of the envelopes of
giant stars \citep[see][]{Alexander2003gbh,Freitag2008IAUS}. The
derived collision probabilities presented in \citet{Freitag2008IAUS}
are observationally supported by the distinct lack of late-type (giant)
stars within $R\approx6''-8''$ of Sgr\,A*
\citep[see][]{Genzel1996ApJ,Haller1996ApJ,Figer2003ApJ,Genzel2003ApJ,Zhu2008ApJ}.
The collision rates reported by \citet{Freitag2008IAUS} are based on
steep ($n(r)\propto r^{-1.5}$) density profiles. While such a steep
profile is not excluded by our analysis and the mass densities of
$\geq10^{7}$\,M$_{\odot}$\,pc$^{-3}$ required by the models of Freitag
et al. may be consistent with our estimates, it is nevertheless
important to note that flatter mass profiles are preferred by our
Jeans modeling. Also, the profiles derived from stellar number counts
suggest rather shallow profiles near Sgr\,A* (see above).

It may appear surprising that our models also allow for a
  decreasing mass density toward Sgr\,A*.  As far as we are aware, no
  one in the past has even allowed the possibility of "centrally
  evacuated" mass models. The standard approach (e.g. in the recent
  paper by Trippe et al.) is to assume a monotonically- decreasing
  mass density at the outset. So, it is quite possible that other
  modellers would have obtained this result if they had looked for it.
  The basic reason why models with central "holes" in the mass density
  are preferred, is that the observed velocity dispersion profile is
  essentially Keplerian in the inner 0.1 pc, leaving little room for a
  distributed component in this region. But we emphasize that the
  preference for such models is weak.

  An inverted mass model would be expect theoretically if (a) a binary
  SMBH existed in the past, and (b) the time scale for dynamical
  regeneration of a density cusp after it has been destroyed by a
  binary SMBH is longer than 10 Gyr
  \citep[e.g.,][]{MerrittSzell2006ApJ}.

\subsection{Mass-to-light ratio}

\begin{figure}
\includegraphics[width=.9\columnwidth]{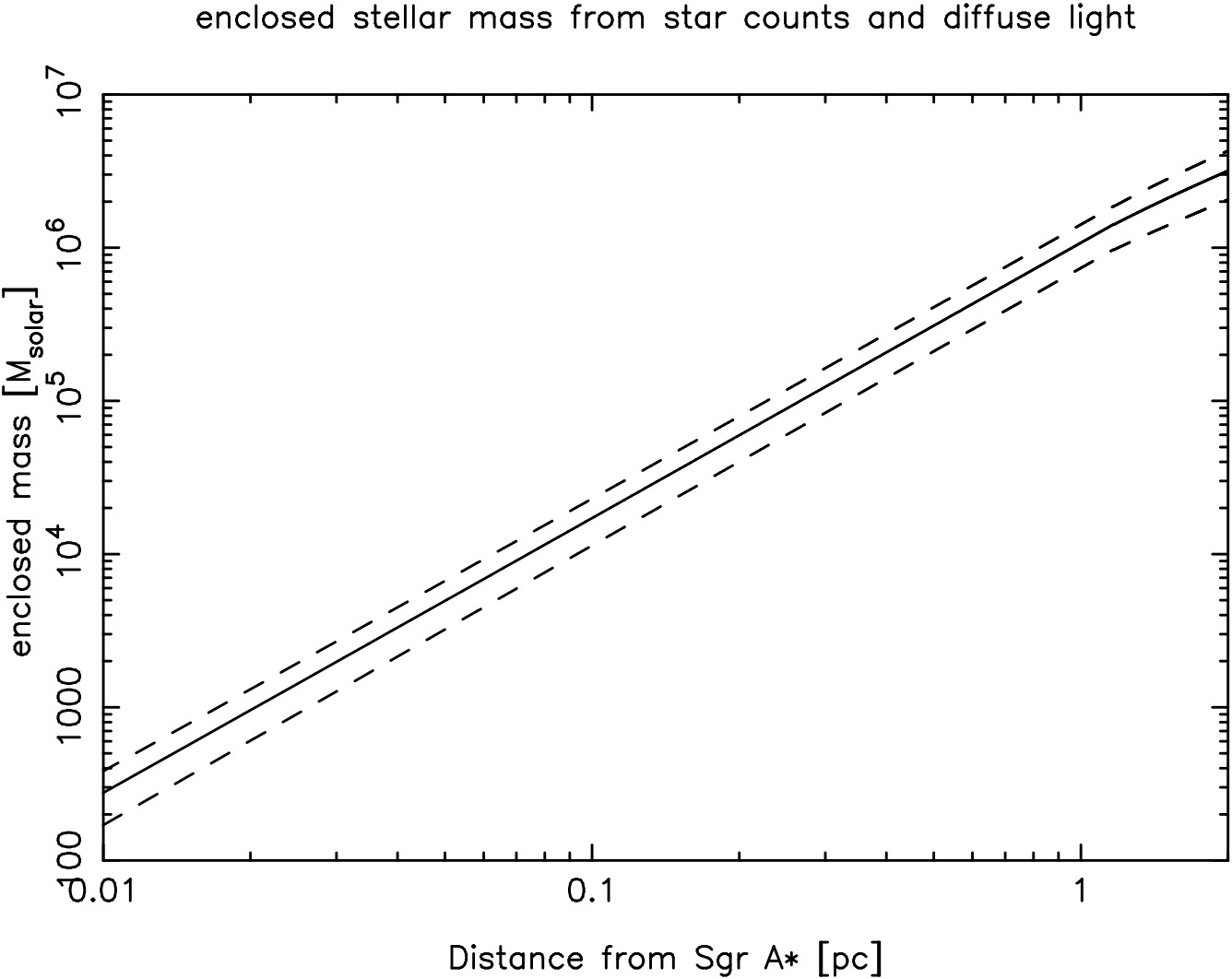}
\caption{ \label{Fig:starmass} Enclosed stellar mass vs.\ distance
  from Sgr\,A*, derived from star counts and diffuse light density in
  the central parsec of the GC, using the broken-power law structure
  of the cluster from \citet{Schoedel2007A&A}. The dashed lines
  indicate the statistical $1\,\sigma$ uncertainties. We estimate that
  the systematic error of this simple model is of the order a factor
  of $\sim2$ (normalization of diffuse light density and
  extinction). }
\end{figure}

How does our finding of $0.5-2.0\,\times10^{6}\,M_{\odot}$ of extended
mass in the central parsec compare with an estimate derived from the
stellar light? We produced a simple model of the enclosed stellar mass
vs.\ distance from Sgr\,A* by using the parameters of the NSC
structure determined by \citet{Schoedel2007A&A}. A broken power-law
was assumed, with an inner power-law index of $a_{in}=1.2\pm0.05$ and
outer power-law index of $a_{out}=1.75\pm0.1$.  The break radius of
$R_{br}=0.22\pm0.04$\,pc given in \citet{Schoedel2007A&A} was
de-projected, using the given values of $a_{in}$ and $a_{out}$
together with the number counts of \citet{Schoedel2007A&A} obtaining
$r_{br}=0.8\pm0.2$\,pc.  The enclosed mass of individually visible
stars in the central parsec was then calculated with the broken
power-law and using the number counts of \citet{Schoedel2007A&A}. An
average mass of 3\,M$_{\odot}$ was assigned to each star \citep[see
  Fig.\, 16 in][]{Schoedel2007A&A}. For the contribution of the
unresolved stellar population we used the same cluster parameters. The
diffuse light due to the unresolved stars was normalized to a value of
$2.0\times10^{-3}$\,Jy\,arcsec$^{-2}$ at a projected distance from
Sgr\,A* of $R=10''$. This value was determined from the Ks-band image
of 13 May 2005 (see Table\,\ref{Tab:Obs}), using a field of low
stellar density at $R=10''$ and avoiding known field of high
extinction or strong dust emission. The diffuse light density was
converted to a mass estimate by using the Ks-band luminosity of the
sun, at a distance of 8\,kpc and assuming an extinction of 3.3
magnitudes in the K-band.

The resulting plot of enclosed stellar, i.e. luminous mass
vs.\ distance from Sgr\,A* is shown in Fig.\,\ref{Fig:starmass}. The
dashed lines indicate the $1\sigma$ uncertainty that results from the
uncertainties of the parameters of the cluster structure. The mass due
to the unresolved stellar population (responsible for the diffuse
light) makes up $>98\%$ of the stellar mass in this simple
estimate. We estimate that there is an systematic uncertainty of
$30\%$ related to our normalization of the diffuse light density at
$R=10''$.  An additional source of systematic uncertainty is the
extinction toward the GC, for which we estimate an absolute
uncertainty of about 0.5\,magnitudes. Hence, there may be a systematic
uncertainty of a factor 2 related to the plot shown in
Fig.\,\ref{Fig:starmass}.

A detailed estimate of the stellar mass in the central parsec of the
NSC via its luminosity is non-trivial and beyond the scope of this
paper. However, the simple model in Fig.\,\ref{Fig:starmass}
demonstrates that the visible stellar mass (through star counts and
diffuse light density) can easily account for an extended/stellar mass
around $1\times10^{6}$\,M$_{\odot}$ in the central parsec of the NSC.

Alternatively, we can use the stellar mass estimate within the central
parsec as given in sub-section \,\ref{sec:clustermass} above (assuming
that the mass density is proportional to the stellar number density
and fixing the BH mass to $4.0\times10^{6}$\,M$_{\odot}$): $\sim1.5
(1.1) \times10^{6}$\,M$_{_\odot}$ for the case of isotropy
(anisotropy). With the above given normalization of the diffuse light
density we obtain $M/L = 1.4 (1.1)$\,M$_{\odot}/$L$_{\odot, Ks}$ for
for the isotropic (anisotropic) case. Here, L$_{\odot, Ks}$ is the
luminosity of the sun in the Ks-band.  As described in the paragraph
above, this result is uncertain by a factor $\sim2$ (normalization of
diffuse light density and extinction; uncertainty of the mass estimate
has not been taken into account here).  Note that this analysis refers
only to the {\it diffuse} light density, i.e.\ after subtraction of
the individually detected stars. As shown in \citet{Schoedel2007A&A},
the detected stars, with mag$_Ks\lesssim17.5$ emit more than 99\% of
the total light. If they were included in the analysis, the $M/L$
ratio would be a factor of $\sim100$ lower.

There are observationally supported claims for the existence of the
order $10^{4}$ neutron stars and stellar mass black holes in the
central parsec \citep{Muno2005ApJ}. The large uncertainties involved
in the determination of the mass-to-light ratio in the central parsec
and the small total mass of these stellar remnants
($<10^{5}$\,M$_{\odot}$) compared to the enclosed total stellar mass
($\sim10^{6}$\,M$_{\odot}$) in the central parsec means that the claim
of \citep{Muno2005ApJ} cannot be tested (or can only be tested with
great difficulty) via estimates of the mass-to-light ratio in the
central parsec.

\subsection{Comparison with similar work}

\citet{Trippe2008A&A} also modelled the kinematics of the old stellar
population using proper motion velocities, and obtained estimates of
$M_{\rm BH}$ and of the distributed mass.  Like us, Trippe et
al. assumed a spherical model for the NSC, but their approach differed
from ours in essential ways.  Trippe et al. did not construct a
self-consistent model for the nuclear star cluster. Instead, they
considered only the intrinsic velocity dispersions parallel to the
Galactic latitude and longitude, which they called $\sigma_l(r)$ and
$\sigma_b(r)$, and assumed that both were functions only of the
radius.  The observed velocity dispersions, $\sigma_L(R)$ and
$\sigma_B(R)$, were then fit with projected, parametrized
representations of the two intrinsic functions.  This approach allowed
Trippe et al. to deal in a natural way with the observed alignment of
the proper motion velocities with Galactic coordinates beyond $\sim
6''$.  They were also able to take into account the overall rotation
of the cluster, which we ignored (for reasons that were justified
above).  However, Trippe et al.'s description is incomplete since it
leaves the third component of the velocity ellipsoid -- the component
along the line of sight -- unspecified.  The symmetries, if any, of
the velocity ellipsoid, or the orientation of its principle axes with
respect to the coordinate axes, were likewise unspecified.  Because
Trippe et al. did not construct complete models, it is impossible to
convert their fitted functions into unique estimates of the enclosed
mass.

These differences may account for the much lower BH mass found by
Trippe et al., $M_{\rm BH}\approx 1.2\times 10^6M_\odot$, although it
is not clear to us which of their assumptions most crucially affected
the results of the modelling.

Trippe et al. cited an argument by \citet{Kormendy1995ARA&A} to
explain their very low inferred value of $M_{\rm BH}$.  However we
were unable to follow the reasoning.  Kormendy and Richstone noted
that the effects of anisotropy on masses inferred from the Jeans
equation can be more significant if the central density gradient is
small.  We agree; but since Trippe et al. explicitly assumed isotropy
when writing the Jeans equation, the Kormendy \& Richstone argument
would not seem to apply.  In fact, we would argue that the expected
bias in the Trippe et al. mass should go in the opposite direction.
Trippe et al. assumed a perfectly flat core for the number density
distribution.  This is inconsistent with their assumption of isotropy,
since an isotropic population must have a density that rises at least
as fast as $r^{-0.5}$ near a BH.  Had they used a steeper central
profile, they would likely have inferred an even {\it smaller} BH
mass, for the reasons discussed above.

\section{Summary}

We have analyzed several years of adaptive optics assisted imaging of
the Milky Way nuclear star cluster in order to examine the cluster
kinematics.  The mass of the supermassive black hole Sgr\,A* and the
extended mass within 1\,pc of Sgr\,A* were estimated via isotropic and
anisotropic Jeans models. Our main results can be summarized as
follows.

\begin{enumerate}
\item The proper motions of more than 6000 stars could be measured
  over a FOV of $\sim\,40''\times40''$ centered on Sgr\,A*. The
  uncertainties of the proper motion velocities in both coordinates
  are $<25$\,km\,s$^{-1}$ for $80\%$ of the sources. The complete list
  is included as online material. Stars that have been identified as
  early-type are marked in this list because of their peculiar
  dynamical properties.
\item From a comparison of the proper motions of maser stars as
  measured in the infrared (stellar cluster at rest) and radio
  (Sgr\,A* at rest) reference frames we infer a non-significant
  relative motion of the radio frame relative to the IR frame of
  $0.4\pm6.4$\,km\,s$^{-1}$ toward east and $8.4\pm6.4$\,km\,s$^{-1}$
  toward south. The given errors are 1\,$\sigma$ uncertainties. This
  means that there is no detectable proper motion -- within the
  uncertainties of our analysis -- between the stellar cluster and the
  central black hole.
\item The projected radial and tangential velocity dispersions show a
  clear Keplerian dependence only in the innermost $\sim\,0.3$\,pc.
\item The velocity dispersion of the stars, after exclusion of the
  early-type stars, is consistent with being isotropic and constant
  within the measurement uncertainties at projected distances $R>10''$
  out to at least $30''$ from Sgr\,A*. A caveat is that the
    rotation signature of the cluster may mask underlying anisotropy.
\item  The NSC rotates parallel to Galactic rotation. This
  rotation implies that it must have at least partly formed by
  accretion of gas or star clusters from the galactic disk
  \citep[see][]{Seth2008ApJ}.
\item We analyze the proper motion data with the aid of isotropic and
  anisotropic models. The early-type stars with their well known
  peculiar dynamical behavior (rotation within disk-like structures)
  are excluded from this analysis. Both models lead to results that
  are consistent with each other. This supports the conclusion that
  the NSC is close to isotropic.
\item The proper motion data imply a best-fit black hole mass of
  $3.6^{+0.2}_{-0.4}\times10^{6}$\,M$_{\odot}$ (68\% confidence).  A
  black hole mass of $4.0\times 10^6M_\odot$ is consistent with the
  proper motion data at the 90\% level if the stellar velocities are
  modelled as isotropic, and at the 99\% level when anisotropy is
  allowed. This is the first time that a proper motion mass estimate
  of Sgr\,A* is consistent with direct mass estimates from individual
  stellar orbits.
\item The point mass of the black hole is {\it not} sufficient to
  explain the dynamics of the stars in the central parsec. An
  additional, extended mass is required.
\item The influence of the extended mass on the gravitational
  potential becomes notable at $R\gtrsim0.4$\,pc. When excluding --
  probably unphysical -- solutions in which the mass density decreases
  toward the black hole, our isotropic and anisotropic Jeans models
  require an extended mass of at least $0.5\times10^{6}º,M_{\odot}$
  within $r\leq1$\,pc of Sgr\,A*. If the mass density is proportional
  to the stellar number density then the stellar mass within 1\,pc of
  Sgr\,A* preferred by our model is $1.5\times10^{6}$\,M$_{\odot}$ in
  the isotropic case and $1.1\times10^{6}$\,M$_{\odot}$ in the
  anisotropic case, respectively.  The extended mass can be explained
  by the stellar mass of the cluster.
\item No strong statement can be made on the \emph{distribution} of
  the extended mass in the central parsec. Shallow or even declining
  mass densities are preferred by the modelling, but only weakly.  A
  mass density that declines as rapidly as $\rho\sim r^{-2}$ in the
  central parsec can be securely ruled out.
\end{enumerate}

%                                     Two column figure (place early!)
%______________________________________________ Gamma_1 (lg rho, lg e)

\begin{acknowledgements}
RS would like to acknowledge the Ram\'on y Cajal programme of the
Ministerio de Ciencia e Innovaci\'on, Spain.  DM was
supported by grants AST-0807910 (NSF) and NNX07AH15G (NASA). RS would
like to thank Andrea Ghez for enlightening discussions and helpful
comments on an early version of this paper.

\end{acknowledgements}

\bibliography{gc}

\appendix
\section{Alignment of stellar positions to a common reference
  frame \label{app:alignment} \label{app:align}}

\subsection{Reference frame}

We chose to base the reference frame on stellar positions obtained
from imaging observations on 1 June 2006. The data are characterized
by (a) the largest FOV of all observations used in this work due to
large dither offsets and (b) a very low number of saturated stars
because of the short integration time used. The data set consists of
80 randomly dithered frames, which allows, in addition, a thorough
determination of average stellar positions and their uncertainties via
multiple measurements.

In a first step, the lists of point sources identified in the
individual exposures (see section\,\ref{sec:astrometry}) were
registered relative to the first exposure in the series (which is
centered on Sgr\,A*) by identifying stars common to the frames and
determining the relative offsets between the exposures via a least
squares fit (uncertainty of the offsets: $< 0.01$\,pixel). The
individual lists of stars were subsequently merged to a preliminary
common list with average positions and corresponding standard
deviations. In order for a star to be included in the common list, it
had to be detected in at least 8 independent exposures, so that its
positional uncertainty could be reliably estimated and in order to
eliminate spurious detections.

\subsection{Selection of reference stars}

In a second step, transformation reference stars were selected from
the preliminary common list of stars. Reference stars were identified
by a four-step process: (a) brightness selection ($10.0 < \rm mag_{Ks}
< 17.0$); (b) selection of stars with positional uncertainty
$<0.15$\,pixel; (c) isolation of the stars: the magnitude difference
between a potential reference star and any other star within 8 pixels
($0.216''$, corresponding to $\sim\,3.5$ times the FWHM of the PSF)
must be at least 3 magnitudes; (d) uniform distribution: final
selection of reference stars on a $50\times50$\,pixel
($1.35''\times1.35''$) grid (in case there are several potential
reference stars per grid field, the star with the smallest positional
uncertainty was selected). We thus obtained 932 reference stars for a
combined FOV of $\sim40''\times40''$. The brightness selection
  excludes the brightest stars because they are saturated in some of
  the frames. The cut-off at the faint end was chosen because of the
  limited sensitivity of the frames of some observing epochs. The
  faintest stars may also be frequently affected by systematic offsets
  due to confusion with nearby resolved and unresolved sources in the
  dense GC field \citep[see][]{Ghez2008ApJ}.

\begin{figure}[!bh]
\centering
\includegraphics[width=\columnwidth]{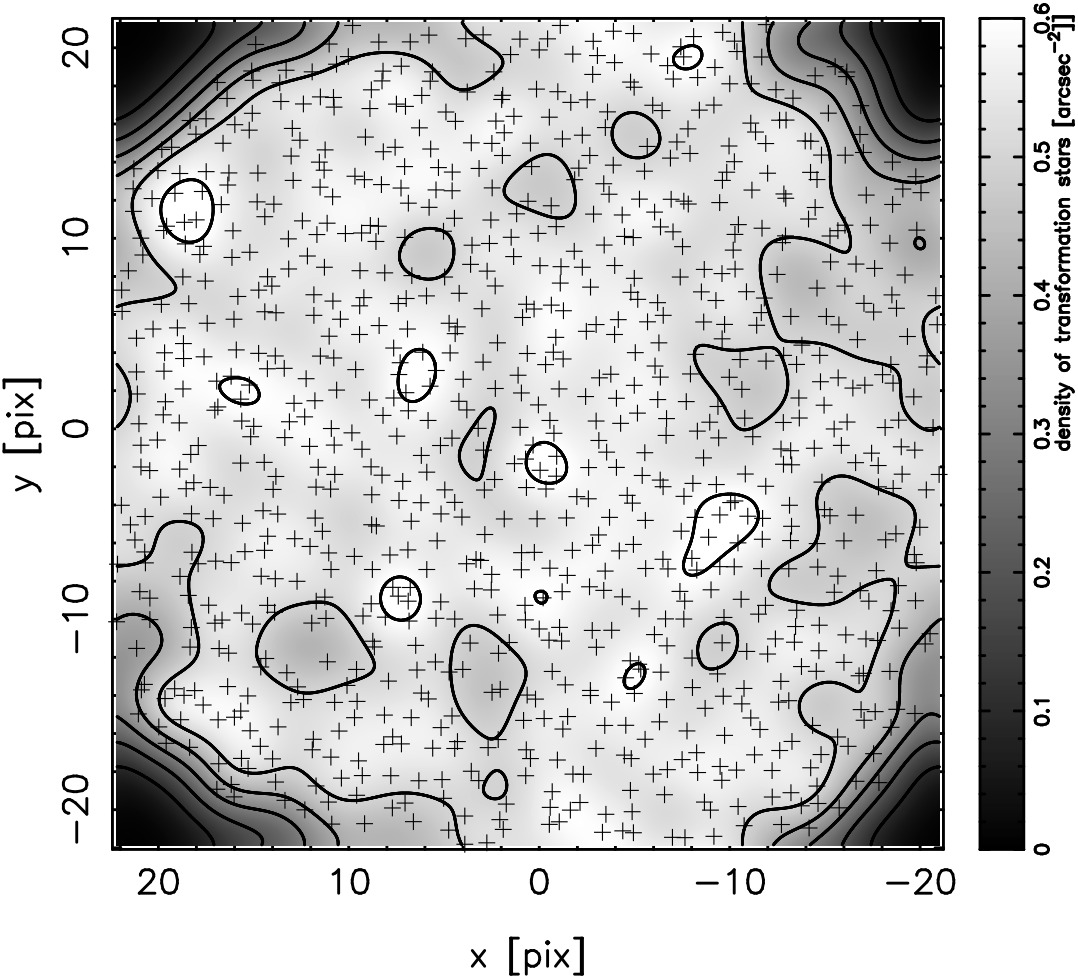}
\caption{\label{Fig:transdens} Transformation stars selected from the
  1 June 2006 observations are marked by crosses on the FOV obtained
  after combining the dithered exposures. The average density of
  transformation stars is indicated by the gray shading. Contour lines
  are drawn at $0.1\ldots 0.6$ stars per arcsec$^{2}$.}
\end{figure}

The last two steps in the selection process are important in order to
avoid being biased by fields of increased stellar density. In the case
of the GC, for example, the density of stars increases toward Sgr~A*
\citep[see][]{Schoedel2007A&A}. It is reasonable to assume that
  the camera distortion can be described by a well-behaved
  function. We believe that uniform sampling helps to avoid possible
  systematic errors. Non-uniform sampling of reference stars would for
  example lead to a significantly larger number of reference stars
  detected near Sgr\,A*. In the central arcseconds there exists a
  population of early-type stars with known rotation in the plane of
  the sky \citep[e.g., ][]{Paumard2006ApJ}. Also, stars in the central
  arcseconds are subject to larger systematic uncertainties of their
  positions because of the high stellar density and the presence of
  numerous bright stars in this region. \citet{Ghez2008ApJ} have
  demonstrated clearly the problem of astrometric errors due to the
  influence of the unresolved stellar population. The sampling of the
reference sources selected by the above described process is highly
uniform, with an average density of 0.5 reference sources per
arcsec$^{2}$ (see Fig.\,\ref{Fig:transdens}).

Due to possible camera distortions, which have at this point not yet
been taken into account, the positions of the reference stars can
still be biased and have increased uncertainties. For example, the
pixel scale may increase or decrease systematically toward the edges
and corners of the detector FOV.  Therefore, two effects can be
expected: (a) systematic deviations of the relative stellar positions
measured on different dithered exposures from the positions as they
would be measured with a distortion-free camera; and, as a
consequence, (b) increased uncertainty of the positions after
combining the source lists.

%We compared the relative positions of stars that are common to all
%exposures. Due to the dithering, they are positioned on different
%detector areas in each exposure. In this way we could estimate that
%the pixel scale of the NaCo S27 camera (epoch June 2006) changes in a
%systematic way by about $0.5\%$ across the detector, decreasing
%systematically toward the lower (south in the used alignment) and left
%(east) edges. Across a $1024\times1024$\,pixel detector this can lead
%to relative changes in position of a few pixels.

The systematic deviations of the stellar positions are difficult to
correct without precise knowledge of the camera distortions. However,
fortunately, if we relax the constraints on \emph{absolute}
astrometry, we can still get accurate measurements of the proper
motions. If we accept that there may be some residual distortion
present in the reference frame, all we have to do is to apply this
distortion \emph{consistently} to all epochs. We chose a 3rd order
polynomial transform for this purpose (see the following section).

The positional accuracy of the reference stars can be improved by
aligning the stellar positions from the 80 exposures of the reference
epochs via a 3rd order polynomial transform, instead of a simple shift
as done to obtain the preliminary combined list. Hence, the
astrometric lists for each exposure were aligned with the first
exposure in the series via such a transform.  The parameters of the
transform were established via a least squares fit, using the
reference stars common between the given frames.  
Finally, all astrometric lists of the individual exposures were
combined again to a common list. 

\begin{figure*}[!htb]
\includegraphics[width=\textwidth]{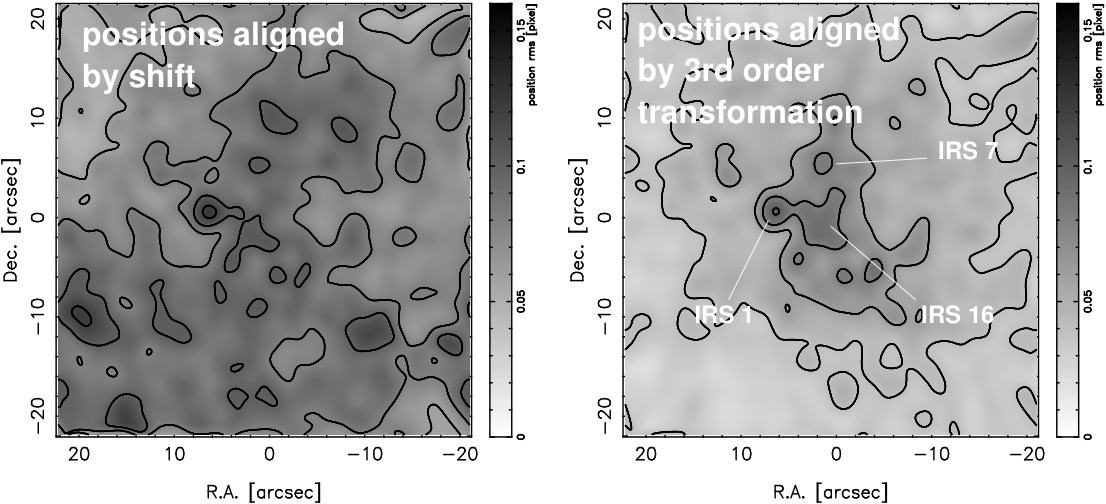}
\caption{\label{Fig:drmaps06} Smooth maps of the positional
  uncertainty of the stars detected in the reference frame from 1 June
  2006. After assigning the positional uncertainty of each star to its
  position, the map was smoothed by applying a Gaussian filter with a
  FWHM of $2.7''$. The left panel shows the uncertainty of the stellar
  positions in the FOV after combining the astrometric lists from the
  individual exposures by applying just shifts in x and y coordinates
  (preliminary combined list, simple mosaicing). The right hand panel
  shows the uncertainties after aligning the stellar positions via a
  3rd order polynomial fit (final combined list). Contour lines are
  drawn from $0$ to $0.16$\,pixel at intervals of 0.02\,pixel. The
  transformation reduces the overall uncertainty of the positions and
  eliminates systematic changes of the positional uncertainty due to
  camera distortions. The average combined uncertainty of all
  positions is $0.081$\,pixel before and $0.045$\,pixel after the
  polynomial fit. After applying a polynomial transformation to the
  stellar positions, the uncertainties correlate with the light
  density in the field. Light density variations arise from the
  combined effect of stellar density and the presence of bright
  stars. This can affect the astrometry via the photometric noise
    in the halos of the bright stars, especially since the used PSF
    was truncated (see section\,\ref{sec:astrometry}). The unresolved
    stellar population will contribute as well to increased
    photometric noise.  This can be seen when comparing the
  uncertainty map with the mosaic image shown in
  Fig.\,\ref{Fig:mosaic}. One can, for example, see that the areas
  around IRS\,7, IRS\,1, or the IRS\,16 cluster are clearly visible in
  the uncertainty map. The uncertainty of the stellar positions
  decreases with distance from Sgr\,A*, in correlation with the
  decreasing stellar density \citep[see][]{Schoedel2007A&A}.}
\end{figure*}

Note that the last step does not necessarily eliminate camera
distortions. The final reference frame may therefore still have
residual astrometric distortions.  However the last step increases
the precision of the stellar positions in the reference frame
significantly. We created smooth maps of the positional uncertainties
of the stars in the preliminary and final combined lists.  As we show
in Fig.\,\ref{Fig:drmaps06}, the map of the uncertainty of the stellar
positions in the final combined list reflects closely the
two-dimensional stellar light density. This is exactly what one would
expect after possible systematic offsets due to camera distortions
have been eliminated successfully and the astrometry of the sources is
almost exclusively affected by the structure of the stellar cluster
itself. This provides and important cross-check for the validity of
our determination of stellar positions in the reference frame.

\subsection{Transformation of positions into reference frame}

Having established a reference frame with a list of transformation
stars, the next step is to transform the stellar positions of the
stars detected in each exposure of each epoch to the reference frame.
About 400 (the FOV of individual exposures being smaller than the
combined FOV of the reference frame) reference stars could be
identified in each exposure of our data set. The positions of those
stars were then compared with the positions of the corresponding stars
in the reference frame in order to determine the parameters (via a
least squares algorithm) of a polynomial transformation into the
reference frame.

The IDL routine \emph{POLYWARP} was used to calculate the
transformation parameters between the stellar positions. The
\emph{POLYWARP} procedure is based on a least squares algorithm to
solve the coefficients of the following polynomial functions:
\begin{equation}
X_{i} = \sum_{i,j}Kx_{i,j}\cdot X_{o}^{j}\cdot Y_{o}^{i}
\label{eq:polyX}
\end{equation}
and
\begin{equation}
Y_{i} = \sum_{i,j}Ky_{i,j}\cdot X_{o}^{j}\cdot Y_{o}^{i}.
\label{eq:polyY}
\end{equation}
In these equations $X_{i}$ and $Y_{i}$ correspond to the positions of
the reference stars in the June 2006 reference frame, and $ X_{o}$ and
$Y_{o}$ are the corresponding positions in the given exposure.  We
chose a maximum value of 3 for the parameters $i$ and $j$.  The result
of the transform is not satisfying for $i,j\leq1$, with strong
systematic effects visible in the error maps (corresponding to the one
shown in Fig.\,\ref {Fig:drmaps06} ). The values $i,j\leq3$ adopted
here produce slightly more accurate results than $i,j\leq2$ (a
reduction of the average positional uncertainties by $5-10\%$). However, this
choice does not alter any of the results of this work in a significant
way. Choosing an even higher order for the transform is therefore
unnecessary and would reduce the precision with which the
transformation parameters can be computed.

After transformation of the lists of point sources for each individual
frame, the lists corresponding to a given observing epoch were
combined. Average fluxes and positions as well as the corresponding
uncertainties (the errors on the mean values) could be directly
derived from the multiple measurements for each star. Sources that
were not detected in multiple exposures were rejected as spurious
detections.

\begin{figure*}[!htb]
\includegraphics[width=\textwidth]{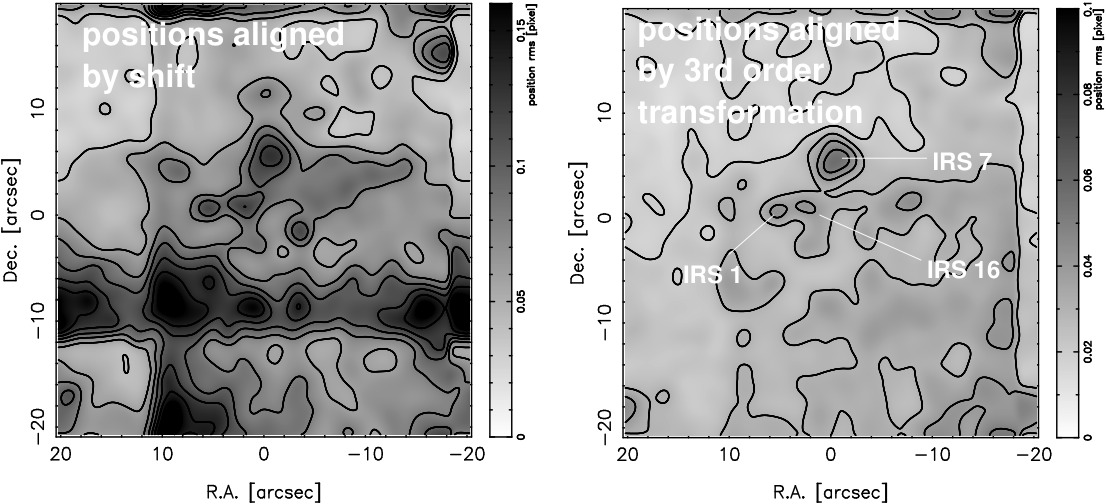}
\caption{\label{Fig:drmaps04} Smooth maps of the positional
  uncertainty of the stars detected in the reference frame from 12
  June 2004. The maps were created in the same way as the maps shown
  in Fig.\,\ref{Fig:drmaps06}. The left panel shows the uncertainty of
  the stellar positions in the FOV after combining the lists of
  detected stars by applying just shifts in x and y coordinates
  (simple mosaicing). The right hand panel shows the uncertainties
  after aligning the stellar positions with the reference frame via a
  polynomial fit. The transformation reduces the overall uncertainty
  of the positions and eliminates systematic changes of the positional
  uncertainty due to camera distortions. The average combined
  uncertainty of all positions is $0.073$\,pixel before and
  $0.027$\,pixel after the polynomial fit. The overall uncertainty is
  lower than for the stars in the reference frame because the quality
  of the AO correction was much better in the observing run on 12 June
  2004. Therefore, the correlation of the uncertainty after the
    polynomial transform with the density of stars and the presence of
    bright stars is less obvious than in the 2006 data. However, areas
    of increased positional uncertainty are clearly associated with
    the IRS\,7, IRS\,16, and IRS\,1. Please note that the different
    scaling of the maps in the left and right panels. The vertical
    strip of low uncertainties at the right edge of the FOV is an
    artifact due to the smoothing and the lack of sources in this area
    in the data from 12 June 2004 }
\end{figure*}

Similar to Fig.\,\ref{Fig:drmaps06}, Fig.\,\ref{Fig:drmaps04} shows
the astrometric uncertainties of the sources detected in the 12 June
2004 epoch after (a) just applying shifts in x and y to combine the
lists, and (b) after a full third order transformation into the
reference frame. Again, it can be seen how the full polynomial
transformation reduces the overall positional uncertainties and their
systematic changes across the field considerably.

\section{Offset field \label{app:offset}}

\subsection{Proper motions}

The methodology applied to extract proper motions from the offset data
set was identical to the one for the central field, except that no
astrometric positions for the stars were derived. The analysis of the
offset proper motions is based on three epochs only (August 2004, July
2005, and May 2008). Stars that were not detected in all three epochs
were rejected from the sample. The stellar positions were just
transferred roughly into the radio reference frame by applying
appropriate offsets.  We estimate that the uncertainties of the
stellar positions in the offset field may reach up to several $0.1''$.
Therefore the proper motions for the offset field are not included in
the list in Table\,\ref{Tab:list}. The distances of the stars from
Sgr\,A* were measured by using the NaCo pixel scale.

\begin{figure*}[!bh]
\includegraphics[width=\textwidth]{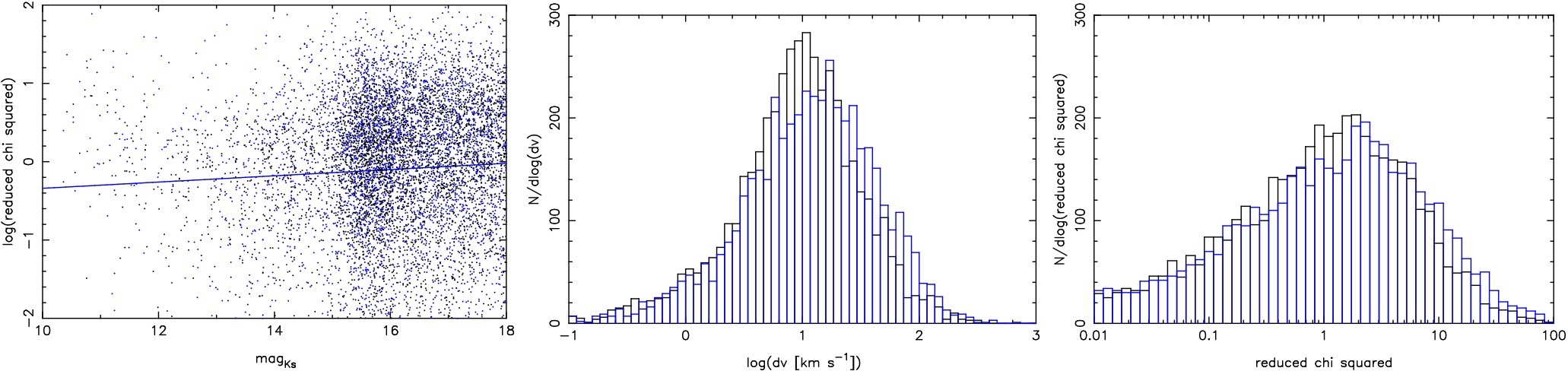}
\caption{\label{Fig:chi2dvoff} Error analysis for offset field
  data. Left: Plot of log($\chi^{2}$) vs.\ Ks-band magnitude. The
  straight line is a least square linear fit.  Middle: 
  Distribution of the velocity uncertainties, black for right
  ascension and blue for declination.   Right: Distribution of
  the reduced $\chi^{2}$ values for the linear fits of the data of
  position vs.\ time. The black histogram is for the fits in right
  ascension, the blue histogram for the fits in declination.}
\end{figure*}

The number stars with proper motions measured in the offset field,
after applying the selection criteria, is 4308.  In the left panel of
Fig.\,\ref{Fig:chi2dvoff} we show a plot of reduced $\chi^{2}$
vs.\ Ks-band magnitude for the proper motion fits of the offset field
data. There is not such a clear correlation between
$\chi^{2}_{red}$  and magnitude as in case of the center field data. We
believe that this is mainly caused by (a) the low number statistics
of the offset field data (only 3 data points per fit) and (b) by the a
factor $\sim\,2$ \citep[see][]{Schoedel2007A&A} lower stellar surface
number density in the offset field. The middle and right panels of
Fig.\,\ref{Fig:chi2dvoff} show plots of the distribution of 
$\chi^{2}_{red}$ and velocity uncertainty for the offset field data.

\begin{figure*}[!htb]
\includegraphics[width=\textwidth]{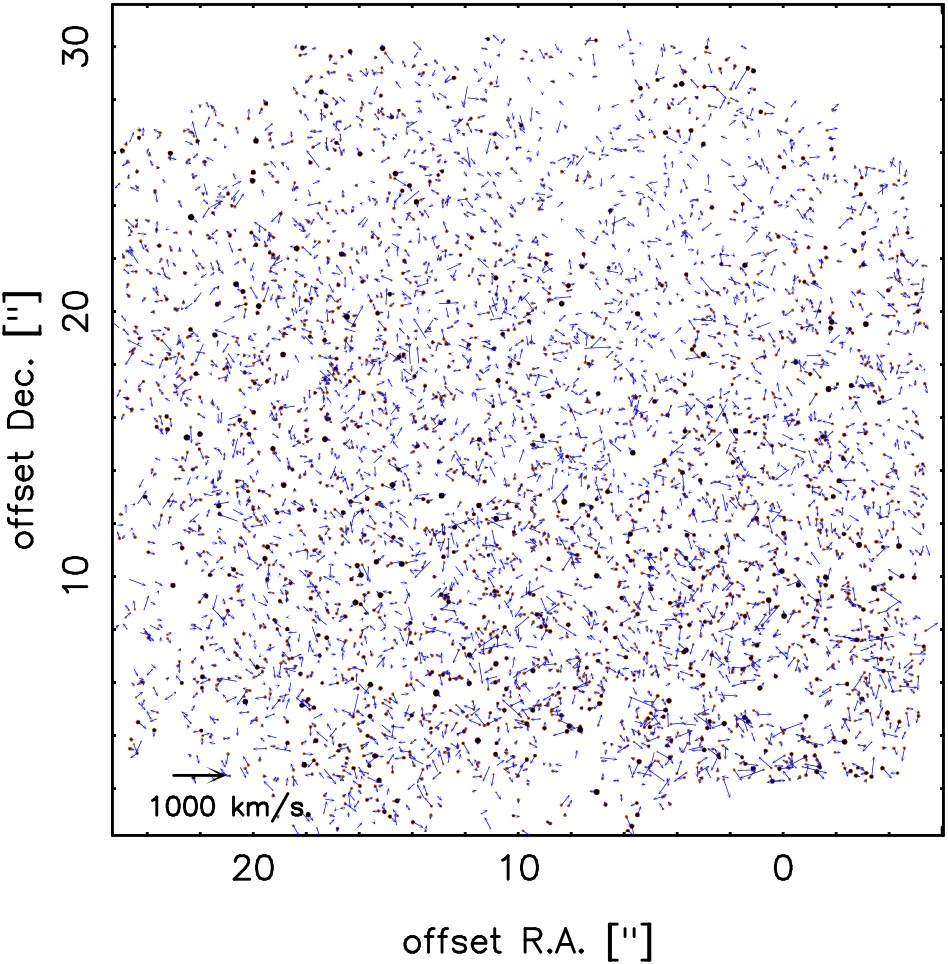}
\caption{\label{Fig:velmapoff} Map of stars and measured proper
  motions of stars in the GC offset field. North is up and east is to
  the left. Arrows indicate magnitude and direction of the proper
  motion velocities. The black arrow in the lower left corner
  indicates the length of a 1000\,km\,s$^{-1}$ arrow. Please note that
  the map is not strictly astrometric.}
\end{figure*}

Figure\,\ref{Fig:velmapoff} shows an image of the offset field with
measured stellar velocities indicated by arrows, after rejecting the
stars with the 5\% highest reduced $\chi^{2}$-values in order to avoid
outliers due to the small number of measured positions and thus rather
noisy statistics.

\subsection{Velocity dispersion}

\begin{figure*}[!htb]
\centering
\includegraphics[width=14cm]{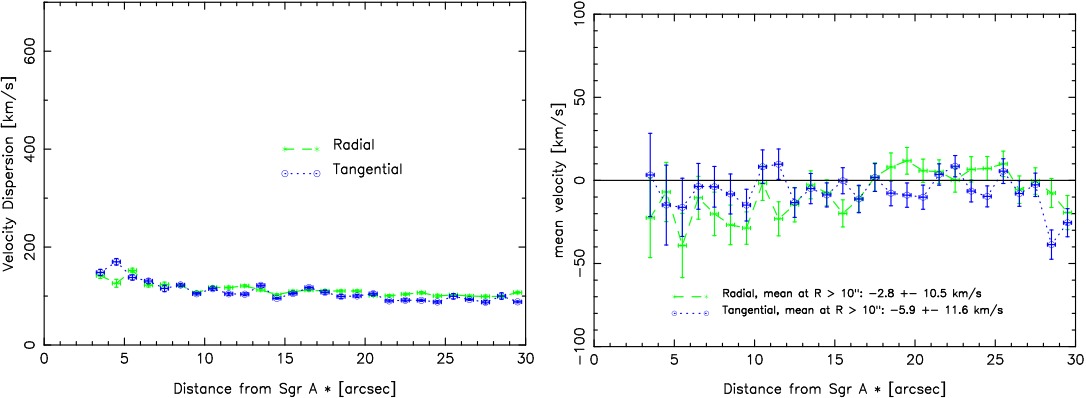}
\caption{\label{Fig:sigmaoff} Left: Projected radial (green) and
  tangential (blue) velocity dispersions in the GC nuclear star
  cluster for the offset field (see Fig.\,\ref{Fig:mosaicoff}). Right:
  Mean projected radial and tangential velocities vs.\ projected
  distance from Sgr\,A* for the offset field. }
\end{figure*}

We analysed the proper motion data for the offset field in the same
way as the data for the central field. The mean projected radial and
tangential velocities and the corresponding velocity dispersions are
shown in Fig.\,\ref{Fig:sigmaoff}. The offset data set covers only a
small angular section of the NSC and is highly incomplete at small
$R$. Also, there are only three epochs available for the offset
field. Therefore the quality of the offset data does not match the
quality of the data on the central field. They have not been used for
the further analysis in this work. The mean velocities show a larger
scatter than in case of the center field data.

However, we believe that it is important to include the offset data in
this work for two reasons. (a) They are centered on a different region
of the cluster and can thus serve to detect systematic errors related
to the alignment of the astrometric data with the reference epoch. We
find that the proper motions for all stars common to the center and
offset data agree within their $1\,\sigma$ uncertainties. (b) The
offset data sample slightly larger distances from Sgr\,A*. They
support the image of a close to isotropic cluster with a constant
velocity dispersion out to projected distances of $R=30"$
($1.14$\,pc).

\Online

{\footnotesize
\begin{longtable}{lrrrrrrrrrrrr}
\caption{\label{Tab:list} List of stars with measured proper motion in
  the GC. The last column contains the value 1 if a star is contained in the
  list of spectroscopically identified early-type stars of
  \citet{Paumard2006ApJ} (quality 1 and 2, their Table~2). It contains
  the value 2 if the star has been identified as an early-type
  candidate by the photometric analysis of Buchholz, Sch\"odel and
  Eckart (2009, submitted to A\&A).}\\
\hline
\hline
 ID & R$_projected$ & R.A. & $\Delta$R.A. & Dec.  & $\Delta$Dec. & mag$_{\rm Ks}$ &  $\Delta$mag$_{\rm Ks}$ & v$_{\rm R.A.}$ & $\Delta$v$_{\rm R.A.}$ &  v$_{\rm Dec.}$  &  $\Delta$v$_{\rm Dec.}$ & type \\
 & [$''$] & [$''$] & [$''$]& [$''$] &  &   & [km\,s$^{-1}$] & [km\,s$^{-1}$] &  [km\,s$^{-1}$] &  [km\,s$^{-1}$] & \\ 
\hline
\endfirsthead
\caption{continued.}\\
\hline
\hline
 ID & R$_projected$ & R.A. & $\Delta$R.A. & Dec.  & $\Delta$Dec. & mag$_{\rm Ks}$ &  $\Delta$mag$_{\rm Ks}$ & v$_{\rm R.A.}$ & $\Delta$v$_{\rm R.A.}$ &  v$_{\rm Dec.}$  &  $\Delta$v$_{\rm Dec.}$ & type \\
 & [$''$] & [$''$] & [$''$]& [$''$] &  &   & [km\,s$^{-1}$] & [km\,s$^{-1}$] &  [km\,s$^{-1}$] &  [km\,s$^{-1}$] & \\ 
\hline
\endhead
\hline
\endfoot
\end{longtable}
}

\end{document}